\documentclass[aps,preprint,nofootinbib,preprintnumbers,eqsecnum]{revtex4-1}
\pdfoutput=1

\usepackage{color}

\usepackage[
      pagebackref=false,
      colorlinks=true,
      linkcolor=blue,
      urlcolor=blue,
      filecolor=black,
      citecolor=red,
      pdfstartview=FitV,
      pdftitle={},
        pdfauthor={},
        pdfsubject={},
        pdfkeywords={},
        pdfpagemode=None,
        bookmarksopen=true
      ]{hyperref}

\usepackage[normalem]{ulem}
\usepackage{amsmath}
\usepackage{enumerate}
\usepackage{amsfonts}
\usepackage{yfonts}

\usepackage{subfigure}
\usepackage{psfrag}

\usepackage{epsfig}
\usepackage[latin1]{inputenc}
\usepackage{float}
\usepackage{graphicx}
\usepackage{cancel}
\usepackage{mathrsfs}
\usepackage{amssymb}
\usepackage{amsfonts}
\usepackage{amsmath}
\usepackage{slashed}

\usepackage{graphicx}
\usepackage{bm}

\def\({\left(}
\def\){\right)}
\def\[{\left[}
\def\]{\right]}
\def\<{\langle}
\def\>{\rangle}

\newcommand\half{{\ensuremath{\frac{1}{2}}}}

\newcommand\abs[1]{\ensuremath{\left\lvert{#1}\right\rvert}}

\newcommand\ket[1]{\ensuremath{\lvert{#1}\rangle}}

\newcommand{\be}{\begin{equation}}
\newcommand{\ee}{\end{equation}}
\newcommand{\bea}{\begin{eqnarray}}
\newcommand{\eea}{\end{eqnarray}}
\newcommand{\bwt}{\begin{widetext}}
\newcommand{\ewt}{\end{widetext}}
\newcommand{\nn}{\nonumber\\}
\newcommand{\bi}{\begin{itemize}}
\newcommand{\ei}{\end{itemize}}
\newcommand{\ben}{\begin{enumerate}}
\newcommand{\een}{\end{enumerate}}
\newcommand{\bca}{\begin{cases}}
\newcommand{\eca}{\end{cases}}
\newcommand{\bln}{\begin{align}}
\newcommand{\eln}{\end{align}}
\newcommand{\bst}{\begin{split}}
\newcommand{\est}{\end{split}}

\newcommand\al{{\alpha}}
\newcommand\ep{\epsilon}
\newcommand\sig{\sigma}
\newcommand\Sig{\Sigma}
\newcommand\lam{\lambda}

\newcommand\om{\omega}
\newcommand\Om{\Omega}

\newcommand\Ga{{\ensuremath{{\Gamma}}}}
\newcommand\de{{\ensuremath{{\delta}}}}
\newcommand\De{{\ensuremath{{\Delta}}}}

\def\th{{\theta}}

\newcommand\ov{\over}
\newcommand\ha{{\half}}

\def\le{\left}
\def\ri{\right}

\newcommand\sA{{\ensuremath{{\mathcal A}}}}

\newcommand\sD{{\ensuremath{{\mathcal D}}}}

\newcommand\sH{{\ensuremath{{\mathcal H}}}}

\newcommand\sM{{\ensuremath{{\mathcal M}}}}
\newcommand\sN{{\ensuremath{{\mathcal N}}}}

\newcommand{\vx}{{\vec x}}

\newcommand{\fR}{\mathfrak{R}}

\newcommand{\Vee}{v_{E}}

\begin{document}

\title {Spread of entanglement and causality
}

\author{ Horacio Casini}
\affiliation{Centro At\'omico Bariloche, 8400-S.C. de Bariloche, R\'io Negro, Argentina }
\author{ Hong Liu}
\affiliation{Center for Theoretical Physics,
Massachusetts
Institute of Technology,
Cambridge, MA 02139, USA }
\author{M\'ark Mezei}
\affiliation{Princeton Center for Theoretical Science,
Princeton University, Princeton, NJ 08544, USA }

\begin{abstract}

We investigate causality constraints on the time evolution of entanglement entropy after a global quench in relativistic theories. We first provide a general proof that the so-called tsunami velocity 
is bounded by the speed of light. 
We then generalize the free particle streaming model of~\cite{Calabrese:2005in} to general dimensions and to an arbitrary entanglement pattern of the initial state. In more than two spacetime dimensions the spread of entanglement in these models is highly sensitive to the initial entanglement pattern, but we are able to 
prove an upper bound on the normalized rate of growth of entanglement entropy, and hence the tsunami velocity. The bound 
is smaller than what one gets for quenches in holographic theories, which highlights the importance of interactions in the spread of entanglement in many-body systems. 
We  propose an interacting model which we believe provides an upper bound on the spread of entanglement 
for interacting relativistic theories. In two spacetime dimensions with multiple intervals, this model and its variations are able to reproduce  intricate results exhibited by holographic theories for a significant part of the parameter space. 
For higher dimensions, the model bounds the tsunami velocity at the speed of light. Finally, we construct a geometric model for entanglement propagation based on a tensor network construction for global quenches.

\end{abstract}

\today

\maketitle

\tableofcontents

\section{Introduction and summary}

Understanding the evolution of quantum entanglement in non-equilibrium processes such as thermalization is 
a question of much interest. Entanglement could reveal quantum correlations not easily accessible by other observables such as thermodynamic quantities or correlation functions, and thermalization provides a dynamical setting to study the generation and spread of entanglement between subsystems. 

A simplest physical context to probe this question 
is the evolution of entanglement entropy $S_{\Sig} (t)$ after a global quench in a conformal field theory (CFT), where one injects 
a uniform energy density in a very short time interval at $t=0$ and then lets the system evolve. 
Here $\Sig$ denotes the entangling surface, whose characteristic size $R$ will be taken to be much larger than the inverse equilibrium temperature, i.e.~$R \gg {1 / T}$.   
Recent studies of evolution of $ S_\Sig (t)$ for quench processes in $(1+1)$ dimensions as well as 
in holographic systems of general dimensions have revealed a ``universal''  linear regime~\cite{Calabrese:2005in,Hartman:2013qma,Liu:2013iza}\footnote{See~\cite{Ho:2015woa} for a proposed theory explaining this behavior.}
 \be \label{onne}
 \De S_\Sig (t)= \Vee s_{\rm eq} A_\Sig t , \qquad R \gg t \gg \ell_{\rm eq} \ .
 \ee 
Here $ \De S_\Sig (t)$ is the change of $S_\Sig (t)$ from its value at $t=0$, 
 $A_\Sig$ is the area of the entangling surface $\Sig$, and $s_{\rm eq} $ is the equilibrium 
 entropy density. $\ell_{\rm eq} \sim {1 /T}$ is the local equilibration time. More explicitly, we expect for a region
$A$ whose size $R_A$ is comparable to, but larger than $\ell_{\rm eq}$, that its entanglement entropy saturates at the thermal value after the local equilibration time, i.e.~
\be \label{smsta}
S_A = s_{\rm eq} V_A, \qquad   {\rm for} \quad  R_A   \gtrsim {1 \ov T} \quad {\rm and} \quad t \gg \ell_{\rm eq} 
\ee
with $V_A$ the volume of $A$.

In~\eqref{onne} $v_E$ is a constant of dimension velocity and depends on macroscopic properties of the state. With the speed of light set to $c=1$, for $(1+1)$ dimensions it was found~\cite{Calabrese:2005in} that
 \be \label{d2}
 v_E = 1, \qquad d=2 \,,
 \ee
 while for higher dimensional holographic systems at zero chemical potential~\cite{Hartman:2013qma,Liu:2013iza}, 
\be \label{emrp}
\Vee^\text{holo}  = {(\eta -1)^{\ha (\eta -1)} \ov \eta^{\ha \eta}} 
= \bca 
          {\sqrt{3} \ov 2^{4 \ov 3}} = 0.687 & d=3 \cr
          {\sqrt{2} \ov 3^{3 \ov 4}} = 0.620 & d=4 \cr
          \ha & d = \infty 
          \eca \ , \qquad \eta = {2(d -1)\ov d} \ .
\ee

\begin{figure}[!h]
\begin{center}
\includegraphics[scale=0.45]{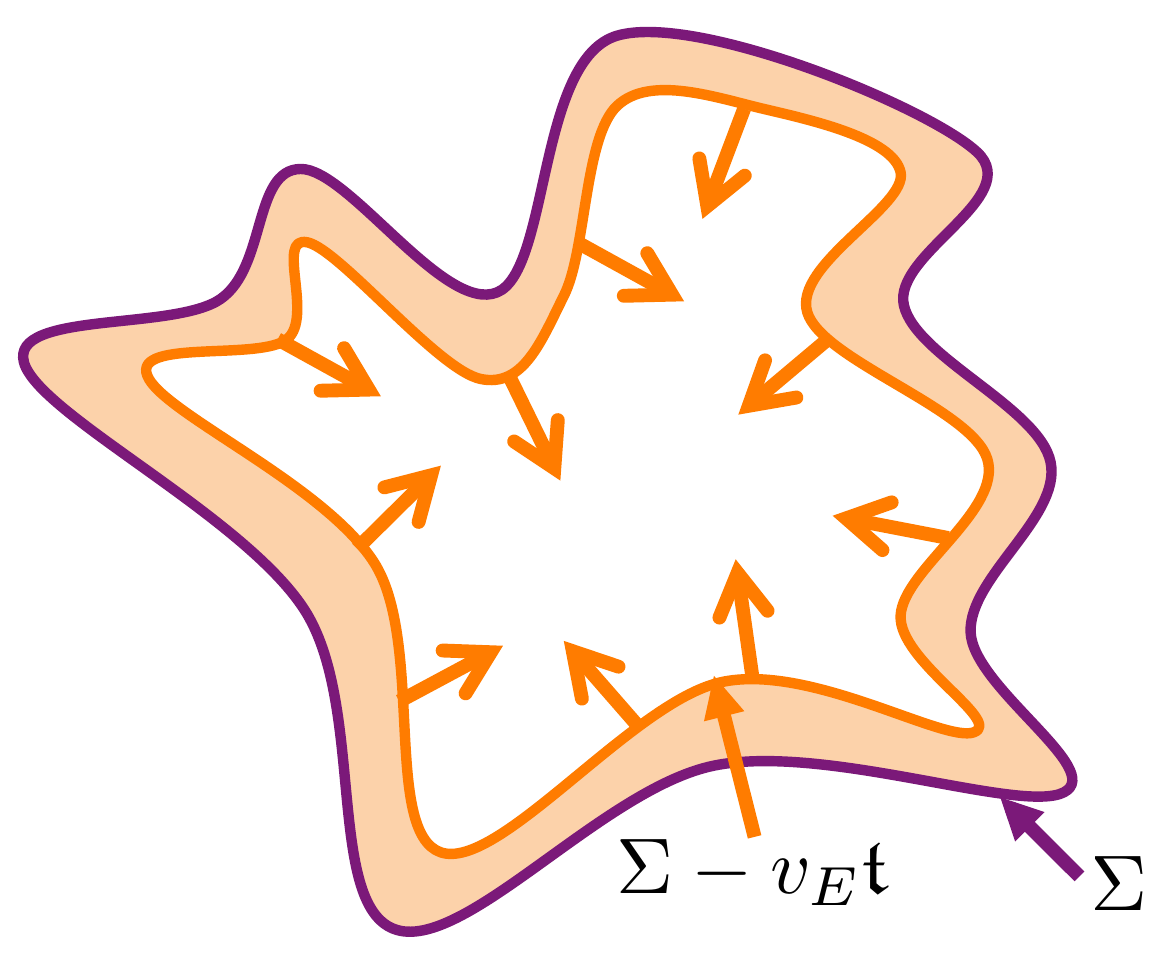} \quad
\end{center}
\caption{
The growth in entanglement entropy can be visualized as occurring via an ``entanglement tsunami" wave carrying entanglement inward from $\Sig$.  The region that has been covered by the wave (i.e.~the orange region in the plot) is entangled with the region outside $\Sig$, while the white region is not yet entangled. 
 }\label{fig:tsunami}
\end{figure}

Equation~\eqref{onne} suggests a simple heuristic picture for the growth of entanglement:  an entanglement wave propagates inward from the boundary of the entangled region, with the region covered by the wave becoming entangled with the outside region, see Fig.~\ref{fig:tsunami}. This was dubbed an ``entanglement tsunami"~\cite{Liu:2013iza} with $v_E$ interpreted as the velocity of the tsunami wave front. 
It was further observed in~\cite{Liu:2013iza} that for holographic models after local equilibration the normalized rate of growth
\be \label{groR}
\fR_{\Sig} (t) \equiv {1 \ov s_{\rm eq} A_{\Sig}} {d S_{\Sig} \ov dt}
\ee
appears to be bounded by~\eqref{emrp} 
\be \label{ineq}
\fR_{\Sig} (t)  \leq \Vee^\text{holo} , \qquad  t \gg \ell_{\rm eq} \ .
\ee
In the linear regime~\eqref{onne}, $\fR_\Sig$ is constant given by $v_E$, but in general it can have a complicated
time dependence. Note that  $d S_\Sig/dt$ cannot be compared meaningfully across different systems or regions of different size as it generally scales with the geometric size of $\Sig$ and the number of degrees of freedom of a system.  $\fR_\Sig$ was designed to provide an intrinsic measure for the rate of growth. With dimension of velocity, we expect that in relativistic systems $\fR_\Sig$  should be constrained from causality by some multiple of the speed of light.\footnote{Equation~\eqref{ineq} implies ${d S_\Sig \ov dt} \leq \# s_{\rm eq} A_\Sig$, which is reminiscent of the small incremental  entangling theorem for spin systems~\cite{Increment}. See~\cite{Avery:2014dba} for an attempt at a field theory approach to the problem.}

The simplicity and universality of the linear growth~\eqref{onne}  begs for an underlying physical 
mechanism. In particular, it would be desirable to relate  $v_E$ and $\fR_\Sig (t)$ to the speed of light,
and to develop some intuition about the physical origin of the value of $v_E$ in~\eqref{emrp}.

In this paper we first derive a formula which relates the tsunami velocity $v_E$ to the mutual information of certain spacetime regions.  The positivity of mutual information then leads to a proof that in relativistic theories $v_E$ is bounded by speed of light in all dimensions, i.e.~
\be \label{vbou}
v_E \leq 1 \ ,
\ee
although, as we will discuss in later sections, likely for $d >2$ the inequality  cannot be saturated. 
A different proof of~\eqref{vbou} has been found by  T.~Hartman~\cite{hartman}, which 
 uses the monotonicity of relative entropy.

We then consider various explicit models for entanglement propagation. 
Calabrese and Cardy~\cite{Calabrese:2005in}  proposed a simple free particle streaming model 
to explain the linear behavior~\eqref{onne} in $(1+1)$ dimensions. In this model, the injected energy density 
due to a global quench at $t=0$ is assumed to create EPR pairs of entangled quasiparticles which subsequently propagate freely. 
 At $t=0$,  entanglement correlations among quasiparticles are assumed to be local,\footnote{At the onset of linear regime, the scale of entanglement correlations should be controlled by $1/T$. It is a good approximation to treat them as strictly local when considering regions with $R \gg 1/T$.} which eventually spread to large distances
via free propagation of quasiparticles. In this model~\eqref{d2} comes from quasiparticles 
traveling at the speed of light. That~\eqref{onne} can be captured by such a simple model is remarkable and surprising. It appears to indicate 
that interactions do not play a role in the growth of entanglement, with the 
long-range entanglement of the final state solely coming from the spread of initial short-distance correlations. 
An indication that this success is likely an accident  is that 
for more than one intervals, the model fails to reproduce the qualitative behavior of both holographic and CFT results~\cite{Asplund:2013zba,Leichenauer:2015xra,Asplund:2015eha}.

The free streaming model can be generalized straightforwardly to a more general entanglement pattern of quasiparticles and 
to higher dimensions.  With no interactions, the wave function of the full system factorizes into those associated with each spatial point at $t=0$, and the 
propagation of entanglement is determined by the entanglement measure $\mu[A]$ for a subsystem $A$ in the Hilbert space of a single  point at $t=0$.
 In $d=2$, due to the special kinematics of one spatial dimension, propagation of entanglement is independent of the choice of $\mu[A]$, and results from the EPR 
 model of~\cite{Calabrese:2005in} are in fact general. This is no longer so in higher dimensions. Essentially all aspects of propagation depends on $\mu[A]$.\footnote{We study these aspects in detail for various choices of $\mu [A]$ in Appendix~\ref{sec:details}.
 For example, while one again finds linear growth~\eqref{onne} at early times, the tsunami velocity $v_E$ now depends on $\mu[A]$.}

A particularly interesting choice of $\mu[A]$ is what we will refer to as a random pure state measure (RPS), for which the entanglement entropy for a subsystem $A$ is proportional to its size, i.e.~
\be
\mu_\text{RPS}(A) \equiv s \,{\rm min}(V_A,V_{\bar{A}})\,, \label{RPSd}
\ee
where $V_A, V_{\bar A}$ denote the volume for $A$ and $\bar A$ (complement of $A$) 
respectively, and $s$ is a constant (which depends on the specific system).  The measure is motivated from the result of~\cite{Page:1993df} where it was found that the average entropy for a subsystem (with size smaller than half of the total system) in a random pure state is to a very good approximation given by its size. The rough intuition behind RPS is that for a  sufficiently ``equilibrated'' system, all degrees of freedom are entangled with one another in an equal way, and thus the entanglement entropy of a subsystem is proportional to the number of degrees of freedom it contains. 

Using the strong subadditivity property one can show that the RPS measure in fact provides an upper bound for the propagation of entanglement 
among all free streaming models, leading to an upper bound for $\fR_\Sig (t)$ (thus an upper bound for $v_E$)
\be  \label{groRBound}
\fR_{\Sig} (t) \leq v^{\rm free}_E  \,,
\ee
where 
\be \label{vstrem}
v^{\rm free}_E = { \Ga({d-1 \ov 2}) \ov \sqrt{\pi}  \Ga ({d \ov 2}) } 
=\bca 1 &  d=2 \cr
          {2 \ov \pi} = 0.637 & d=3 \cr
          \ha & d=4 \cr
           \sqrt{2 \ov \pi d}  & d = \infty 
          \eca 
\ 
\ee
is calculated using the RPS measure. 
Note that $v^{\rm free}_E$ is smaller than $1$ for $d > 2$, because quasiparticles propagate in different directions. Comparing with~\eqref{emrp} note that 
\be 
 v^{\rm free}_E <  \Vee^\text{holo} , \qquad d \geq 3 \ . 
 \ee
In other words, in higher dimensions, the spread of short-distance correlations limited by causality
cannot account for the result~\eqref{emrp} for strongly coupled systems. Thus 
interactions must play a role. 
The ratio ${v^{\rm free}_E / \Vee^\text{holo}}$ decreases with $d$ for $d \geq 3$, i.e.~the higher the 
spacetime dimension, the more significant role of interaction is. In particular, 
as $d \to \infty$, $v^{\rm free}_E \to 0$, while $\Vee^\text{holo} \to \ha$. 

This then motivates to introduce interactions in quasiparticle propagation. 
With interactions we then immediately face the problem of characterizing
the quantum state of an interacting many-body system.  Instead of confronting this very difficult problem
directly, here we seek a qualitative understanding of how linear growth can arise in an interacting system
and how interactions can enhance the spread of entanglement. For this purpose, we will consider the infinite scattering 
limit, i.e.~we assume that scatterings among constituents of a system are so efficient that the typical scattering 
time can be taken to be zero compared to the scales of entangling region size $R$ and time $t$ of interests.  
Holographic systems after a quench in the limit of  equilibrium temperature $T \to \infty$ can be considered 
as an example, as there the local scattering rate is controlled by $T$. In the infinite scattering limit, the 
evolution simplifies as interactions do not introduce additional scales into the problem. 

We propose a very simple model which applies  the RPS measure~\eqref{RPSd} to certain spatial regions 
determined from the causal structure associated with the entangling surface. The use of RPS measure is natural in the infinite 
scattering limit as in this limit interactions are extremely efficient in redistributing entanglement. 

The model, to be referred to as the maximal RPS 
model, appears to capture gross features of entanglement spread, including the linear growth, of holographic systems in the $T \to \infty$ limit. 
In fact in $d=2$, it does much better than expected.\footnote{For multiple intervals, results from our model coincide with a phenomenological formula recently proposed in~\cite{Leichenauer:2015xra} to capture holographic results.}  Applying it to a single interval in $d=2$ we again find~\eqref{d2}. For two intervals, the model precisely recovers holographic results. For three and four intervals, we find the model (and its slight variations) reproduces intricate entanglement patterns exhibited by holographic systems for a significant part of parameter space. For general $d >2$, this model gives in all dimensions $v_E = 1$.

We also show that the failure in reproducing holographic results for certain regions of parameter space for three and four intervals 
in $d=2$, as well as $v_E =1$ for $d>2$ can be attributed to the fact that our model can violate the strong subadditivity condition and/or 
does not take full account of causality constraints.

Given the already remarkable success of the maximal RPS model, in an attempt to better understand how quantum information is organized in holographic systems, we construct another model for the evolution of entanglement entropy that is inspired by the tensor network description of state evolution after a quench.
This interacting model satisfies many natural geometric criteria on the entropy function, including strong subadditivity, and reproduces the holographic results for multiple intervals in $d=2$. However, it does not provide a stronger constraint than $v_E \leq 1$ on the tsunami velocity for $d> 2$.
 It would be very interesting to find further criteria on the entropy function for the time evolution generated by a local Hamiltonian. 
  
The plan of the paper is as follows. In the next section we provide a proof that the tsunami velocity is bounded by the speed of light. In particular, we derive a formula which relates the tsunami velocity $v_E$ to the mutual information of certain spacetime regions. 
As a preparation for later discussions, in Sec.~\ref{sec:rps} we introduce the RPS measure for the entropy of subsystems and show that it is an upper bound for the entropy of systems of particles homogeneous in space. In Sec.~\ref{sec:bound} we introduce free streaming models of entanglement propagation and discuss some explicit examples. A general upper bound on ballistic propagation of entanglement in these free streaming models is proven in Sec.~\ref{sec:bound1}. The proof demonstrates that these models cannot account for the velocity of entanglement propagation in holographic theories. In Sec.~\ref{interacting} we present an interacting model 
and apply it to various examples. It indicates  that interactions can increase the tsunami velocity in higher dimensions and gives an upper bound for this velocity in relativistic theories. We also discuss shortcomings 
of the model. 
In Sec.~\ref{sec:tensor} we develop yet another description of entanglement evolution that is inspired by tensor network constructions. This model has the advantage of satisfying the strong subadditivity constraints and reproducing the right entanglement pattern for holographic theories in $d=2$. In this context we also show
 the holographic result gives an absolute upper bound on the entropy after a quench for relativistic theories in $d=2$.
In Sec.~\ref{sec:tus} we further examine a relation discovered in Sec.~\ref{sec:vproof} between tsunami velocity and mutual information in the context of various models discussed in earlier sections. 
Some more technical results and proofs are given in the appendices.  
In particular in Appendix~\ref{sec:details} we work out many aspects of ballistic propagation of entanglement in general dimensions, including mutual information and finite volume effects, for different entanglement patterns
of the initial state.

\section{Proof of a general upper bound on the tsunami velocity}\label{sec:vproof}

For $d > 2$, so far the linear growth~\eqref{onne} has only been established for holographic systems~\cite{Hartman:2013qma,Liu:2013iza}. In particular, both the linear growth~\eqref{onne} and tsunami velocity $v_E$ are independent of the shape of entangling surface $\Sig$~\cite{Liu:2013iza}. 
Assuming this behavior persists  for general systems, here we prove that 
that the tsunami velocity $v_E$  is bounded by the speed of light.

\begin{figure}[!h]
\begin{center}
\includegraphics[scale=0.8]{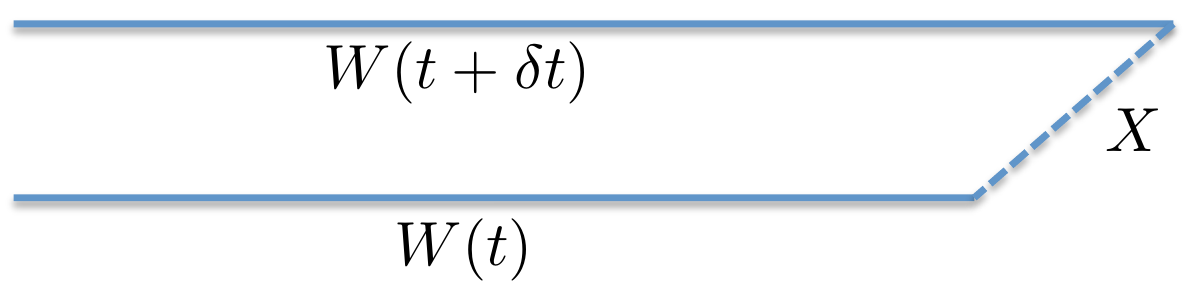}
\end{center}
\caption{A half space at time $t$ and another at time $t+\de t$. The null region $X$ connects the boundaries of these two regions. The horizontal direction is $x_1$ and vertical direction is time. Directions parallel to the boundary are not shown. }
 \label{uno}
\end{figure}

Since~\eqref{onne} is shape-independent, it is enough to consider $\Sig$ a straight hyperplane, i.e.~the entangling region is a half space.  
 Let us consider at time $t$ a half-space $W(t)$ whose boundary is perpendicular to the $x_1$-direction and another half space $W(t+\de t)$ at time $t+\de t$. Since entropy in the quench evolution is translation invariant, we can take $W(t+\de t)$ infinitesimally displaced in the $x_1$ coordinate such that the boundaries of these two regions $W(t)$ and $W(t+\de t)$ are connected by a strip of a null plane $X$ (see Fig.~\ref{uno}). 
Consider the mutual information between $W(t)$ and $X$
\bea
I(W(t),X) & =&   S(W(t))+S(X)-S(W(t)\cup X) \cr
&=&  S(X) - \le[S(W(t+\de t))-S(W(t)) \ri] \cr
&=&  S(X) - v_E s_{\rm eq} A_\Sig \de t \,,
 \label{mumu}
\eea
where in the second equality we used $S (W(t)\cup X) = S (W(t + \de t))$ and in
the third line the linear regime formula~\eqref{onne} for $S (W (t))$ and $S (W(t + \de t))$. 

Here we are interested in an excited state with vacuum entanglement subtracted, thus all quantities
in~\eqref{mumu} should be understood as so. 

Equation~\eqref{mumu} then gives 
\be \label{nerc}
v_E = {S (X) - I(W(t),X) \ov s_{\rm eq} A_\Sig \de t }\,,
\ee
and from non-negativity of the mutual information
\be
v_E \leq {S (X)  \ov s_{\rm eq} A_\Sig \de t } \ .
\label{prono}
\ee   
It should be emphasized that here we are saying the mutual information with the vacuum part subtracted 
should also be non-negative. This can be justified since 
we are considering only contributions linear in $s_{\rm eq}$. We can take 
$s_{\rm eq}$ large so the term proportional to $s_{\rm eq}$ in the mutual information 
has to be non-negative by itself.\footnote{ Note that in the vacuum, there are divergences associated with the sharp corner between $W(t)$ and $X$, which could ruin an inequality like~\eqref{prono}.  In the same spirit, peculiarities for the entropy of sharp null surfaces in interacting theories~\cite{bousso} do not apply here since  in the present context all geometries can be thought of as having a minimum curvature radius $1/T$.}

\begin{figure}[!h]
\begin{center}
\includegraphics[scale=0.8]{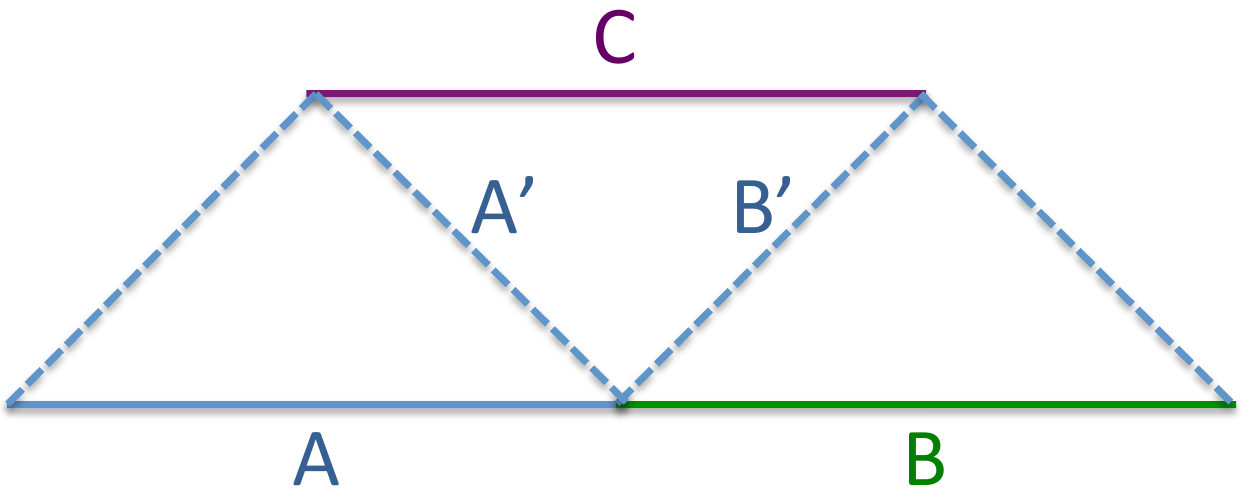}
\end{center}
\caption{The small strip regions $A$, $B$, and $C$ have all the same width. The null regions $A^\prime$ and $B^{\prime}$ are included in the 
causal development of regions $A$ and $B$ respectively.}
 \label{two}
\end{figure}

Now consider the three strip like regions $A$, $B$ and $C$ of figure \ref{two}. We take these regions to have small width $1/T \ll \de x = 2 \de t \ll t$. Hence, from~\eqref{smsta} the entropies for these regions are given by 
\be
S(A)=S(B)=S(C)=s_{\textrm{eq}} A_\Sig \de x\,, 
\ee   
with $A_\Sig$ the area of the plane bounding the strips. 
From this volume law and the corresponding one for the strip $AB$, $S(A\cup B)=2 s_{\textrm{eq}} A_\Sig \de x$ it follows that 
\be
I(A,B)=S(A)+S(B)-S(A\cup B)=0\,.
\ee
The null regions $A^\prime$ and $B^\prime$ are included in the causal domain of dependence of $A$ and $B$ respectively. Then, because of the monotonicity of  mutual information, the mutual information between the null strips $A^\prime$ and $B^\prime$ must also vanish
\be \label{u1}
0 \leq I(A^\prime, B^\prime)=S(A^\prime)+S(B^\prime)-S(A^\prime\cup B^\prime) \leq I(A,B) =0\,.
\ee  
Now given that $S(A^\prime\cup B^\prime)=S(C)$ and that by symmetry the entropy in the two null strips is the same, 
from~\eqref{u1} we find that 
\be \label{null1}
S(A^\prime)=S(B^\prime)=s_{\textrm{eq}} A_\Sig \frac{\de x}{2} = s_{\textrm{eq}} A_\Sig \de t 
\quad \implies \quad S (X) = s_{\textrm{eq}} A_\Sig \de t ,
\ee
i.e.~the entropy of a small null strip is equal to the thermal entropy of its projection to a constant time slice. 
Plugging~\eqref{null1} into~(\ref{prono}) we then get
\be
 v_E \leq 1 \,, \label{vbou2}
\ee
which proves that for any relativistic system $v_E$ must be bounded by the speed of light. A different proof of~\eqref{vbou2} has been found by  T.~Hartman~\cite{hartman}, which uses the monotonicity of relative entropy.

Plugging the expression for $S(X)$ into~\eqref{nerc}, we also find 
\be
v_E=1-\frac{I(W(t),X)}{s_{\textrm{eq}} A_\Sig \de t} \label{ve} \ . 
\ee
Equation~\eqref{ve} is an instructive formula which relates the deviation of a tsunami velocity from the speed of light to the entanglement between $W(t)$ and the null surface $X$. 
In particular, it says for $v_E$ to be equal to $1$, $W(t)$ and $X$ have to be unentangled.

\section{Random pure state measure} \label{sec:rps}

For the rest of the paper we investigate the propagation of entanglement in various free streaming and 
interacting models. Before doing that, we digress here to discuss the random pure state measure, which will 
play a crucial role in our later discussions. 

\begin{center}
\begin{figure}[!h]
\includegraphics[scale=0.4]{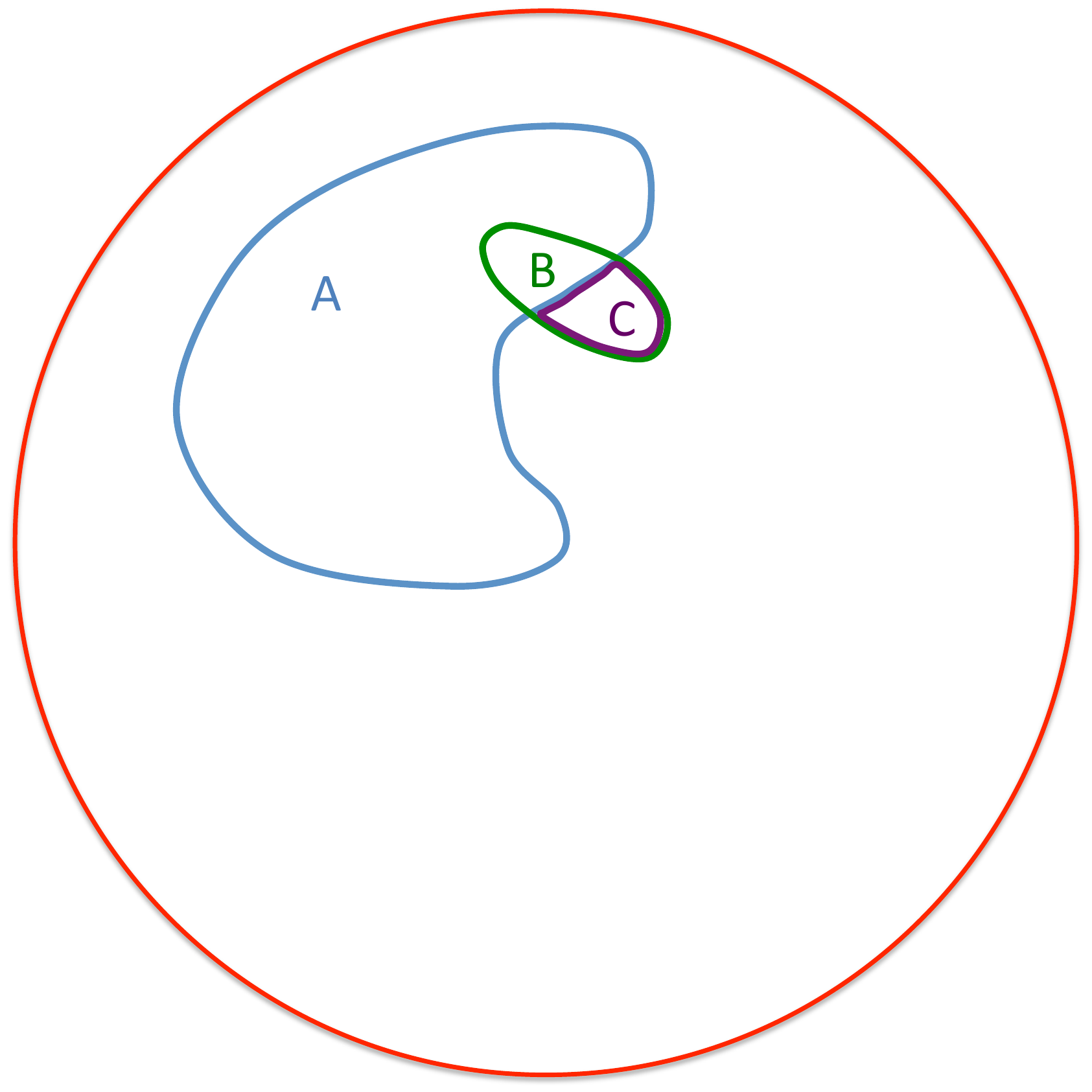}
\caption{ Regions $A, B, C$ on $\sM$. \label{fig:SSA}
}
\end{figure}
\end{center}

Consider a gas of particles on a spatial manifold $\sM$ in some pure state.
We assume that the state is homogeneous, i.e.~the particles are uniformly distributed on $\sM$. Now consider the entanglement entropy $S (A)$ for a subregion $A$. We assume that $S (A)$ satisfies the condition that as the size of $A$ goes to zero 
\be \label{cond1}
\lim_{A \to 0} S (A)= s V_A   \,,
\ee
where $s$ is a constant. We will comment on the motivations for this condition a bit later. 
We  first show that given~\eqref{cond1} the entanglement entropy $S(A)$ is bounded by 
\be
S (A) \leq \mu_\text{RPS}(A) \equiv s \,{\rm min}(V_A,V_{\bar{A}})\,, \label{asser}
\ee
where $V_A, V_{\bar A}$ denote the volume for $A$ and $\bar A$ (complement of $A$) 
respectively. We will refer to $ \mu_\text{RPS}[A]$ as a random pure state measure. 
We apply the strong subadditivity condition~\cite{Lieb:1973cp} to regions $A,B,C$ shown in Fig.~\ref{fig:SSA} to get the inequality:
\be 
S(A ) + S(B\cup C) \geq S (A \cup C) + S(B)\,.
\ee
Using that $B,C$ are infinitesimal, 
we then have 
\be \label{nnp}
S (A \cup C)  - S (A) \leq S(B\cup C) - S(B) = s V_C  \,,  
\ee
where we have used~\eqref{cond1} on right hand side of the equality. 
This inequality holds for any region $A$ and any other infinitesimal region $C$ outside $A$. It implies that as we increase the size of a region the entropy cannot increase faster than the volume of the region times $s$.\footnote{ 
Similarly, from $S(A)=S(\bar{A})$ it also follows that $S(A) - S (A \cup C)  \leq s V_C$,  
i.e.~as we increase the size of a region the entropy cannot decrease faster than the volume  times $s$.} 
Therefore, for any region $A$ with $V_{A}\leq V/2$ where $V$ is the total volume 
we necessarily have $S(A)\leq \mu_\text{RPS}(A)=s \,V_A$, while  for a region with $V_{A}>V /2$ we have $S(A)=S(\bar{A})\leq \mu_\text{RPS}(\bar{A})=\mu_\text{RPS}(A)$, proving our assertion (\ref{asser}). 

Let us now elaborate on the condition~\eqref{cond1}. Suppose that as $V_A \to 0$, $S_A$ is instead given by 
$S_A \sim V_A^\al$ with $\al \neq 1$. Then for $V_B , V_C \sim \ep \to 0$, 
we find that 
\be \label{scl}
S (A \cup C)  - S (A) \leq S (B \cup C) - S(B) \propto \ep^{\al}   \ .
\ee
For the case of super-volume behavior, $\al > 1$, since $\ep^{\al}/\ep \to 0$ as $\ep \to 0$, we conclude that 
$S(A)$ is identically zero for all $A$. Clearly, this is unphysical. 

For $\al < 1$, for small $A$, $S_A$ can increase as $V_A^\al$ which is faster than~\eqref{cond1}. This is indeed possible, but if this power law behavior continues to large sizes no volume law or equilibrium entropy after a quench can be achieved. In fact, $S_A/V_A\rightarrow 0$ for large volume if $\al<1$. Hence, our condition \eqref{cond1} simply means that for the quenched systems we are interested in, which have large energy density with respect to subsystems sizes and times involved in the dynamics, we are assuming a local equilibrium has already taken place for small sizes and times of order $1/T$.    

The condition~\eqref{cond1} and the random pure state measure~\eqref{asser} are also reminiscent of~\cite{Page:1993df} where it was found that the average entropy for a subsystem (smaller than half size of the total system) in a random pure state is to a very good approximation given by its size.

\section{Free streaming models in general dimensions}
\label{sec:bound}

\subsection{Setup}

Here we consider a model of free propagation of entanglement after a quench at $t=0$. We assume the system is homogeneous, isotropic and is in a pure state after the quench. 
We take the initial state at $t=0_-$ as unentangled, and require that in equilibrium (defined as $t \to \infty$) the entanglement entropy is proportional to the volume of a region.  
We make the following assumptions on the quench and the subsequent propagation of entanglement:

\ben

\item The quench generates a large amount of 
short-distance entanglement correlations which subsequently propagate freely.  
We will assume that the quench time interval is negligible and the initial entanglement correlations can be taken 
to be local. In other words, at $t=0$, each point acts as an {\it independent} source of entanglement correlations, which then spread freely, 
limited only by causality. At time $t$, the entanglement relations from $\vec x =0$ spread at most to 
the sphere $|\vx | = t$. 

\item For simplicity, we will assume that the correlations are concentrated on the light cone. It should be straightforward to generalize the discussion to include entanglement correlations inside the light cone, which we expect not to change qualitatively the physical picture and 
the upper bound on the propagation derived below.

\item  There is no interactions/interference among light cones 
from different points. This implies that throughout the evolution different light cones can be associated with independent wave function factors inherited from those at the origin of each light cone at $t=0$.  Furthermore,  entanglement correlations on each light cone do not evolve 
with time. More explicitly, consider a region $A$ on a light cone from $\vec x =0$ with {\it fixed angular extension}
with respect 
to the origin.  Denote the entanglement entropy of $A$ with respect to the rest of the light cone as $\mu [A]$. We assume that $\mu [A]$ is independent of time. 

\een
Note that the model does not assume propagation of particles, only free 
propagation of entanglement correlations. The model is fully specified by the entanglement measure
$\mu [A]$ on a light cone. When considering  certain specific realizations of $\mu [A]$ it will often  be useful to 
consider quasiparticles such as the EPR pair and GHZ block examples discussed below. We could also simply postulate a $\mu [A]$ which satisfies all the properties of entanglement entropy as in the example of random pure state measure discussed below.

Given that each light cone is independent, the time evolution of the entanglement entropy of the region $\sA$ enclosed by a surface $\Sig$ can be obtained by summing over the entanglement entropy of the parts of the light cones intersecting this region, i.e.~
\be \label{masf}
S_\Sig (t) = \int d^{d-1} x \, \mu [L_{\Sig} (\vx; t)]\,,
\ee
where $L_{\Sig} (\vx, t)$ denotes the region(s) of the light cone with center $\vx$ lying inside $\Sig$ at time $t$. $\mu [L]$ is zero if $L$ is an empty set (i.e.~no intersection).

The  entanglement measure $\mu [A]$ for a region $A$ on a light cone from $\vec x$ can be interpreted as the entanglement entropy for $A$ by tracing out degrees of freedom outside $A$ within the Hilbert 
space of $\vec x$.  It should satisfy 
all the properties of entanglement entropy for a pure state, including for example,
\be 
\mu [A] = \mu [\bar A] \,, \label{purity}
\ee
where $\bar A$ denotes the complement of $A$ on the light cone, and the strong subadditivity condition
\be \label{strm}
\mu [A] + \mu [B] \geq \mu [A \cap B] + \mu [A \cup B]
\ee
for any region $A$ and $B$. 

We also assume that measure $\mu [A]$ for small $A$ (even with several disconnected pieces) is proportional to the normalized area $\xi_A$ of that region, which we already motivated around~\eqref{scl} in the last section.
More precisely, for a region $A$ included in a spherical cap of angular size $\Delta \theta$ 
\be \label{apro}
\lim_{\Delta \theta \rightarrow 0}\frac{\mu [A]}{\xi_A} =s\,, \hspace{1cm}  \xi_A \equiv {\om_A   \ov \om_{d-2}}\,. 
\ee 
Here $s$ is a constant and $\om_A$ denotes the volume of region $A$ on a unit sphere and $\om_{d-2}$ is the volume of a unit $(d-2)$-sphere. In our current context, we will see in Sec.~\ref{sec:eqv} that imposing~\eqref{apro}  is equivalent to the requirement of a final equilibrium state with a volume law entropy distribution. In fact, the final equilibrium entropy density is precisely given by the constant in (\ref{apro}), i.e.~$s_{\rm eq}=s$.  

By definition, $\mu[A]$ and thus $s$ has the dimension of an entropy density, i.e.~$1/{\rm volume}$. Given that $s$ is the only scale of the system, for a scalable surface $\Sig$, on dimensional grounds we can write $S_\Sig$ in a scaling form
\be \label{scaling0}
S_\Sig (t) = s R^{d-1} f (t/R)\,,
\ee
where $R$ is a characteristic size of $\Sig$, or equivalently
\be
S_{\lam\Sig}(\lam t)=\lam^{d-1} S_{\Sig}( t)\,, \label{scaling}
\ee
where $\lam\Sig$ is $\Sig$ rescaled by a factor $\lam$. This scaling relation is also satisfied by holographic 
systems in the large size and long time limit, as we will discuss more in Sec.~\ref{sec:infinite}. 
Equation~\eqref{scaling} implies that for $t$ small, when $S_\Sig$ should be proportional to the area of $\Sig$, $S_\Sig$ must grow linear with $t$. At large times as $t \to \infty$, if the system has an equilibrium, i.e.~$f (t/R)$ has a well defined $t \to \infty$ limit, then $s$ must be proportional the equilibrium entropy density.

\subsection{Some examples} \label{sec:exk}

It is instructive to look at some specific examples of $\mu [A]$. 

\subsubsection{Entanglement carried by EPR pairs} 

One assumes that the quench creates a uniform density of independent EPR pairs of quasiparticles which are entangled within each pair and subsequently travel in opposite directions at the speed of light. The distribution of the directions of pairs is isotropic. This is a higher dimensional generalization of the model of~\cite{Calabrese:2005in}.
Under these assumptions, at time $t$, a point $\vx$
is entangled with another point $\vec y$ if and only if 
\be
|\vx - \vec y| = 2t \ . 
\ee
If $A$ consists of a region included in one half of the light cone it immediately follows that
\be \label{bel1}
\mu_{\rm EPR} [A] = s \, \xi_A \,, 
\ee
where $s$ is a normalization constant and $\xi_A$ is the area of region $A$ normalized by the area of the 
whole light cone~\eqref{apro}. The normalization constant $s$ can be written more explicitly in terms of the particle density $n$ as 
\be \label{endi}
s = {n } \nu\,,
\ee
where $\nu$ the entanglement entropy within the pair from tracing out one of the particles. 
For general $A$, not necessarily included in a half light cone sphere the measure is more complicated and depends on relative locations of different parts of $A$. It can be written formally as 
\be \label{fjr}
\mu_{\rm EPR} [A] = {s \ov \om_{d-2}} \int_A d^{d-2} \Om \int_{\bar A} d^{d-2} \Om' \,  \de^{(d-2)}  (\vec n + \vec n') \,,
\ee
where $\vec n$ and $\vec n'$ denote unit vectors on a unit sphere with $\Om$ and $\Om'$ the respective associated angular measure.  
 We can give a more compact expression for this measure as follows. We define $A^\prime$ as the set of antipodal points in $A$, that is, $A^\prime$ is the set of unit vectors $\vec{n}$ such that $-\vec{n}\in A$. Then, as only those quasiparticles in $A$ contribute to the entropy, whose pair (at the antipodal point) is in $\bar{A}$, we have
 \be
 \mu_{\rm EPR} [A]= s\, \xi_{A\cap \bar{A}^\prime}\,.\label{epr2}
 \ee 
If $A$ is included in one half of the light cone, $A\cap \bar{A}^\prime=A$, and we get back~\eqref{bel1}.

\subsubsection{$2m$ particles forming a GHZ block}

The EPR pair example has only bipartite entanglement among particles. We now consider an example 
with multipartite entanglement. 
One again assumes that the quench creates a uniform density of quasiparticles, but  
now particles from each point  separate 
into uncorrelated blocks each of which consists of $2 m$ particles, with $m$ an integer. Within each block, 
the $2m$ particles are entangled. To satisfy momentum conservation, for simplicity we will take that the $2m$ particles within a block come in pairs with back to back momenta. 

We consider the simplest entanglement relation that when tracing out any subset of particles within the block of $2m$ particles, one gets the same entanglement entropy $\nu$. This is motivated by the so-called GHZ state for $k$ qubits
\be 
\ket{\rm GHZ} = {1 \ov \sqrt{2}} \le(\ket{0}^{\otimes k} +  \ket{1}^{\otimes k} \ri)\,,
\ee
which has this property and thus the name for this example.\footnote{We note that because the tripartite information $I_3\geq 0$ for GHZ states and holographic mutual information is monogamous~\cite{Hayden:2011ag}, the GHZ pattern of entanglement cannot be realized using holographic geometry. Nevertheless, it is a simple example that holographic results can be compared to.} 

Consider a region $A$ which is included in half of the light cone, then we find that 
\be \label{ghzm}
 \mu_{\rm GHZ} [A; m] = {s \ov 2m} \le[1- (1-  2 \xi_A)^m \ri] \,,
\ee
where $s$ is again given by~\eqref{endi}.
To see~\eqref{ghzm}, we note that : (i) ${n \ov 2m}$ is the number of GHZ blocks at each point; (ii) from pairwise momentum conservation there can be at most $m$ particles lying in $A$; (iii) as far as there are particles lying in $A$, the entanglement entropy for any particular configuration is always $\nu$; (iv) 
 $(1- 2 \xi_A)^m$ is the probability that none of the $2m$ particles lies within $A$.
 
For a general region, we have to calculate the probability that none or all particles are inside $A$. These probabilities are given by $(1-  \xi_{A\cap A^\prime})^m$ and $\xi_{A\cap A^\prime}^m$ respectively. Hence, for any region $A$ we have 
 \be \label{ghzmGeneral}
 \mu_{\rm GHZ} [A;m]={s \ov 2m} \le[1- (1-  \xi_{A\cup A^\prime})^m - \xi_{A\cap A^\prime}^m \ri]\,.
 \ee
 This formula gives back~\eqref{ghzm} for a region included in half of the light cone, as $\xi_{A\cup A^\prime}=2\xi_A$ and $A\cap A^\prime=\emptyset$ for this case. The case $m=1$ reproduces the EPR result~\eqref{epr2} since $(\xi_{A\cup A^\prime}-\xi_{A\cap A^\prime})=2 \xi_{A\cap \bar{A}^\prime}$. The formula~\eqref{ghzmGeneral}  also obeys~\eqref{purity}, which is most easily seen by rewriting it using $1-\xi_{A\cup A^\prime}=\xi_{\overline{A\cup A^\prime}}$ and $\overline{A\cup A^\prime}=\bar{A}\cap\bar{A^\prime} $ as
 \be
  \mu_{\rm GHZ} [A;m]={s \ov 2m} \le[1- \xi_{\bar{A}\cap\bar{A^\prime} }^m - \xi_{A\cap A^\prime}^m \ri]\,.
 \ee

\subsubsection{Random pure state measure} 

Another measure is the random pure state measure we discussed in Sec.~\ref{sec:rps}, i.e.~
\be \label{ran0}
\mu_\text{RPS} [A] =  s \, {\rm min} (\xi_A, \xi_{\bar A}) 
\ee
for {\it any} region $A$ and $s$ is a normalization constant. While $\mu_\text{RPS}$ coincides with $\mu_{\rm EPR}$ when $A$ consists of a region included in the half sphere,
 in general~\eqref{epr2} is clearly different from~\eqref{ran0}. Note that~\eqref{ran0} only depends on the area of a region on the unit sphere (or its complement) for any $A$. This is not so for both $\mu_{\rm EPR}$ and $\mu_{\rm GHZ}$. 
As we showed in Sec.~\ref{sec:rps}, given~\eqref{apro}, the RPS measure provides an upper bound for other measures. This will enable to us to establish an upper bound on $\fR_\Sig (t)$ and $v_E$ in Sec.~\ref{sec:bound}.

\subsection{Evolution of the entanglement entropy} \label{sec:Evolution}

We now use~\eqref{masf} to derive some general results for the evolution of $S_\Sig (t)$. 
Here we will focus on universal features which do not depend on the specific form of $\mu [A]$. 
In Appendix~\ref{sec:details} we study many other aspects including mutual information and finite volume effects 
based on the specific examples of the last subsection.

We will consider a compact surface $\Sig$ with a characteristic size $R$. In other words, $R$ collectively denotes the curvature radii of $\Sig$.

\subsubsection{Equilibrium value} \label{sec:eqv}

\begin{center}
\begin{figure}[!h]
\includegraphics[scale=0.4]{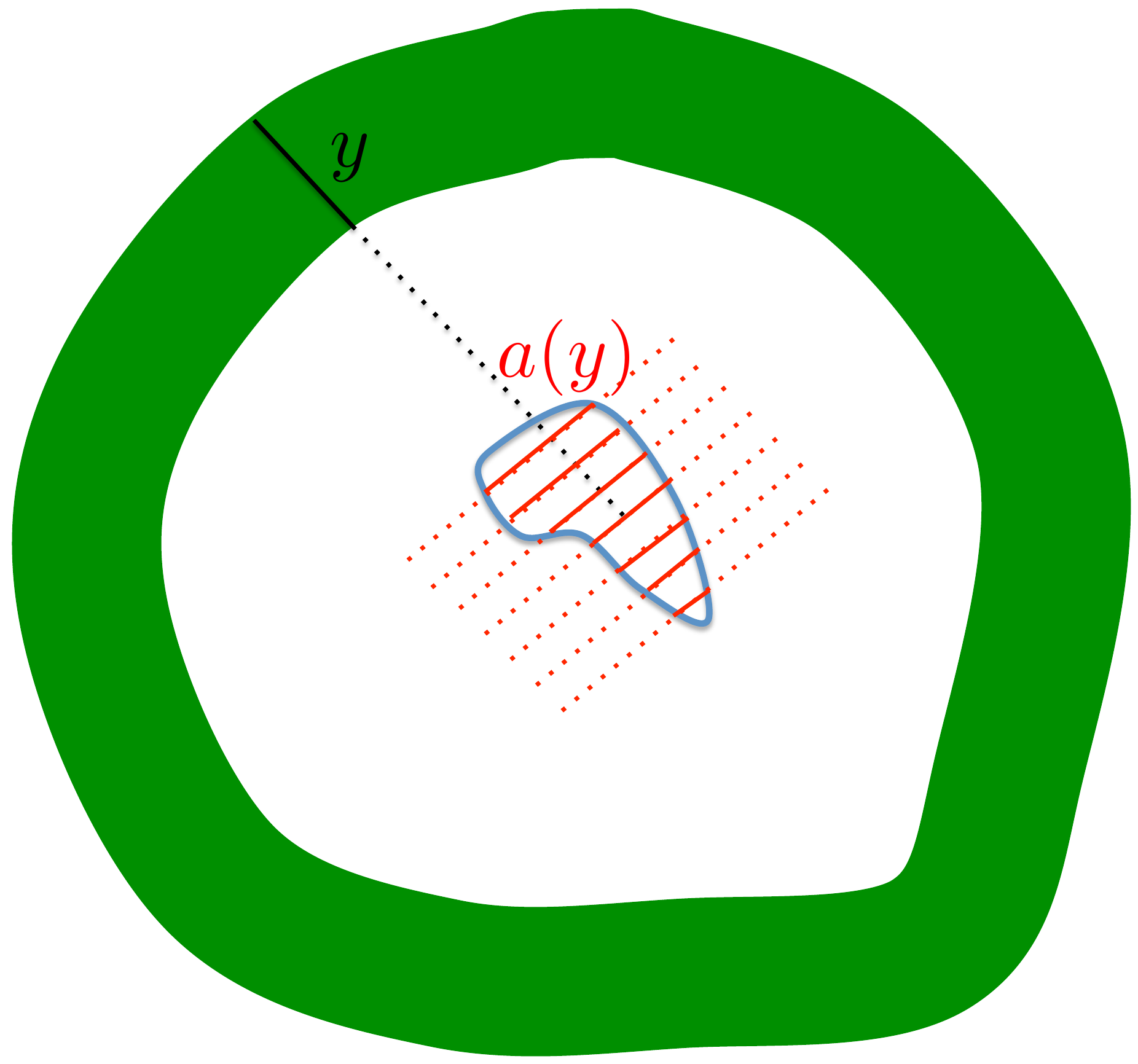}
\caption{Analysis of the late time behavior. 
The collection  of points $\sN_{\Sig} (t)$, whose light cones intersect with $\Sig$ is given by a shell centered around $\Sig$ (green region). The approximate radius of the shell is $t$.  For a fixed $\vec \eta$, the intersections of light cones with $\Sig$  from different values of $y$  provide a foliation of 
region $\sA$ enclosed by $\Sig$. \label{fig:LateTime}}
\end{figure}
\end{center}

The equilibrium value may be defined as 
\be \label{eqv}
S^{\rm eq}_\Sig = S_\Sig (t = + \infty) \ . 
\ee
  As $t \to \infty$, the size of light cones become much greater than that of $\Sig$, i.e.~$t \gg R$. Thus only small fractions of a light cone can be inside $\Sig$. The collection of points whose light cones intersect with $\Sig$ will be denoted as  $\sN_{\Sig} (t)$ and  is given by a shell centered around the region, see Fig.~\ref{fig:LateTime}. The integration over $\sN_{\Sig} (t)$ can be written 
schematically as 
\be \label{neke}
\int_{\sN_{\Sig} (t)} = \int d^{d-2} \vec \eta \int d y \,,
\ee
where $\vec \eta$ denotes directions tangent to shell, while $y$ denotes the integration along the width of the shell as indicated in Fig.~\ref{fig:LateTime}. Clearly, the precise shape of the shell will depend on the shape of $\Sig$ and 
for an irregular shell there may not be a preferred splitting in~\eqref{neke}, but as we will see, such details are not important. 

Now fix an $\vec \eta$ and consider the integral~\eqref{masf} 
over $y$. As we vary over the range of $y$, the corresponding $L_{\Sig} (\vec \eta, y ; t)$ provides a foliation of 
region $\sA$ enclosed by $\Sig$, see Fig.~\ref{fig:LateTime}. In particular, for $t \to \infty$, $L_{\Sig}(\vec \eta, y ; t)$ corresponds to 
a tiny part of the light cone from $\vx = (\vec \eta, y)$ and we can approximate
\be \label{roro}
\mu [L_{\Sig}(\vec \eta, y ; t)] = s \, \xi (L_{\Sig}(\vec \eta, y ; t)) \,,
\ee
where we used~\eqref{apro} for infinitesimal regions. 

The normalized area $\xi (L_{\Sig}(\vec \eta, y ; t))$ for $L_{\Sig}(\vec \eta, y ; t)$ can be further written as 
\be 
\xi (L_{\Sig}(\vec \eta, y ; t)) = {a (y) \ov \om_{d-2} t^{d-2}} \,,
\ee
where $a(y)$ is area of $L_{\Sig}(\vec \eta, y ; t)$. We thus find the integral of~\eqref{masf} over $y$ gives 
\be\label{onei}
\int dy\, \mu [L_{\Sig}(\vec \eta, y ; t)] ={s \ov \om_{d-2} t^{d-2}} \int dy  \, 
a (y) = {s V_{\Sig} \ov \om_{d-2} t^{d-2}} \,,
\ee
where $V_{\Sig} =  \int dy  \, a (y)$ is the volume of the region $\sA$ enclosed by $\Sig$. 
Note that the above integral is independent of $\vec \eta$. Now integrating over $\vec \eta$, to leading 
order in large $t$ approximation we simply get the area of the cross section of the shell, which is 
in turn given by the area of a sphere of radius $t$. Such a factor precisely cancels the denominator of~\eqref{onei} and we conclude that 
\be \label{keke}
S_\Sig (t \to \infty) = s_{\rm eq} V_\Sig  
\ee
with the equilibrium entropy density given by 
\be \label{hjek}
s_{\rm eq} = s\ .
\ee 

For the RPS example $s$ is simply a normalization constant and nothing more can be said. 
For the EPR and GHZ examples, the above equation can be further written in terms of the particle density $n$ as 
\be \label{uief}
s_{\rm eq} = n \nu\,,
\ee
which has a simple physical interpretation. Recall that for both examples $\nu$ is the entanglement entropy 
for a single particle when tracing out the others. As $t \to \infty$, the entangled particles are separated by infinite distances and thus the entanglement entropy for any finite region is given by the particle number density times $\nu$.  For a generic interacting system the equilibrium entropy density $s_{\rm eq}$ is  expected to coincide with 
the thermal entropy density, and~\eqref{uief} is also natural from that perspective.  

It is not difficult to realize from this proof that a uniform volume law for infinitesimal regions {\sl of any shape} on the sphere as in eq. \eqref{apro} is also {\sl necessary} to get the volume law at late times (\ref{keke}).

\subsubsection{Early linear growth} 

\begin{center}
\begin{figure}[!h]
\includegraphics[scale=0.4]{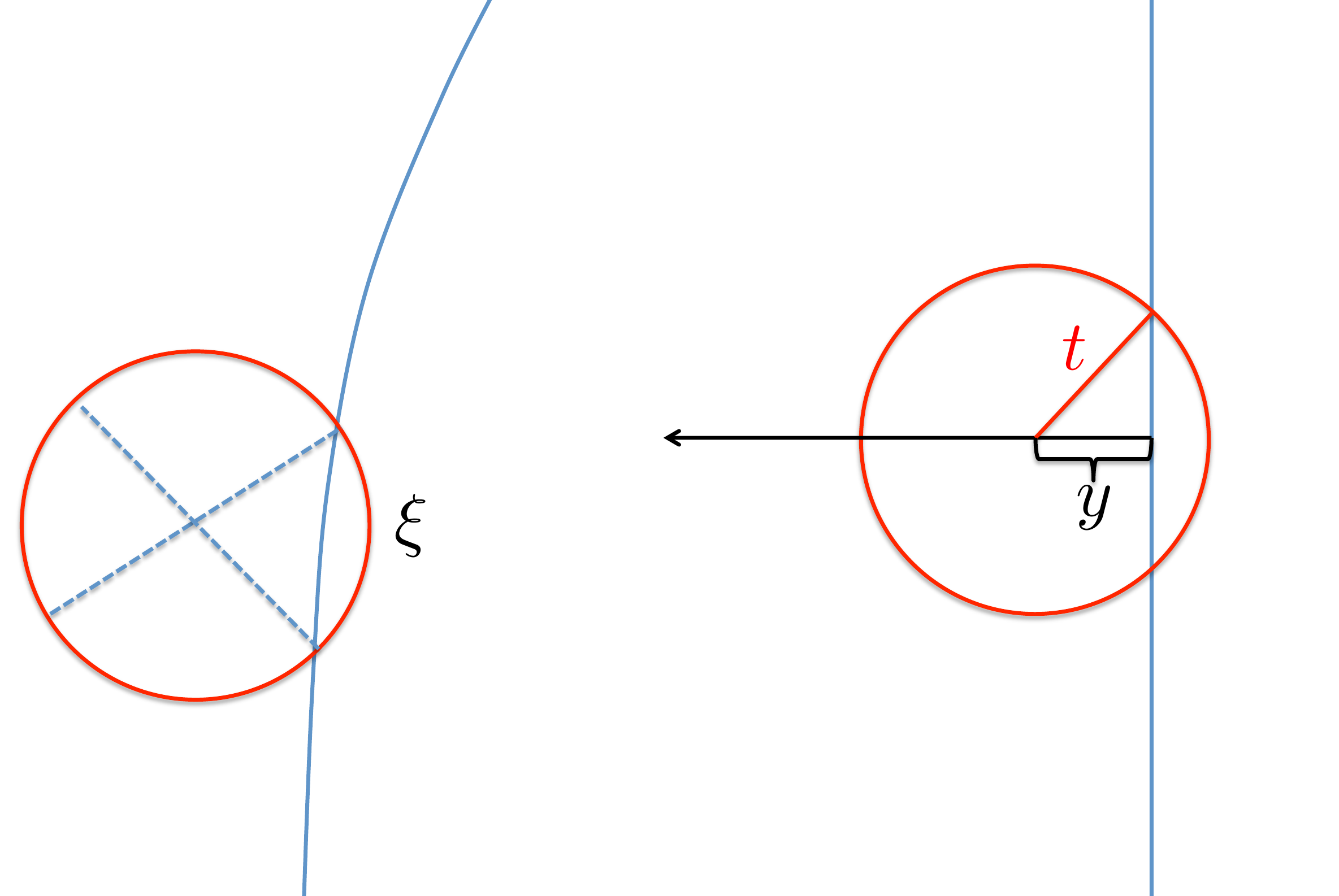}
\caption{Analysis of the early behavior. 
{\bf Left:} For early times the curvature of $\Sig$ is much grater than the radius of the light cone, and $\Sig$ can be approximated by a hyperplane, shown on the right.
{\bf Right:} Variables used in~\eqref{linel} to calculate the early time evolution of entanglement entropy.
\label{fig:linear}}
\end{figure}
\end{center}

Now let us consider the 
early growth, i.e.~for $ t \ll R$.
At such times, the radius of a light cone is much smaller than the curvature radius at any point of $\Sig$, 
see Fig.~\ref{fig:linear}. We can then locally approximate $\Sig$ as a straight hyperplane, with translational symmetries 
along directions tangent to $\Sig$. The integrations in~\eqref{masf} can then be factorized into an integral along $\Sig$, which simply gives a factor $A_\Sig$ (area of $\Sig$), and the relative location $y$ of centers of light cones with respect to $\Sig$ in the perpendicular direction. 
Then the early growth of the entropy will be determined completely by the measure $\mu$ applied to spherical caps. Let us introduce the notation
\be
\mu_{\rm cap} (\xi_{A})\equiv \mu[A]
\ee
for the measure for spherical caps $A$ as a function of their normalized area (note the area of a spherical cap determines it uniquely). Using this definition~\eqref{masf} can be written more explicitly as 
\be \label{linel}
S_{\Sig} (t)= 2 A_\Sig \int_{0}^t dy \, \mu_{\rm cap} (\xi (y/t)) = v_E \, s_{\rm eq} A_\Sig t \,, 
\qquad t \ll R\,,
\ee
where the factor of 2 comes from the domain of integration $y\in(-t,0)$,
\be \label{env}
v_E ={ 2 \ov s_{\rm eq}} \int_0^1 dx \, \mu_{\rm cap} (\xi (x)) \,,
\ee
and $\xi (x)$ is the normalized area of a spherical cap for a unit 
sphere with angular spread defined by the perpendicular distance $x = y / t$ from the center (see Fig.~\ref{fig:linear}). 
In~\eqref{linel} we used the time independence of the entropy measure on the light-cone. 
In~\eqref{env} we chose to normalize the quantity by the equilibrium 
entropy density $s_{\rm eq}$. 

\begin{center}
\begin{figure}[!h]
\includegraphics[scale=1]{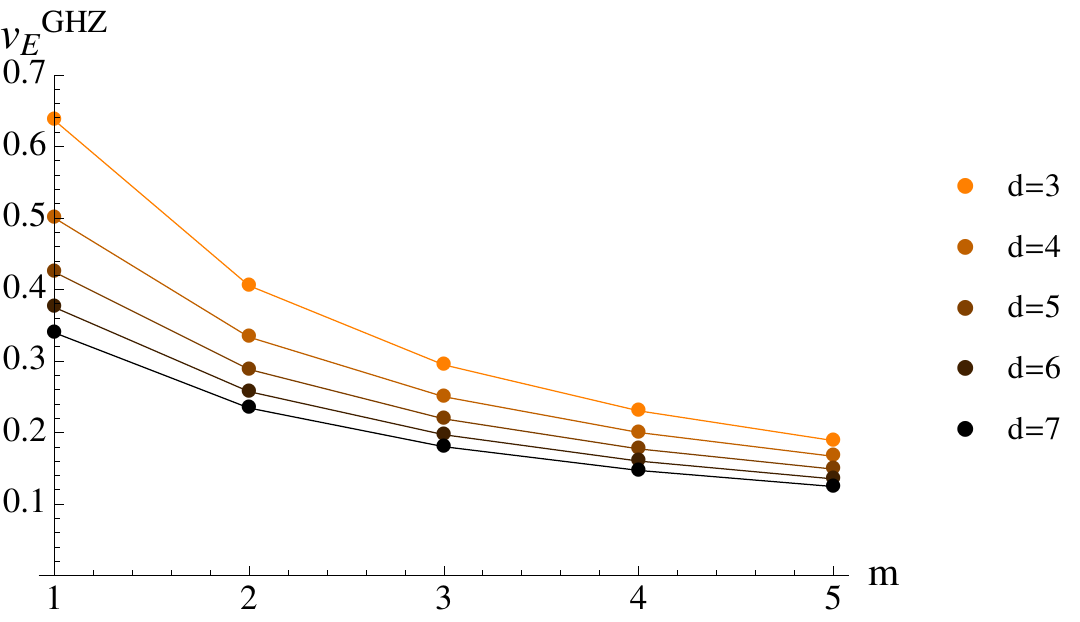}
\caption{$v_E^{\rm GHZ}$ in various dimensions as a function of $m$. For $m=1$ we get the EPR result and for $m \to\infty$ $v_E^{\rm GHZ} \to 1/m$. 
\label{fig:vn}
}
\end{figure}
\end{center}

Since~\eqref{env} only involves a region smaller than half the sphere, $\mu_{\rm EPR}$ and $\mu_\text{RPS}$ give the same result 
\be \label{eprv}
v_E^{\rm EPR} = v_E^{\rm RPS} = 2 \int_0^1 dx \, \xi (x) = 
 {2 \om_{d-3} \ov \om_{d-2}} \int_0^1 dx \int_0^{\arccos x} d \th \sin^{d-3} \th   = 
 {\om_{d-3} \ov \om_{d-2} {d-2 \ov 2}} 
=  {\Ga ({d-1 \ov 2}) \ov \sqrt{\pi} \Ga ({d \ov 2})}  \,. 
\ee
  For $\mu_{\rm GHZ}$, a closed formula is not available, but one can easily 
obtain the numerical values for different values of $m$ and spacetime dimensions, see Fig.~\ref{fig:vn}.
It is clear from the figure that for $m > 1$ and $d > 2$
\be 
v_E^{\rm GHZ}  < v_E^{\rm EPR} \ .
\ee
 In Sec.~\ref{sec:bound} we provide an upper bound for the speed $v_E$ for any entanglement measure.
 
\subsection{Quadratic growth before local thermalization} \label{sec:quad}

In~\cite{Liu:2013iza}, it was found that for fast quenches in holographic theories, there is a 
period of quadratic growth before the linear growth, which sets in only after the local equilibration
time $t_{\rm eq}$. The local equilibration time is defined as the time scale by which  local thermodynamics already applies with a thermal entropy density $s_{\rm th}$, but long range correlations in which we are interested have not been established yet. For strongly interacting systems, various holographic studies~\cite{Sekino:2008he,Balasubramanian:2011ur} indicate 
that the local equilibration  $t_{\rm eq} \sim {1 / T}$ where $T$ is the final equilibrium temperature, 
and by this time the entanglement correlations from the quench 
will have at most spread to a length scale $\ell_{\rm eq} \sim {1 / T}$. For regions with size satisfying $R T \gg 1$, we can treat $t_{\rm eq}$ and $\ell_{\rm eq}$ as zero, which is essentially what we have being doing so far. 
In other words, our above discussion should be interpreted as applying only after $t_{\rm eq}$. 
More explicitly, in our setup we have assumed that at $t=0$ entanglement measure $\mu [A]$ on light cones have already been fully established.  We believe this assumption is reasonable only  after 
local equilibrium has been established, i.e.~after the quench has finished, it still takes some time for a system to build up the local entanglement measure $\mu [A]$, and that time scale can be interpreted to be $t_{\rm eq}$. 

We will now show that with some very simple assumptions, one can easily obtain quadratic growth of entanglement entropy with time. For definiteness of the discussion, we will use the EPR model as an 
example, although the discussion can be easily adapted for a generic measure $\mu [A]$.  
Consider equation~\eqref{bel1}, except that now the prefactor $s$ is taken to be a function of time 
for $t < t_{\rm eq}$. We will take the simplest possibility: a linear function, i.e.~
\be 
s(t) = \bca {s \ov  t_{\rm eq}} \,t  & t < t_{\rm eq}\,, \cr
                        s & t > t_{\rm eq} \,.
                        \eca 
\ee
For $t > t_{\rm eq}$ the discussion is essentially the same as before, recovering the  linear growth. For $t < t_{\rm eq}$, the integral~\eqref{linel} becomes 
\be 
S_{\Sig} = 2 A_\Sig {s \ov t_{\rm eq}} \int_{0}^{t} dt_0 \, t_0 \int_{0}^{t-t_0} dy \, \xi \le({y \ov t-t_0}\ri)
=   {v_E^{\rm EPR} \,s \ov 2\,  t_{\rm eq}}  A_\Sig t^2  \ . \label{tloc}
\ee
In contrast, for holographic systems one finds~\cite{Liu:2013iza} 
\be 
S_\Sig = {\pi\,  \ep \ov d-1} A_\Sig t^2 \,,
\ee
where $\ep$ is the energy density. 
This is not that different from (\ref{tloc}) considering $t_{\rm eq}\sim 1/T$ and that for a CFT $\ep= \frac{d-1}{d}\, s \,T$.

\section{Upper bound on the ballistic propagation of entanglement} \label{sec:bound1}

In Sec.~\ref{sec:rps} we showed that the RPS measure~\eqref{ran0} is an upper bound for all measures, an immediate consequence of which is an upper bound for the entropy of any region at any moment of time in models with ballistic propagation by an explicit geometric function. From~\eqref{masf}
\be \label{ueb0}
S_\Sigma(t)\leq S_\Sigma^{\rm RPS}(t)=s_{\rm eq} \int d^{d-1}x\,\,\min (\xi_{L_{\Sigma}(\vec{x},t)},\xi_{\overline{L_{\Sigma}(\vec{x},t)}})\,.
\ee
 In this section we use this to prove an upper bound on $\fR_\Sig (t)$. 

First let us look at $v_E$, which only involves spherical caps. Using
\be 
\mu_{\rm cap}(\xi(x))\leq \mu_{\rm cap}^{\rm RPS}(\xi(x))= s_{\rm eq} \xi(x)\,, \label{231} 
\ee
in~\eqref{env} we find
\be \label{ubd0}
v_E \leq v_E^{\rm free} \equiv   2 \int_0^1 dx\ \xi (x) = {\Ga ({d-1 \ov 2}) \ov \sqrt{\pi} \Ga ({d \ov 2})} \, .
\ee
Note that the EPR measure~\eqref{eprv} saturates the bound. We note  that there is an alternative proof of 
\be
\mu_{\rm cap}(\xi(x))\leq  s_{\rm eq} \xi(x)\,, \label{below}
\ee
without using~\eqref{asser}. In fact one can use the strong subadditivity  
condition to prove a stronger inequality 
\be \label{cne0}
\mu_{\rm cap}^{\prime\prime}(\xi)\leq 0\,,
\ee
i.e.~$\mu_{\rm cap} (\xi)$ is a concave function. We give the proof of~\eqref{cne0} in Appendix~\ref{appendix1}. Because a concave function always lies below any of its tangents, and $\mu_{\rm cap}'(0) = s_{\rm eq}$, equation~\eqref{below} then follows.

We now show that the normalized rate of growth of entropy $\fR_\Sig (t)$  is bounded by $v_E^{\rm free}$ at all times. From~\eqref{ueb0} we get
\be \label{asser2}
\fR_{\Sig} (t)=\frac{1}{s_{\rm eq} A_\Sig}\frac{dS_\Sigma(t)}{dt}\leq {1 \ov A_\Sig} \int d^{d-1}x\, \left|\frac{d\xi_{L_{\Sigma}(\vec{x},t)}}{dt}\right|\,.
\ee
Therefore a bound on normalized grow rate
\be
\fR_{\Sig} (t)\leq v_E^{\rm free}
\ee
would follow from a purely geometric inequality
\be
\int d^{d-1}x\, \left|\frac{d\xi_{L_{\Sigma}(\vec{x},t)}}{dt}\right|\leq v_E^{\rm free}\, A_{\Sigma}\,.\label{geom}
\ee
We now  prove this last inequality (\ref{geom}). Let us write an integral representation for the normalized area:
\be
\xi_{L_{\Sigma}(\vec{x},t)}=\frac{1}{\omega_{d-2}}\int d^{d-1}y\, \delta(|\vec{y}|-1)\, \Theta_\Sigma(t\, \vec{y}+\vec{x})\,,
\ee
where $\Theta_\Sigma(\vec{z})$ is the characteristic function of $\Sigma$, which takes the value $1$ for $\vec{z}$ inside $\Sigma$ and $0$ for $\vec{z}$ outside of it. The time derivative is given by
\be
\frac{d\xi_{L_{\Sigma}(\vec{x},t)}}{dt}=\frac{1}{\omega_{d-2}}\int d^{d-1}y\, \delta(|\vec{y}|-1)\, \int_\Sigma d\sigma\, \sqrt{g_\Sigma}\,(\vec{y}\cdot \vec{n}(\sigma)) \delta(t\, \vec{y}+\vec{x}-\vec{x}(\sigma))\,,
\ee
where $\sigma$ denote collectively a set of coordinates on the entangling surface $\Sigma$, $\vec{n}(\sigma)$ is the unit vector normal to $\Sigma$, and $\vec{x}(\sigma)$ the position vector on the surface. Then we have
\bea
&&\int d^{d-1}x\, \left|\frac{d\xi_{L_{\Sigma}(\vec{x},t)}}{dt}\right| \nonumber\\
&&=\int d^{d-1}x\, \Bigg|\frac{1}{\omega_{d-2}}\int d^{d-1}y\, \delta(|\vec{y}|-1)\, \int_\Sigma d\sigma\, \sqrt{g_\Sigma}\ (\vec{y}\cdot \vec{n}(\sigma)) \,\delta(t\, \vec{y}+\vec{x}-\vec{x}(\sigma))\Bigg|\\
&&\leq \int d^{d-1}x\, \frac{1}{\omega_{d-2}}\int d^{d-1}y\, \delta(|\vec{y}|-1)\, \int_\Sigma d\sigma\, \sqrt{g_\Sigma}\ \left|\vec{y}\cdot \vec{n}(\sigma)\right|\,  \delta(t\, \vec{y}+\vec{x}-\vec{x}(\sigma))\,, \nonumber
\eea
where we took the absolute value inside the integrals. By performing the integral over $x$, we can get rid of one of the delta functions, and we obtain
\bea
\int d^{d-1}x\, \left|\frac{d\xi_{L_{\Sigma}(\vec{x},t)}}{dt}\right|&\leq&\frac{1}{\omega_{d-2}}\int d^{d-1}y\, \delta(|\vec{y}|-1)\, \int_\Sigma d\sigma\, \sqrt{g_\Sigma}\,\left|\vec{y}\cdot \vec{n}(\sigma)\right| \label{nity}\,.
\eea
By using the rotational symmetry of the integral over $\vec{y}$ in~\eqref{nity}, we can make the replacement $\vec{n}(\sigma)\rightarrow \vec{n}$, with a fixed (and arbitrary) unit vector $\vec{n}$, thus the integral over $\sigma$ can be evaluated to give $A_\Sigma$
\bea
\int d^{d-1}x\, \left|\frac{d\xi_{L_{\Sigma}(\vec{x},t)}}{dt}\right|&\leq&A_\Sigma\,\left(\frac{1}{\omega_{d-2}}\int d^{d-1}y\, \delta(|\vec{y}|-1)\,\left|\vec{y}\cdot \vec{n}\right|\right)\\
&=&A_\Sigma\,v_E^{\rm free} \,, \nonumber
\eea
where in the last line we recognized~\eqref{eprv}. This completes the proof.

Note that the bound for the growth rate can only be saturated if both inequalities~\eqref{asser2} and~\eqref{geom} are saturated. The inequality~\eqref{asser2} is only saturated for all light cones, if: (i) the measure is equivalent to the RPS measure; (ii) all light cones whose intersection area with the region $\sA$ bounded by $\Sig$ is less than half the sphere volume are increasing  their intersection with time; (iii) all light cones with intersection greater than half the sphere volume are decreasing their intersection with time (the latter two conditions are satisfied by any shape at $t=0$). Even for the RPS measure this is not the case for long enough times (except for a planar entangling surface). We also note that $\fR_{\Sig} (t)$ can be negative in some circumstances, as shown in Appendix~\ref{sec:negativeR}.

\section{Interacting models}
\label{interacting}

In this section we present an interacting model.\footnote{Interactions between quasiparticles and their effect on the tsunami velocity was recently investigated in $(1+1)$ dimensions by Cardy in~\cite{Cardy:2015xaa}.  The interactions he considered break the scale invariance of the theory, whereas our setup preserves it. } 
Although the model of~\cite{Calabrese:2005in} captures the time evolution of entanglement entropy for a single interval precisely in two dimensions, we saw that $v^{\rm free}_E <  \Vee^\text{holo}$ in higher dimensions, and qualitative differences also arise for entanglement entropy for multiple 
intervals in two dimensions~\cite{Asplund:2013zba,Leichenauer:2015xra}. Furthermore, in free propagation 
models the spread of entanglement depends sensitively on the entanglement pattern of the initial state as we have studied in detail in Sec.~\ref{sec:bound} and Appendix~\ref{sec:details}. In contrast, in holographic systems, as emphasized in~\cite{Liu:2013iza}, the linear growth emerges after a system has locally equilibrated with thermodynamical concepts such as temperature and entropy density already applying 
at scales of order $1/T$. This implies that by the time the linear growth emerges, the details of the initial 
state should have already been largely erased by interactions, and the linear growth must be a consequence 
of interactions. 

An immediate consequence of interactions is that the quantum state of the system can no longer be described as a tensor product of those resulting from each point at $t=0$. As a result, our fundamental equation~\eqref{masf} 
breaks down, and one has to face the problem of characterizing
the quantum state of an interacting many-body system. Instead of confronting this very difficult problem
directly, here we seek a qualitative understanding of how linear growth can arise in an interacting system
and how interactions can enhance the spread of entanglement.  
We will present a very simple model, which we believe 
gives an upper bound on the spread of entanglement 
in interacting theories. In particular, it gives $v_E=1$ in any dimensions saturating the bound~\eqref{vbou2} in Sec.~\ref{sec:vproof}.

\subsection{The effect of a scattering event}\label{sec:singlescat}

\begin{figure}[!h]
\begin{center}
\includegraphics[scale=0.9]{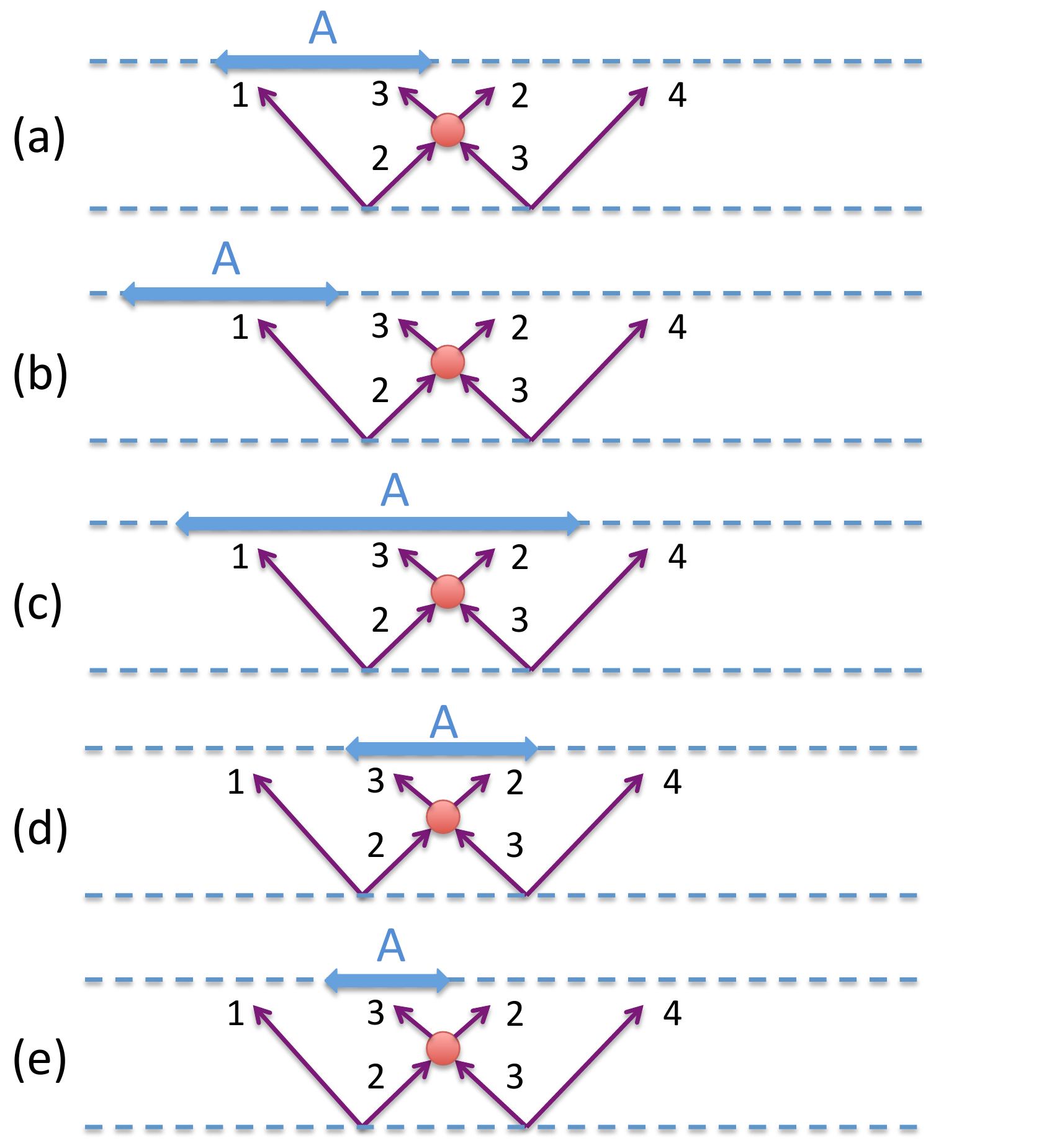}
\end{center}
\caption{The effect of one scattering event on the entanglement of region $A$. The scattering is represented as the red dot, where a unitary operator $U$ acts on the Hilbert space of particles $(2,3)$. The labeling of the indistinguishable particles  $(2,3)$ after the scattering is arbitrary, but it is convenient to choose that the particles maintain their original directions. For a single interval $A$ we show all the nontrivial scenarios, and we analyze the consequences of each configuration in the text. }
 \label{fig:scattering}
\end{figure}

To develop some intuition on the possible consequences of interactions on the propagation of entanglement, 
let us first consider the effects of a single scattering in the EPR model of quasiparticle propagation in $(1+1)$ dimensions.
Again we consider a single interval $A$ of length $2R$, with various different situations of scattering  
depicted in Fig.~\ref{fig:scattering}.  Diagrams (a), (b), (c) can happen for $t < {R \ov 2}$, (a), (b), (c), (d) for $R > t > {R \ov 2}$, and  (a), (b), (d), (e) for $t > R$. We will approximate each particle as a qubit and treat the scattering as 
a unitary transformation. In other words, after scattering, particles $2$ and $3$ maintain their original directions 
but the resulting state is related to the product state before the scattering by a 
unitary transformation. Note that an arbitrary unitary can encode not just forward, but backscattering as well. The latter is implemented by a unitary that swaps the wave functions of $2$ and $3$.

More explicitly, with the notation
\be 
\ket{1} = \ket{00}_{23}, \qquad \ket{2} = \ket{01}_{23}, \qquad  \ket{3} = \ket{10}_{23}, \qquad  \ket{4} = \ket{11}_{23}\,,
\ee
the scattering of $2$ and $3$ can be described by 
\be 
\ket{i} \to \ket{i'} = U \ket{i} \equiv U_{ij} \ket{j}, \qquad i, j=1,2,3,4\,,
\ee
with $U_{ij}$ a unitary matrix. Without loss of generality the state for $1,2,3,4$ before the scattering can be taken to be 
\be \label{ins}
\ket{\psi_i} = {1 \ov 1 + |\al|^2} (\ket{00}_{12} + \al \ket{11}_{12}) \otimes (\ket{00}_{34} + \al \ket{11}_{34})
 \ee
with $\al$ parametrizing the entanglement among the pair.  The entanglement entropy obtained from tracing out one of the particles 
in a single pair is 
\be 
\nu = - p_\al \log p_\al - (1- p_\al) \log (1-p_\al) , \qquad p_\al = {1 \ov 1 + |\al|^2} \ .
\ee
The state after the scattering is 
\be \label{fine}
\ket{\psi_f} = (1 \otimes U \otimes 1) \ket{\psi_i} \ . 
\ee

For different situations depicted in Fig.~\ref{fig:scattering}, we are interested in different reduced density matrices: 

\ben

\item[(a)] In this case the relevant quantity is $S_{13}$, i.e.~the entropy for the reduced density matrix $\rho_{13}$ from tracing over particles $2$ and $4$. We will denote the entropies corresponding to~\eqref{ins} and~\eqref{fine} as $S_{13}^{(i)}$ and $S_{13}^{(f)}$ respectively;  
$S_{13}^{(i)} = 2 \nu$. Clearly for $|\al| = 1$, i.e.~when an EPR pair is maximally entangled, $S_{13}^{(f)}$ is always smaller than the maximal possible value $S_{13}^{(i)} = 2 \log 2$ for any $U \neq 1$. Physically, some of the initial
entanglement between $(1,2) $ and $(3,4)$ is now shared between $1$ and $3$, and $2$ and $4$. 
In the other extreme, for  $|\al| \ll 1$ (or equivalently $|\al| \gg 1$), i.e.~when the original pair is only slightly entangled, the scattering  between $2$ and $3$ should generate new 
entanglement between the two particles and thus enhance the entanglement of $(1,3)$ with the rest of the particles, i.e.~we expect $S_{13}^{(f)} > S_{13}^{(i)}$ for generic $U$. For $\al$ in between, some $U$ could enhance the entropy, and some could reduce it.  
The explicit expression for $S_{13}^{(f)}$ is given in Appendix~\ref{app:a}.

\item[(b)]  The relevant quantity is $S_1$. Since $U$ only acts on the complement of $1$, clearly 
$S_1$ is not modified by the scattering. 

\item[(c)]  The relevant quantity is $S_4 = S_{123}$ and as in (b) the value is not modified. 

\item[(d)]  The relevant quantity $S_{23}$ is again not modified, as $U$ acts on the subspace of $2$ and $3$. 

\item[(e)]  The relevant quantity is $S_3$. When a pair is maximally entangled~($|\al|=1$), $S_3$ is not modified for any $U$, as the entanglement between $3$ and others is simply redistributed. 
For other values of $\al$, the situation is a bit similar to that of $S_{13}$ discussed above in (a). For $\al \ll 1$ or $\al \gg 1$, the entanglement will be enhanced for generic $U$. For generic values of $\al$ in between, then depending on $U$, the entropy can be either enhanced or reduced. The explicit expression for $S_3^{(f)}$ is given in Appendix~\ref{app:a}. 

\een

One should be able to use the above analysis to construct a dilute gas model to obtain quantitative results, 
as there are only a small number of scatterings. We will leave this for the future. 
Here we consider a model for the infinite scattering limit.

\subsection{The infinite scattering limit} \label{sec:infinite}

When including multiple scatterings, one needs to consider states associated with more and more particles 
and clearly we lose control very quickly. Nevertheless based on intuitions obtained from single-scattering results, we will see that one could still draw some qualitative conclusions about the general interacting case. 

On dimensional grounds we can write the evolution of entanglement entropy for a region $A$ enclosed by $\Sig$ in a form 
\be \label{scaling1}
S_\Sig (t) = s_{\rm eq} R^{d-1} f (t/R, \xi/R, TR)\,,
\ee
where $T$ is the equilibrium temperature, $s_{\rm eq}$ is the equilibrium entropy density,  and $\xi$ collectively denote other length scales in the system.
  In  strongly interacting systems such as holographic theories, the local scattering rate is controlled by $1/T$. For weakly interacting systems, the mean free path is typically controlled by other scale(s) $\xi \gg 1/T$. 
To simplify the analysis, we will work in the regime 
\be \label{rop}
\xi/R \to 0, \quad TR \to \infty, \quad t/R ={\rm finite}\,,
\ee
and assume that the scaling function $f$~\eqref{scaling} has a finite limit in this case:
\be \label{scaling2}
S_\Sig (t) = s_{\rm eq} R^{d-1} f (t/R) \ .
\ee
We refer to this regime~\eqref{rop} as the infinite scattering limit. For example, holographic systems after local equilibration are governed by it.  Note that~\eqref{scaling2} is of the same scaling form as~\eqref{scaling0} valid in the free propagation case, even though the underlying physics is very different. In free case~\eqref{scaling0} is a consequence of no scattering and $s$ is determined by the initial state, while in~\eqref{scaling2} is essentially a consequence of infinite number of scatterings (as $t$ is infinite compared with any scattering time) with $s_{\rm eq}$ determined by dynamics. As already mentioned below~\eqref{scaling}, the scaling form~\eqref{scaling2} implies that if there is a regime that $S_\Sig$ is proportional to the area of $\Sig$, then the 
time dependence must be linear. 

We again assume that the quench generates a finite density of identical particles, which then subsequently propagate 
at the speed of light isotropically, and we allow an arbitrary number of scatterings. We will assume that on average scattering events are isotropic and homogeneous in space, implying that both incoming and outgoing particles are uniformly distributed in all directions. As in the $(1+1)$-dimensional example of last section, 
scattering events will be treated as unitary transformations on all particles (which are assumed to be identical) that are at the same point at a given time. The labeling of particles after a scattering event is again arbitrary. Given the isotropy we can choose the labeling so that {\it particles will not change directions after scatterings}. This means that we can trace the whole spacetime trajectory of a particle from its origin even in the presence of interactions. The ability of tracing a particle trajectory will play an important role in our discussion below.

 At any given time, to calculate the reduced density matrix $\rho_A$ for $A$ (and thus the entanglement entropy), we need to simultaneously trace over all particles lying outside $A$, rather than restricting to a subspace like in the non-interacting or single-scattering case. Now note the following:

\ben

\begin{figure}[!h]
\begin{center}
\includegraphics[scale=1]{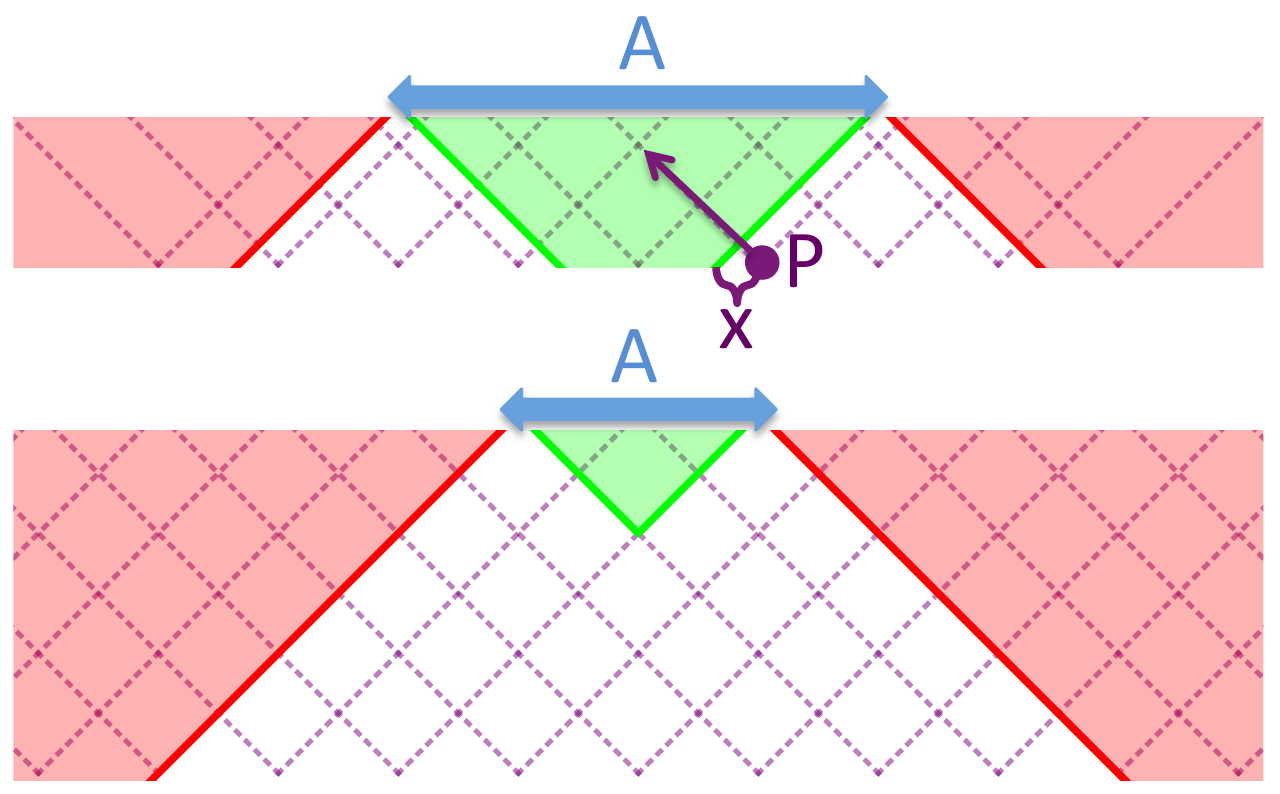}
\end{center}
\caption{Entanglement entropy of an interval $A$ in $(1+1)$ dimensions in the interacting model. The top figure shows a time before saturation, while the bottom figure applies for times after saturation. The dashed purple lines show particle trajectories, and their intersections are scattering events. The green region is  $\sD_- (A)$ and the red region is $\sD_- (\bar{A})$. As explained above scatterings that take place in these colored regions do not change the entanglement entropy. The point $P$ at distance $x$ from $\sD_- (A)$ and the left moving particle emanating from it will play a role in the discussions in Sec.~\ref{sec:conc}.   \label{fig:2dScattering}}
\end{figure}

\item The situation in (b) of last subsection can be immediately generalized to conclude that scatterings that happened in the past domain dependence of $\bar{A}$ (complement of $A$), i.e.~in $\sD_- (\bar{A})$ are not relevant, as they amount to unitary transformations in $\sH_{\bar{A}}$ that do not change $\rho_A$.\footnote{Recall that the past domain dependence $\sD_- (\bar{A})$ of $\bar{A}$ is defined 
as the spacetime region where all future-extended causal curves pass through $A$.} In Fig.~\ref{fig:2dScattering}, where we depict the time evolution for one interval in $(1+1)$ dimensions, these are regions shaded in red.

\item 
Similarly as in 
(c) and (d), scatterings among particles which fall inside region $A$ are also not relevant, as such scatterings act by unitary transformation on $\sH_A$ and thus will not change $\rho_A$.  In other words, scatterings in region $\sD_- (A)$, shaded green in Fig.~\ref{fig:2dScattering} can be neglected. Thus particles that spend their whole life in the green region do not give rise to any entanglement. Note that for $t > R$, the green region no longer intersects with the $t=0$ 
spatial manifold. 

\item Situations like (a) and (e) corresponds to scattering between particles one of which falls into $A$ and the other falls into $\bar A$. We will refer to these as effective scatterings. As in the single scattering case effective scatterings do affect entanglement.  

\een

The above discussion shows that we only need to consider particles originated from the region 
\be \label{eej}
\sN (t) \equiv \sM - (\sD_-(A) \cap \sM) - ( \sD_-(\bar A) \cap \sM)\,,
\ee
where $\sM$ denotes the full spatial manifold at $t=0$.\footnote{Note that the region $\sN_\Sig (t)$  we used in the free streaming model  is in general a subset of  $\sN (t)$, see Sec.~\ref{sec:conc} for further discussion.}  Furthermore,  only those scatterings of these particles 
that take place in the white regions in Fig.~\ref{fig:2dScattering} are relevant. This implies that disconnected regions of $\sN (t)$
can be treated independently of one another.

In the regime~\eqref{rop} all the particles will have scattered essentially an infinite number of times.  
In such a situation we expect that any memory of the initial state will be forgotten, and we can simply assign 
a geometric measure for the entanglement entropy. Since now the relevant Hilbert space is that for all the particles in $\sN$, we will simply  postulate a random pure state measure for the entanglement, i.e.~
\bea
S_A (t) &=& \nu_{\rm eq} \sum_i {\rm min} \le(N_A\le[\sN_i(t)\ri],\,N_{\bar{A}}\le[\sN_i(t)\ri]\ri) \label{master}\,, \\
N_A\le[\sN_i(t)\ri]&\equiv&\int_{\sN_i (t)} n_A (x,t)\,, \qquad
 N_{\bar{A}}\le[\sN_i(t)\ri]\equiv  \int_{\sN_i(t)} n_{\bar A} (x,t) \,,\label{masterN}
\eea
where $\nu_{\rm eq}$ may be interpreted as 
average entropy per particle, and $N_A\le[\sN_i(t)\ri]$ is the number of particles originated in $\sN_i(t)$ that fall into $A$ at time $t$. The number of such particles is given by the integral of the density of particles $n_A (x,t)$ that originated from $x$ and
fall into $A$ at $t$.   Taking the smaller value of the number of particles falling into $A$ and $\bar A$ 
has the same rationale as the earlier postulate of random pure state, and ensures $S_A = S_{\bar A}$. 
The sum $i$ is over the disconnected components of $\sN$, i.e.~$\sN = \cup_i\, \sN_i$. As mentioned earlier, 
 disconnected components of $\sN_i$ should be treated independently and thus summed 
separately. 

At small $t$, the number of $\sN_i$ is always the same as the number of disconnected boundaries of $A$.  As time evolves, different $\sN_i$ can join each other. Eventually all $\sN_i$'s will merge into a single connected region
$\sN(t)$.  At sufficiently late times, for $A$ in an infinite space, it will always be the case that $N_A\le[\sN (t)\ri] < N_{\bar{A}}\le[\sN (t)\ri] $ as $\bar A$ is infinite. From homogeneity we can then conclude that 
\be 
S_A = s_{\rm eq} V_A, \qquad s_{\rm eq} = \nu_{\rm eq} n , \quad t \; \text{sufficiently large,}
\ee 
where $n$ is the number density.

We move on to analyze some examples.

\subsection{$(1+1)$-dimensional examples} \label{sec:OnedimInteracting}

Let us first consider various examples in $(1+1)$ dimensions. In this case, the time evolution of entanglement entropy that we obtain~\eqref{master} 
coincides with a phenomenological formula recently proposed to describe holographic results in~\cite{Leichenauer:2015xra}. In the limit $R,t\gg1/T$ the holographic result for $N$ intervals can be obtained from a very simple minimization procedure: take all possible pairings of the  left interval endpoints $\{\ell_1,\ell_2,\dots,\ell_N\}$ with the right interval endpoints $\{r_1,\dots,r_N\}$ and connect them by an extremal surface in the bulk. For the area of one extremal surface we get the one interval result
\be\label{1int}
S_\text{interval}(t,R)=2s_{\rm eq}\begin{cases}
t \quad &\le(t<R\ri)\,,\\
R\quad &\le(t\geq R\ri)\,.
\end{cases}
\ee
We have to add up these contributions and minimize over the pairings:
\be\label{HoloResult}
S_H(t)=\min_{\sig} \le[  \sum_{i=1}^N S_\text{interval}\le(t,{\abs{\ell_i - r_{\sig(i)}}\ov 2}\ri)\ri] \,,
\ee
where $\sig$ is a permutation. Note that for different times a different permutation may realize the minimum, which corresponds to the change of dominance of the extremal surfaces in holography.

\subsubsection{One interval}

Consider a single interval of length $2R$ in $(1+1)$ dimensions, see Fig.~\ref{fig:2dScattering}. For $t < R$, we have two disconnected $\sN_i$ each of width $2t$, 
and for any point $n_A = n_{\bar A} = {n \ov 2}$, where $n$ is the total particle density. 
We then find that  
\be \label{emm0}
S_A (t) =  \nu_{\rm eq} \times 2 \times {n\ov 2} \times 2 t = 2 s_{\rm eq} t, \quad s_{\rm eq} \equiv n \nu_{\rm eq} , \qquad t < R  \ .
\ee
For $t \geq R$, there is only one connected $\sN$, and 
\bea 
N_A\le[\sN(t)\ri]&=&\int_{\sN (t)} n_A (x,t) = 2 \times {n \ov 2} \times 2R = 2 n R\,,\\
N_{\bar{A}}\le[\sN(t)\ri]&=&\int_{\sN (t)} n_{\bar A} (x,t) = 2nR + 2 (t-R) n  = 2n t\,,
\eea
thus we conclude that 
\be \label{emm}
S_A (t) = 2 s_{\rm eq} R = S_A^{\rm (eq)}, \qquad t > R \,.
\ee
We then find that 
\be 
v_E = 1 \ .
\ee
The time evolution of entanglement is in agreement with~\eqref{1int}.

\subsubsection{Multiple intervals}

\begin{figure}[!h]
\begin{center}
\includegraphics[scale=0.65]{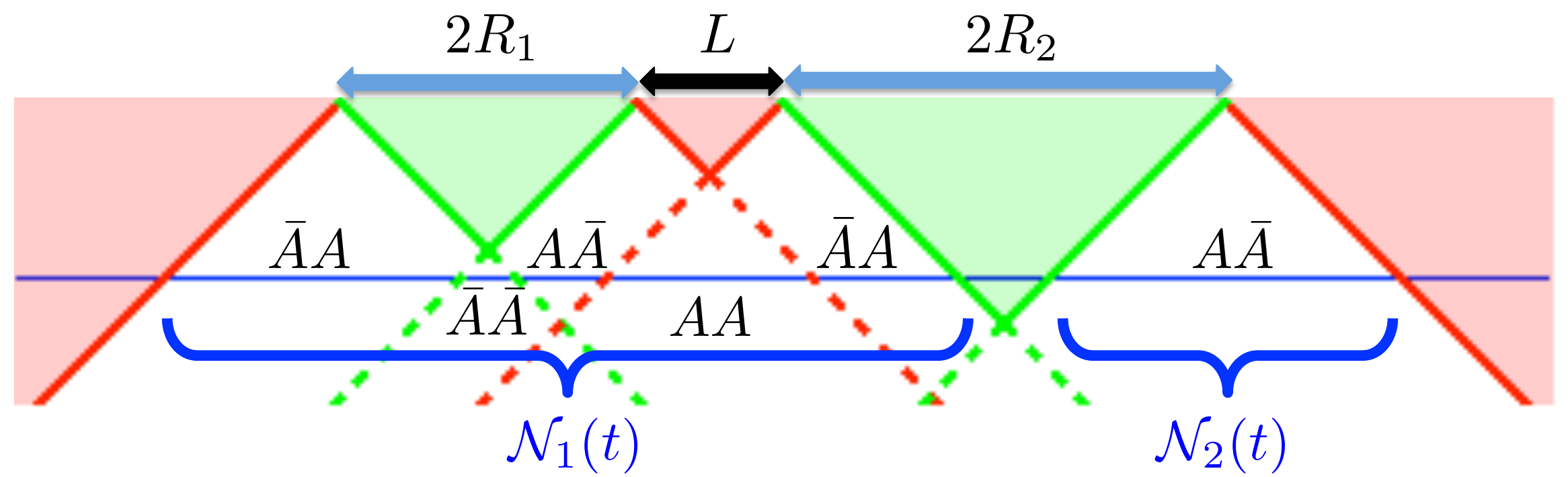}\\
\includegraphics[scale=0.6]{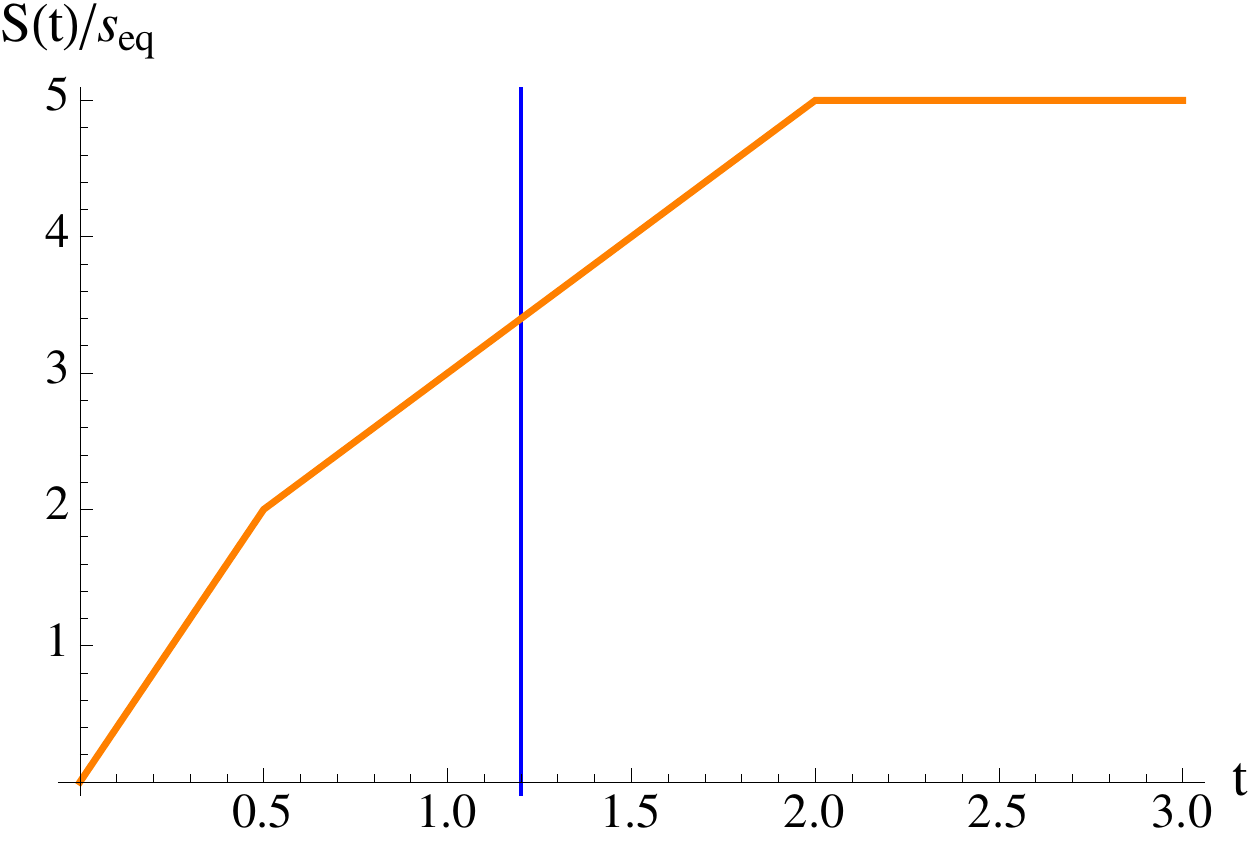}\hspace{1cm}
\includegraphics[scale=0.6]{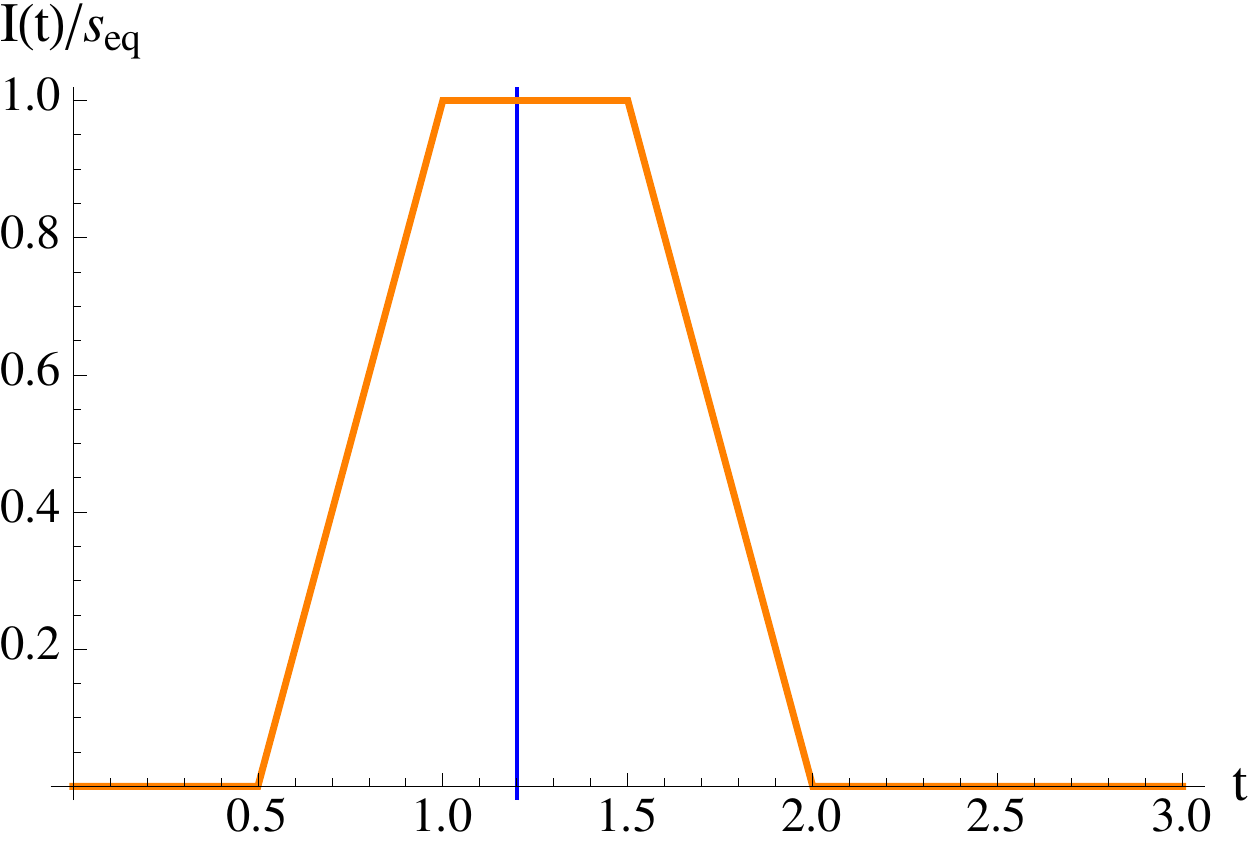}
\end{center}
\caption{{\bf Top:}  Explanation of how to calculate the entanglement entropy in the interacting model for two intervals of length $2R_1=2$ and $2R_2=3$ separated by a distance $L=1$.  Light rays starting at the entangling surface partition $\sN_i$ into multiple pieces characterized by where the left and right movers end up. In the figure we label each such piece by two letters, with the left letter standing for left movers, the right one for right movers.
For example $\bar A A$ means left movers end up in $\bar A$ while right movers in $A$. 
 {\bf Bottom:} Time evolution of entanglement entropy and mutual information for the same two intervals. The time slice considered in the top row is drawn by a solid blue line. The results agree with what one gets from the holographic result~\eqref{HoloResult}. (In the holographic calculations of e.g.~\cite{Balasubramanian:2011at} one  sees the smoothed out versions of these plots, as they do not take  the $R,L\gg 1/T$ limit.)
  \label{fig:TwoInt}}
\end{figure}

Let us first explain how the calculation goes in an example of  two intervals. 
For definiteness, consider intervals of lengths $2R_1,\,2R_2$  separated by a distance $L$ with $L<2R_1<2R_2$,
see Fig.~\ref{fig:TwoInt}.  As in Fig.~\ref{fig:2dScattering}, green light cones indicate $\sD_- (A)$ and red ones $\sD_- (\bar{A})$.  
At a given time, indicated by the horizontal blue line we decompose $\sN(t)$ into disconnected pieces. At the time indicated in the plot we have two disconnected pieces. We then count the number of particles that end up in $A$ and those that end up in $\bar{A}$ for each $\sN_i$, and take the minimum of the two numbers.

From Fig.~\ref{fig:TwoInt} and elementary geometric considerations one readily sees that the number of particles from each $\sN_i$  that end up in $A$ and $\bar{A}$ is:
\bea
N_A\le[\sN_i\ri]&=&n\,{\rm vol}\le(\sN_i\cap A\ri), \qquad 
N_{\bar{A}}\le[\sN_i\ri]=n\,{\rm vol}\le(\sN_i\cap \bar{A}\ri)\,,
\eea
and~\eqref{master} can be rewritten in a simpler form:
\be 
S_A (t) = s_{\rm eq} \sum_i {\rm min} \le[{\rm vol}\le(\sN_i(t)\cap A\ri),\, {\rm vol}\le(\sN_i(t)\cap \bar{A}\ri)\ri]  \ .\label{master2}
\ee
Equation~\eqref{master2} was proposed recently in~\cite{Leichenauer:2015xra} to capture the time evolution of entanglement entropy in holographic systems, motivated from the picture of entanglement tsunami~\cite{Liu:2013iza}. In this interpretation, $\sN_i (t)$'s  are given by the regions covered by the tsunamis,\footnote{Entanglement tsunamis originate from the boundaries of entangled regions and propagate in both 
directions.} and one again applies RPS to each $\sN_i$ which amounts to taking the smaller volume between those of $A$ and $\bar A$ regions within an $N_i$. As time evolves, different $\sN_i$ regions join when their respective tsunamis meet.

One can readily check that~\eqref{master} (and equivalently~\eqref{master2})
reproduces holographic results for two intervals~\eqref{HoloResult} for all parameters, $R_1,R_2,$ and $L$.

\begin{figure}[!h]
\begin{center}
\includegraphics[scale=0.9]{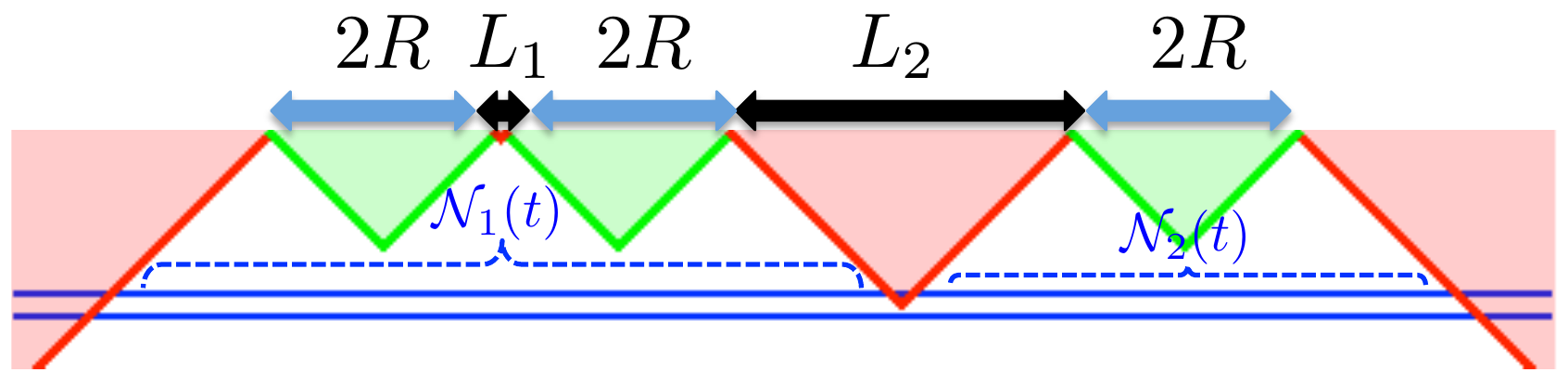}
\includegraphics[scale=0.6]{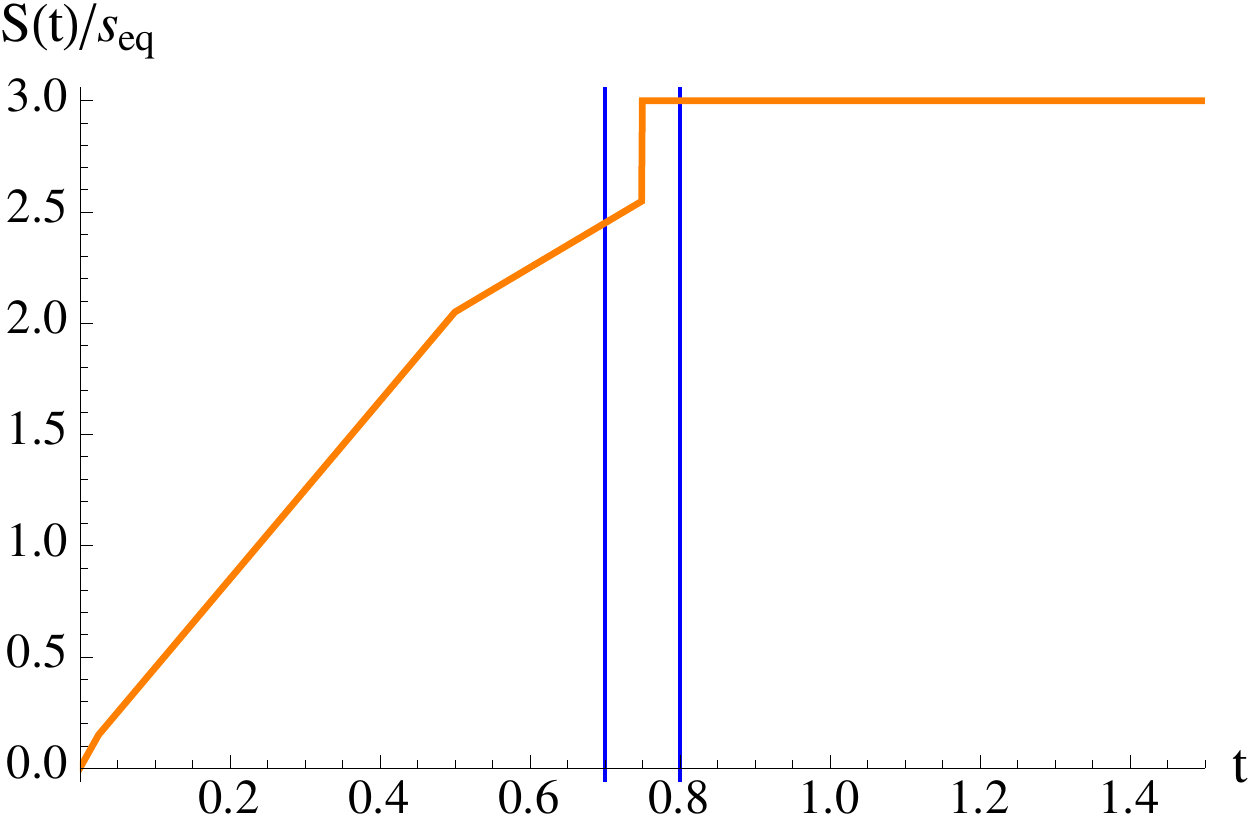}\hspace{1cm}\includegraphics[scale=0.6]{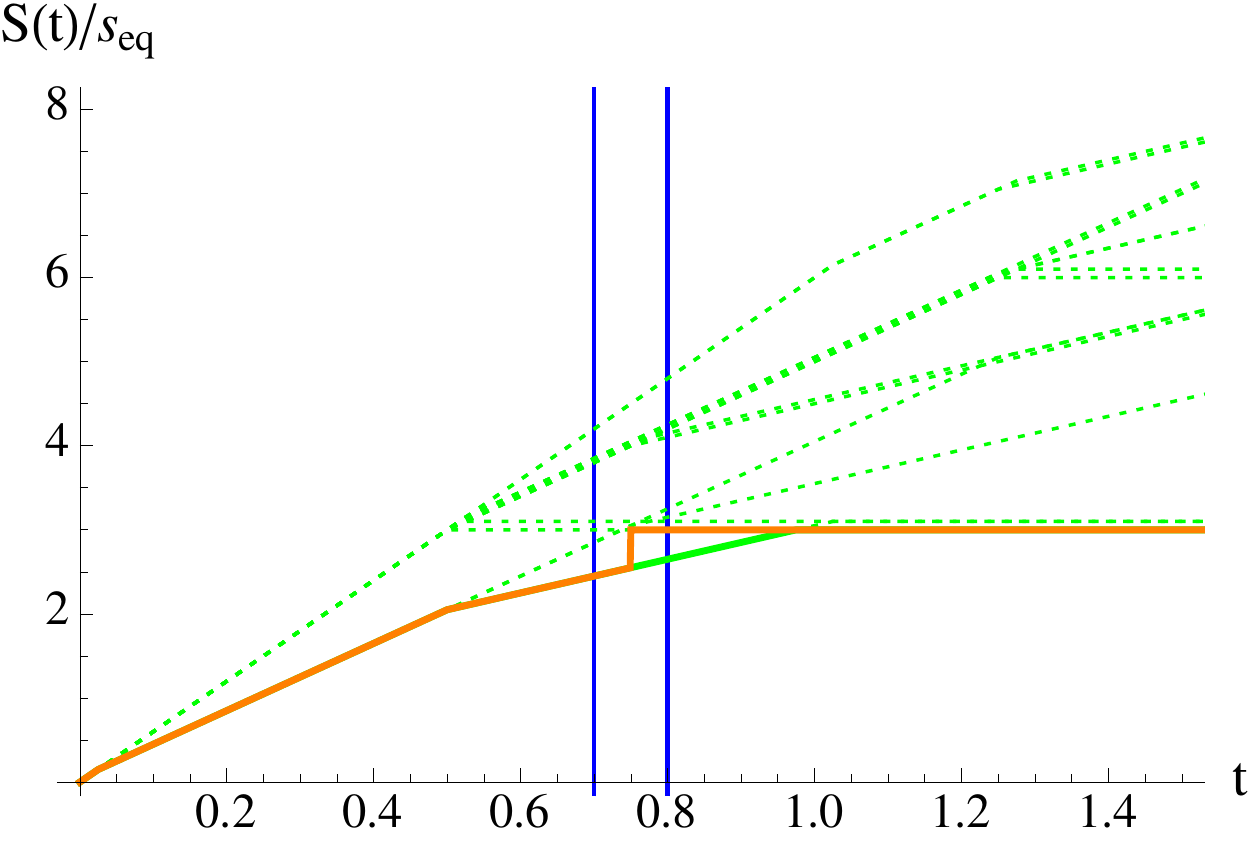}
\end{center}
\caption{{\bf Top:} Setup with three equal length intervals. We chose $R=1/2\,, L_1=1/20\,, L_2=3/2$. We drew the past domain of dependence for our setup, with times before and after the jump marked by solid blue lines. 
 {\bf Bottom:} Time evolution of entanglement entropy for our setup with a jump between the blue solid lines. On the right hand side we illustrate how holographic CFTs behave. We drew the contribution from all possible locally extremal surfaces (i.e.~the contribution from every possible permutation in~\eqref{HoloResult}) by dotted green lines and the smallest one among them by a solid green line. Note that for different times surfaces connecting different endpoints of the intervals dominate. The result from~\eqref{master2} is again drawn by orange. \label{fig:ThreeInt}}
\end{figure}

For three and four intervals, we find the model still reproduces precisely the rather intricate holographic results for a significant  part of the parameter space. With a random sampling of the parameter space with a large number of examples, for $3$ intervals about $21\%$ of the examples and for 4 intervals  $36\%$ deviate from the holographic results.\footnote{We sampled the parameter space by fixing the leftmost and rightmost boundary points and by throwing the other boundary points between them randomly using the uniform distribution. Our sample size was 500.}
Even when~\eqref{master} (and~\eqref{master2}) deviates from the holographic results, the overall trend is still quite similar, but it can give unphysical answers. We give an example in Fig.~\ref{fig:ThreeInt}, where the entanglement entropy develops a discontinuous jump at some time. The mathematical reason for the jump is as follows. 
Let us denote the time of the jump as $t_c$. At $t_c - \ep$ (with $\ep \to 0$) we have two disjoint regions 
$\sN_1$ and $\sN_2$: for $\sN_1$, $N_{\bar{A}}\le[\sN_1\ri]$ is smaller than $N_{{A}}\le[\sN_1\ri]$, while for $\sN_2$, $N_{A}\le[\sN_2\ri]$ is smaller, thus 
\be \label{gh0}
S (t_c -\ep) = \nu_{\rm eq} (N_{\bar{A}}\le[\sN_1\ri] + N_{A}\le[\sN_2\ri]) \ .
\ee
At $t_c + \ep$, $\sN_1$ and $\sN_2$ join into a single region and we should count the particles of $N_A$ and $N_{\bar A}$ for $\sN_1 \cup \sN_2$, i.e.~
\be \label{gh1}
S (t_c +\ep) = \nu_{\rm eq} \, {\rm min} (N_{\bar A} [\sN_1] +  N_{\bar A} [\sN_2 ] ,N_A [\sN_1] + N_A [ \sN_2]) \ .
\ee
There is a discontinuity from~\eqref{gh0} no matter which is chosen in~\eqref{gh1}. 
This phenomenon is likely generic as one increases the number of intervals: whenever two regions where $A$ and $\bar A$ dominate respectively join together, there will be  
a discontinuous jump. In Fig.~\ref{fig:ThreeInt} we also plot the corresponding holographic result 
which is continuous and is smaller than~\eqref{master2} for a period after $t_c$. We will further discuss the physical origin of the jump in Secs.~\ref{sec:conc} and~\ref{sec:rpsdiss}.

Given its simplicity, it is quite remarkable that the maximal RPS model manages to reproduce
intricate holographic results for multiple intervals for a significant part of the parameter space. As already alluded to in the Introduction and at the beginning of this section, the ballistic picture of the spread of entanglement discussed in Sec.~\ref{sec:bound}--\ref{sec:bound1} of this paper cannot account for the holographic results for more than one interval in $(1+1)$-dimensions. This was analyzed in detail in~\cite{Asplund:2013zba,Leichenauer:2015xra}, and we do not repeat this comparison between the results of the ballistic and scattering pictures. Recently, a CFT analysis showed that rational CFTs behave according to the ballistic model, while non-rational CFTs are expected to interpolate between the free streaming and the holographic behavior, which is reproduced by~\eqref{master2}~\cite{Asplund:2015eha}.\footnote{The holographic behavior is universal in large central charge CFTs with a sparse low-lying operator spectrum.}

Finally,  note that there is a simple ``phenomenological fix'' to the discontinuity problem, as follows.  
When two RPS regions with different dominance join, say $\sN_1$ with $\bar A$ and $\sN_2$ with $A$ as in the example of Fig.~\ref{fig:ThreeInt}, immediately after the two regions start overlapping we still keep $\sN_1$ and $\sN_2$ as independent, i.e.~the contribution from them is still given by $S_{\rm RPS} (\sN_1) + S_{\rm RPS} (\sN_2)$ rather than the bigger value $S_{\rm RPS} (\sN_1 + \sN_2)$. As time evolves we  merge them into a single RPS region when $S_{\rm RPS} (\sN_1 + \sN_2) = S_{\rm RPS} (\sN_1) + S_{\rm RPS} (\sN_2)$. If before these two regions merge, other RPS regions start overlapping with either of them, we follow the same rule in deciding whether they should merge with other regions. 
Sampling over the parameter space, we found that in the improved model the failure rate in reproducing the holographic answer~\eqref{HoloResult} is only $2\%$ for three intervals and $8\%$ for four intervals.

\subsection{Higher dimensions}

\begin{figure}[!h]
\begin{center}
\includegraphics[scale=0.5]{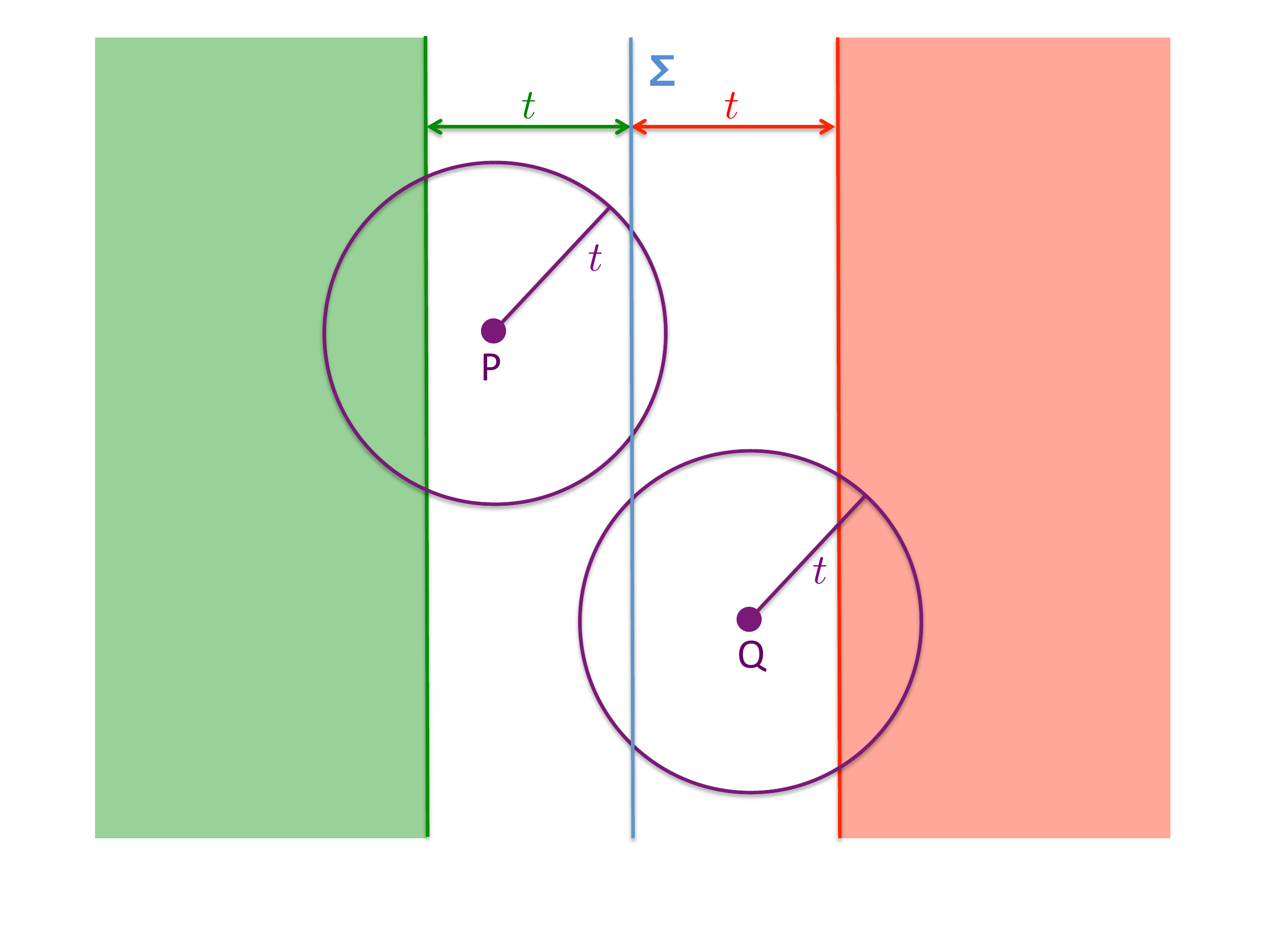}
\end{center}
\caption{Illustration of the computation of entanglement entropy for early times. We suppress the time direction and look at the configuration from ``above". Region $A$ is to the left of the entangling surface $\Sig$. The green region is  $\sD_- (A)$, the red region is $\sD_- (\bar{A})$, and the white region in between them is $\sN(t)$. Points $P$ and $Q$ are on the $t=0$ time slice and the purple light cones show where the particles that started out in $P$ and $Q$ end up at time $t$.  }
 \label{fig:region}
\end{figure}

 Let us first consider early times $t \ll R$, for which we can approximate the boundary as a straight-line, then $\sN(t)$ has is a strip of width $2t$ with $\Sig$ lying in the middle as indicated in Fig.~\ref{fig:region}. 
 For this geometry the total number of particles from $\sN$ falling into $A$ or $\bar A$ are the same, so we can choose either of them.  It then immediately follows from~\eqref{master} that 
 \be \label{vehi}
S_\Sig (t)=  {n\nu_{\rm eq}\ov 2}  \times 2A_\Sig t = s_{\rm eq} A_\Sig t \quad \implies \quad v_E = 1\,,
\ee
where in the above equation $2A_\Sig t $ is the volume of $\sN(t)$ and $n\nu_{\rm eq}$ is divided by two, as
exactly half of all particles from $\sN(t)$ will fall into region $A$ because of isotropy. Thus, in this model 
the tsunami velocity $v_E$ is precisely given by the speed of light in all dimensions!

It is interesting to contrast this computation with the earlier free propagation calculation of $v_E$ in 
Sec.~\ref{sec:Evolution}. We take the EPR (or equivalently the RPS) measure for the free streaming model. In both the free and interacting models we associate a measure to points on the $t=0$ time slice. Consider the contributions from points $P$ and $Q$ in Fig.~\ref{fig:region}. In the earlier free propagation
calculation, for  light cones originating from $P$ or $Q$ we took the area of the smaller spherical cap of the intersection of the light cone with $\Sig$, while for~\eqref{master}, we simply use the spherical caps inside $A$, which for $P$ is the bigger spherical cap. This thus leads to an enhancement in $v_E$. Physically, in free particle models entangled particles originating from the same point and ending up in $A$ do not contribute to the entanglement, while in the interacting model
the initial entanglement pattern is forgotten and all particles ending up in $A$ contribute.

\begin{figure}[!h]
\begin{center}
\includegraphics[scale=0.5]{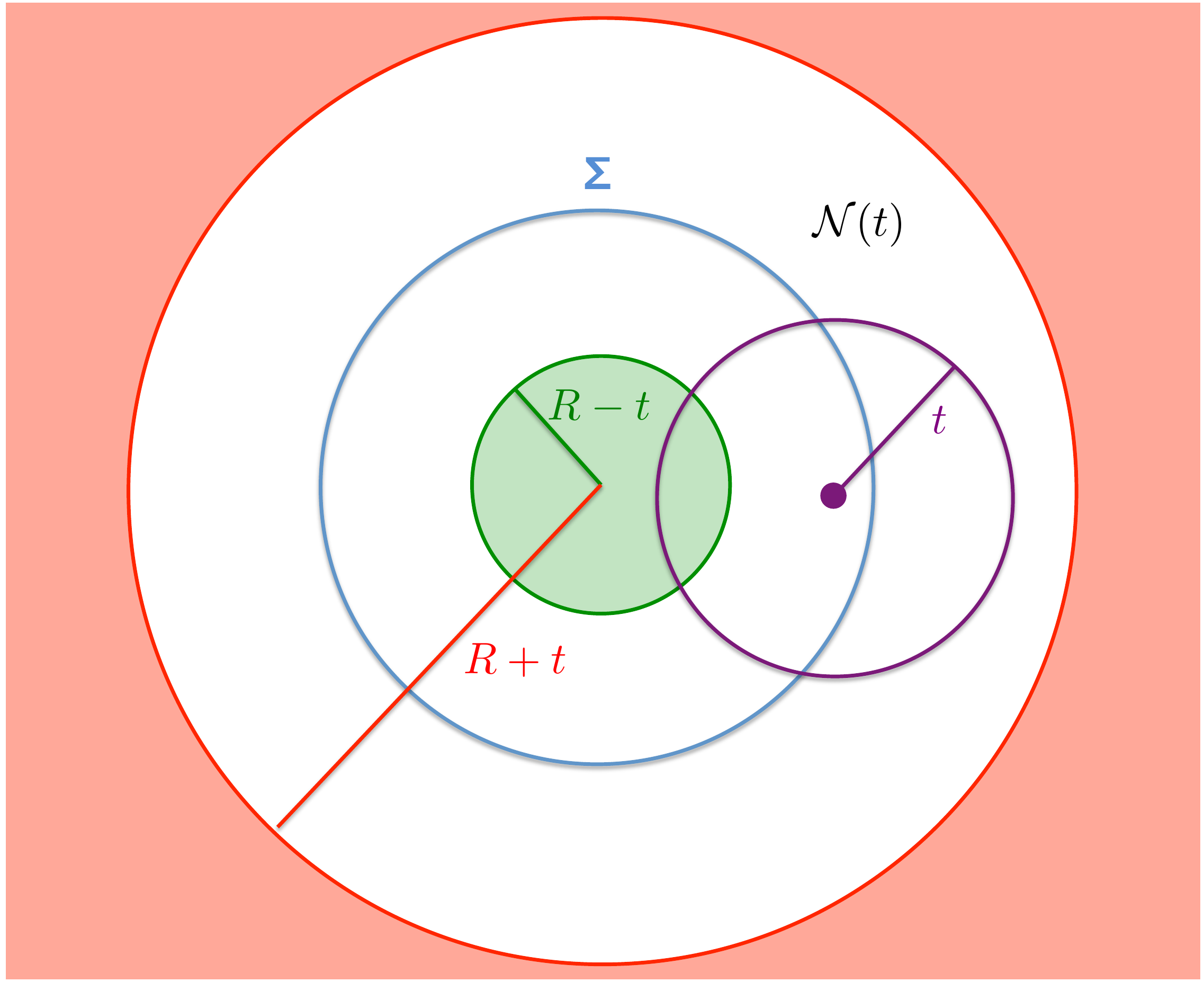}
\end{center}
\caption{A disk in $d=3$ before saturation. The green region is  $\sD_- (A)$, the red region is $\sD_- (\bar{A})$, and  the white region in between them is $\sN(t)$. According to~\eqref{master} by summing up the contributions of light cones like the one drawn with purple, we obtain the time evolution of entanglement entropy. Of course, it is a lot easier to use the simplification provided by~\eqref{master2}. }
 \label{fig:t0slice}
\end{figure}

Now consider a spherical region of radius $R$. Region $\sN (t)$ is now the white annulus region indicated in Fig.~\ref{fig:t0slice}. This annulus is cut into two parts by $\Sig$, with the inner part having smaller volume. Then applying~\eqref{master2} gives that the entanglement entropy is equal to the volume of the inner white annulus in in Fig.~\ref{fig:t0slice}:
\be
S(\tau)= s_{\rm eq} R^{d-1} \begin{cases}
\om_{d-2} \, {1-\le(1-\tau\ri)^{d-1}\ov d-1}& (\tau<1)\,,\\
 V_{B^{d-1}} & (\tau>1)\,,
\end{cases}\label{3dTimeScattering}
\ee
where we have introduced $\tau \equiv {t / R}$ and denoted the volume of the unit ball by $V_{B^{d-1}}={\om_{d-2} \ov d-1}$.\footnote{A nice check of~\eqref{master2} is that~\eqref{master} gives the same result. } By taking the early time behavior we see that $v_E=1$. This result can be contrasted with the ballistic propagation calculation for spheres presented in Appendix~\ref{sec:Fulltime}.

\subsection{A family of RPS models}\label{sec:conc}

Recall that in the free streaming model the wave function $\ket{\psi}$ factorizes into 
those of each spatial point at $t =0$, i.e.~
\be \label{woo}
\ket{\psi}= \otimes_{\vec x} \ket{\psi_{\vx} }
\ee
and the upper bound on the entanglement propagation is achieved when using RPS for each $\ket{\psi_{\vx} }$. 
When including interactions clearly~\eqref{woo} does not apply. Nevertheless due to constraints 
from causality, within a finite interval $t$ not all degrees of freedom can interact with one another. The basic idea behind~\eqref{master} 
is that at time $t$, the full wave function can be factorized based on the casual structure of $A$, i.e.~
\be \label{woo1}
\ket{\psi} = \le(\otimes_i \ket{\psi_{\sN_i (t)}} \ri) \otimes (\cdots)\,,
\ee
where $\cdots$ denotes the factor of the wave function which is irrelevant for the entanglement of $A$.
We then apply RPS to each $\ket{\psi_{\sN_i (t)}}$. 

Instead of~\eqref{woo1} one can in principle
consider a finer partition of $\sN(t)$ than connectivity,
\be \label{finer}
\ket{\psi} = \le(\otimes_\al \ket{\psi_{\sM_\al (t)}} \ri) \otimes (\cdots) \,,
\ee
where 
\be 
\cup_\al \sM_{\al} (t) = \sN (t) = \cup_i \sN_{i} (t)\,,
\ee
and for any $\al$, there exists an $i$ such that $\sM_\al (t) \subseteq \sN_i (t)$. We can obtain a general class of RPS models of entanglement 
propagation by applying RPS to each $\sM_\al$, i.e.~
\be 
S_A^{\{\sM_\al\}} (t) = \nu_{\rm eq} \sum_\al  {\rm min} \le(N_A\le[\sM_\al(t)\ri],\,N_{\bar{A}}\le[\sM_\al (t)\ri]\ri) \ .
\ee

The free streaming RPS model~\eqref{woo} is a special case of~\eqref{finer} with $\sM_\al$ given by a point.\footnote{Note that in free streaming model only the points in $\sN_\Sig (t)\in\sN (t)$ contribute, whose light cone can intersect with $A$.} In other words, the free streaming RPS model is the finest division of $\sN(t)$, it is the minimal RPS model. In contrast, the model~\eqref{master} is the coarsest division, and thus we will refer to it as the maximal RPS model.  

Now using the same argument as demonstrating the discontinuity in~\eqref{gh0}--\eqref{gh1}, we can show that the entropy $S_A^{\{\sM_\al\}} (t)$ always decreases with a finer partition of $\sN (t)$. 
To see this, let us consider dividing some block $\sM$ within a partition into $\sM_1 \cup \sM_2 = \sM$. We then have 
\be 
N_{\sM} (A) = N_{\sM_1} (A) + N_{\sM_2} (A) \,, \qquad N_{\sM} (\bar A) = N_{\sM_1} (\bar A) + N_{\sM_2} (\bar A) 
\ee
and thus 
\bea
 \min(N_{\sM}(A),N_{\sM}(\bar{A}))&=&\min(N_{\sM_1}(A)+N_{\sM_2}(A),N_{\sM_1}(\bar{A})+N_{\sM_2}(\bar{A}))\nonumber \\
 &\ge& \min(N_{\sM_1}(A),N_{\sM_1}(\bar{A}))+\min(N_{\sM_2}(A),N_{\sM_2}(\bar{A}))\,. 
\eea
The fact that a subdivision decreases the entropy implies that with the class of all RPS models, the free streaming and the maximal RPS model provide respectively the lower and upper bounds,
\be 
S_A^{\rm free} (t)  \leq S_A^{\{\sM_\al\}} (t) \leq  S_A^{\{\sN_i \}} (t) \ .
\ee

Furthermore from the discussion of Sec.~\ref{sec:rps}, the RPS measure provides an upper bound for the 
entanglement entropy among all possible entanglement measures with a given partition. We thus conclude that the maximal 
RPS model~\eqref{master} should provide an upper bound on entanglement propagation for all relativistic interacting systems after a global quench. This conjecture is consistent with our empirical observation that when the maximal RPS result deviates from the holographic one~\eqref{HoloResult}, it always lies above it.\footnote{In Sec.~\ref{sec:holomax} we prove that in $(1+1)$ dimensions holographic theories give the fastest  possible entanglement spread. This then implies that whenever the maximal RPS result deviates from holography, it overestimates the entropy. } That for $d > 2$ the maximal RPS gives $v_E =1$ is also consistent with the result of Sec.~\ref{sec:vproof}.

\subsection{Further discussion  of the maximal RPS model} \label{sec:rpsdiss}

We expect that the RPS 
measure applies when all degrees of freedom in the relevant region (i.e.~within each disconnected $\sN_i$) 
have fully ``equilibrated,'' i.e.~have interacted sufficiently with one another. Otherwise it provides an overestimate.
Consider at some time $t_c$, there are two regions $\sN_1$ and $\sN_2$ joining into a single connected region.~\eqref{master} then dictates that at $t_c + \ep$ (with $\ep \to 0$) we must apply the RPS measure to the whole $\sN_1 \cup \sN_2$. But physically in going from $t_c - \ep$ to $t_c + \ep$, there is just not enough time for this ``total equilibration'' to happen. When  $\sN_1$ and $\sN_2$ are dominated by $A$ and $\bar A$  respectively before joining, such ``lack of equilibration'' will lead to a discontinuity, as in the example of Fig.~\ref{fig:ThreeInt}.  When $\sN_{1,2}$ are both dominated by $A$ (or $\bar A$), or for one of them $A$ and $\bar A$ give equal contributions,  there will not be a discontinuity. 
Nevertheless, one may have expected that even in such cases the ``lack of equilibration'' may also lead to deviations from holographic results during the subsequent evolution after the joining. It is then rather curious that we do not observe such deviations at least in our sampling of the parameter space. This  indicates that holographic systems
``equilibrate'' remarkably efficiently. 

The discontinuity in the example of Fig.~\ref{fig:ThreeInt} also means that the strong subadditivity (SSA) condition 
is violated, as SSA implies that the time evolution should be continuous as can be seen from the results of Sec.~\ref{sec:vproof}. The violation of SSA can also seen
from the behavior of the entanglement entropy at a fixed time as follows. Consider at time $t = t_c - \ep$ another three-interval region $\tilde A$ which differs from $A$ only by having a slightly smaller $\tilde L_2 = L_2 - \de$ with $\de \to 0,\, \ep \to 0,\, \de > \ep$. Then at time $t_c - \ep$ 
for $\tilde A$, the corresponding $\sN_1$ and $\sN_2$ have already joined, hence at $t_c - \ep$ the entanglement entropies  for $\tilde A$ and $A$ differ by a finite amount despite the fact that the two regions only differ infinitesimally, which violates SSA as expressed in~\eqref{nnp}. Thus, if a model can be constructed which guarantees SSA, then such discontinuities cannot arise.

The issue of ``lack of equilibration'' for $\sN (t)$ becomes more significant in higher dimensions.  
Recall the calculation of $v_E$ illustrated in Fig.~\ref{fig:region}, for which $\sN (t)$ has only one connected 
component.  Clearly 
degrees of freedom far separated in directions parallel to the boundary of the region cannot be in direct causal 
contact.  Phrased slightly differently, 
the maximal RPS model fails to take into account of causal constraints along longitudinal directions. 
Note that in this particular setup SSA does not appear to be violated. Thus longitudinal causality constraints (which arise 
only in $d > 2$) should be considered as an independent requirement from SSA. 

That applying RPS to $\sN(t)$ is suspect, makes one wonder whether a tighter bound 
than $v_E =1$ can be found, if longitudinal causality constraints are properly taken into account. Incorporating them would provide a better understanding of holographic result~\eqref{emrp}. In Sec.~\ref{sec:tensor} we propose another approach --- different from the family of RPS models --- for the computation of entropy in the infinite scattering limit, which again gives $v_E =1$. In this regard, it is also interesting to note that in the Floquet systems discussed in~\cite{2015arXiv150101971C}, there is an exact causal light cone for any local operator and $v_E=1$ is actually achievable, at least for a certain class of initial states. But we should note that the 
systems of~\cite{2015arXiv150101971C} do not appear to have a continuum limit and thus may not be describable 
as a relativistic system with a local Hamiltonian.

Finally, let us note that~\eqref{master} is based on tracing over a local Hilbert space at $t=0$. The tracing is not local at time $t$. This is a consequence of unitary transformations we performed so as to track the trajectories of particles. 
This is certainly not ideal. In $(1+1)$-dimension, the equivalent proposal~\eqref{master2} motivated from the tsunami picture is based on tracing out a local Hilbert space at time $t$, and thus is conceptually more appealing. But it appears not easy to generalize~\eqref{master2} to higher dimensions for regions of general shapes as the behavior of entanglement tsunamis become complicated at late times.

From the success of the maximal RPS model in reproducing intricate holographic results for multiple intervals in
$(1+1)$ dimensions, and the discussion of Sec.~\ref{sec:rpsdiss}, it is tempting to speculate that a model which incorporates RPS, SSA, and full causality constraints (including the longitudinal causality constraints) may go a long way toward describing entanglement propagation in interacting systems, and in particular should provide a microscopic physical 
understanding of holographic results. Such a model, however, appears hard to come by from the scattering picture. For example,  
it is not clear how to formulate a precise set of causality conditions on the partition of Hilbert space (on which RPS is based). In the next section we use intuition derived from tensor networks to construct an entropy function  which  coincides with  holographic results for any number of intervals.

\section{A new model inspired by tensor networks}\label{sec:tensor}

\subsection{Tensor network interpretation of the scattering picture}

\begin{figure}[!h]
\begin{center}
\includegraphics[scale=0.68]{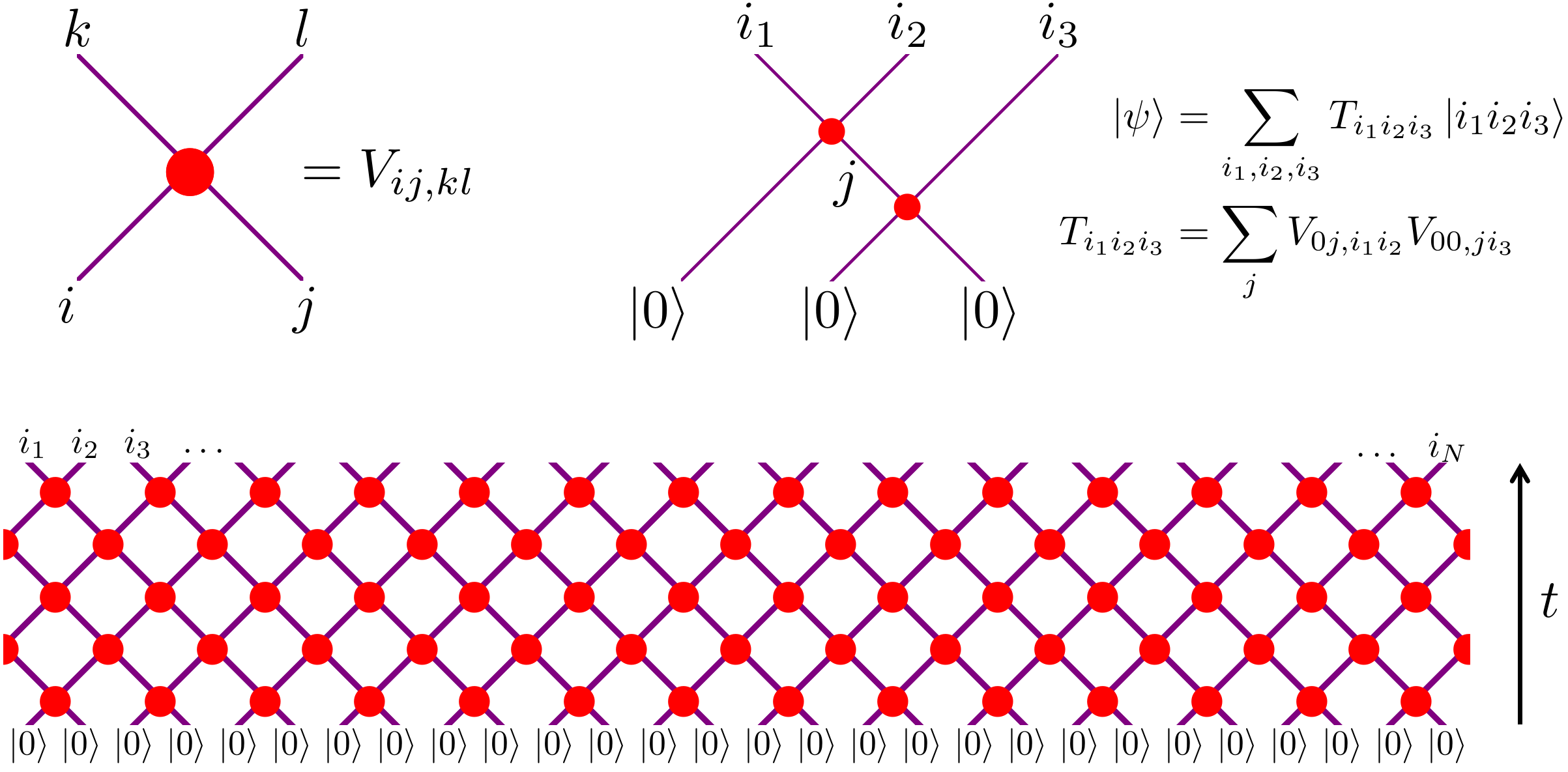}
\end{center}
\caption{{\bf Top:} Explanation of the notation $V_{ij,kl}$, and an example of the wave function created by three particles undergoing two scattering events. The two scattering events can be regarded as two unitary gates in a quantum circuit, and the wave function it prepares from the product state $\ket{000}$ is represented by the tensor network. The tensor $T_{i_1i_2i_3}$ can be obtained by performing one contraction on the internal index $j$. {\bf Bottom:} The quantum circuit of depth $t$ composed of the scattering matrices $V$ prepares $\ket{\psi(t)}$ from the product state $\prod_{\alpha=1}^N \ket{0}_\alpha$.  $\ket{\psi(t)}$ can be decomposed according to the basis $\ket{i_1i_2i_3\dots i_N}$~\eqref{psiDef}, and the coefficient tensor $T_{i_1i_2i_3\dots i_N}(t)$ is given by the tensor network on the figure.
}
 \label{fig:tensor}
\end{figure}

We can view the wave function $\ket{\psi(t)}$ prepared by the interacting model of Sec.~\ref{sec:infinite} as a tensor network, and abandon the idea of associating entropy to the particles ending up in the region $A$. Instead we can regard the time evolution as a quantum circuit of depth $t$ that prepares an entangled state from a product state through the action of the unitary scattering matrices $U$ introduced in Sec.~\ref{sec:singlescat}. It will be easier for us to convert $U$ into a four-indexed tensor $V_{ij,kl}$, and to be agnostic about the dimension of the Hilbert space associated to the particles scattered, and simply denote it by $\chi$; the indices run over $i,j,\dots\in(1, \chi)$. We explain our notation in Fig.~\ref{fig:tensor}.

The unitary time evolution prepares us a state
\be
\ket{\psi(t)}=\sum_{\{i_j=1\}}^\chi T_{i_1i_2i_3\dots i_N}(t)\,\ket{i_1i_2i_3\dots i_N}\,, \label{psiDef}
\ee
where $T_{i_1i_2\dots i_N}(t)$ can be obtained from contracting $V_{ij,kl}$ according to the pattern described in Fig.~\ref{fig:tensor}. This description provides a convenient interpolation between the free streaming and  the infinite scattering pictures. In the free streaming case $V_{ij,kl}=\de_{il}\de_{jk}$\footnote{Of course, in this case one has to supply a nontrivial locally entangled initial state.}, while for strong scattering we expect $V$ to be random. By tuning $V$, we should be able to learn how the behavior of entanglement spread interpolates between the two.

 The network of Fig.~\ref{fig:tensor} resembles 
that for global quench described in~\cite{Hartman:2013qma}, but there is a fundamental difference. Here the vertical direction is physical time, i.e.  Fig.~\ref{fig:tensor} is a quantum circuit, while in~\cite{Hartman:2013qma} the vertical direction is an auxiliary RG time. Accordingly, 
here the state of the system at a given time $t$ is described by a single slice of the network, while 
there the state is described by the whole network. Nevertheless, in the linear growth regime 
there appears some isomorphism between the network here and that of~\cite{Hartman:2013qma}. It would be interesting to understand this better.

\begin{figure}[!h]
\begin{center}
\includegraphics[scale=0.68]{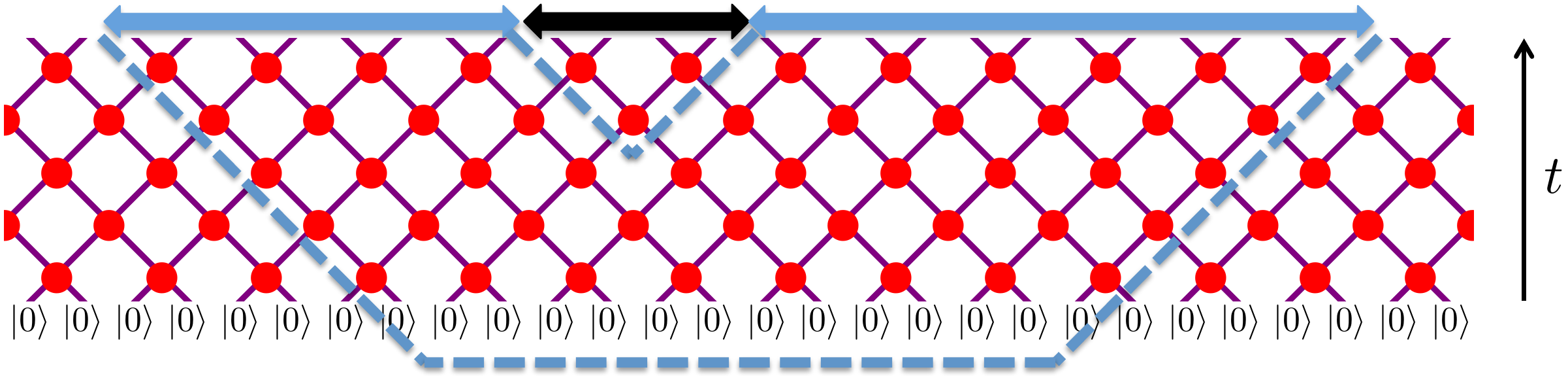}
\end{center}
\caption{We draw the minimal length cut at time $t$ corresponding to two intervals in $(1+1)$ dimensions at some intermediate time $t$. The that cut is not unique due to the discreteness of the tensor network. The number of links the cut intersects, $\ell_\text{cut}$ is what appears in~\eqref{cutbound}. Note that the horizontal section of the cut does not intersect with any links, and it readily follows that $\ell_\text{cut}$ is proportional to the holographic result~\eqref{HoloResult}.
}
 \label{fig:mincut}
\end{figure}

For tensor networks there exists a bound on the entanglement entropy of a region $A$:
\be
S_A\leq \ell_\text{cut}\log\chi\,, \label{cutbound}
\ee
where $\ell_\text{cut}$ is the length of the minimal cut through the tensor network that separates $A$ from $\bar A$. An example of a minimal cut for two intervals in  $(1+1)$ dimensions is presented in Fig.~\ref{fig:mincut}. For large enough $\chi$ and for a generic unitary scattering matrix $V$ the bound~\eqref{cutbound} is expected to be saturated. We then point out that from Fig.~\ref{fig:mincut} it follows immediately that in $(1+1)$ dimensions our tensor network exactly reproduces the holographic result~\eqref{HoloResult} for arbitrary number of intervals. 
 In the next section we discuss a higher dimensional model inspired by the minimal cut bound in tensor networks.

We emphasize that the tensor network of Fig.~\ref{fig:tensor} is a purely field theory construct, which we obtained from the picture of scattering quasiparticles, and {\it a priori} it has nothing to do with holography. The minimal cut in Fig.~\ref{fig:mincut} is also a tensor network concept. There are tantalizing connections with holography though. With physical time replaced by RG time, the network here resembles that of~\cite{Hartman:2013qma} which in turns looks like a ``nice slice" inside an eternal black hole. The minimal cut prescription is also reminiscent of the holographic extremal surfaces~\cite{Swingle:2012wq}.

\subsection{A geometric model inspired by tensor network}

Working with a discrete tensor network in higher dimensions is inconvenient due to the breaking of rotational symmetry. 
Using the physical insight from the tensor network picture, we now propose a simple continuum model for the time dependence of the entropy of an arbitrary region that satisfies all the physical criteria we are aware of that such a function should satisfy. We emphasize that we do not know of any local Hamiltonian that would produce the entropy given by this model, and it would be very interesting to find further criteria that the entropy function should satisfy that would constrain or rule out the model we propose in this section.

The entropy function after a global quench should satisfy the following geometric requirements, some of which we have discussed in previous sections, others were implicit:
 \begin{itemize}
\item[(a)] The entropy $S(A)$ is finite and positive for any spacelike region $A$ for $t>0$.

\item[(b)] $S(A)$ is the same for any Cauchy surface for $A$, or phrased in another way, $S(A)$ is a function of the domain of dependence of $A$. This is a requirement for all Lorentz invariant theories.

\item[(c)] Take any Cauchy slice of the spacetime that contains $A$, and define $\bar{A}$ as the complement of $A$ on this slice. Then $S(A)=S(\bar{A})$.

\item[(d)] $S(A)$ is invariant under translations and rotations of $A$. This condition follows from homogeneity and isotropy of the state.

\item[(e)]  For any space or lightlike region $A$ the scaling relation $S(A^{(t,x)}_\lambda)=\lambda^{d-1} S(A)$ holds, where we defined the  (in time and space) scaled region $A^{(t,x)}_\lambda=\{\lambda x\, \vert\, x\in A\}$ and  $\lambda>0$. This relation is just the generalization of~\eqref{scaling} to regions which do not lie on a constant time slice.

\item[(f)] For any region $A$ lying on a constant $t$ slice define the spatially scaled region at the same time by $A^{(x)}_\lambda=\{(t,\lambda \vec{x})\, \vert\, (t,\vec{x})\in A\}$. We have the volume law for small regions, 
\be
\lim_{\lambda\rightarrow 0} {S(A^{(x)}_\lambda)\ov \textrm{vol}(A^{(x)}_\lambda)}=s\,.
\ee

\item[(g)] For any region $A$ lying on a constant $t$ slice define the time scaled region by $A^{(t)}_\lambda=\{(\lambda t, \vec{x})\, \vert\,(t,\vec{x})\in A\}$. We have the tsunami law for small times, 
\be
\lim_{\lambda\rightarrow 0} {S(A^{(t)}_\lambda)\ov {\rm area}(\partial A^{(t)}_\lambda)\, \lam}=s\, v_E \,.
\ee

\item[(h)] Strong subadditivity $S(A)+S(B)\geq S(A\cap B)+S(A\cup B)$ applies for any two regions $A$ and $B$ on the same Cauchy surface. 
\end{itemize}

Using some insight from the tensor network picture, below we will produce a family of entropy functions $S(A)$  --- including one with $v_E=1$ --- with these properties in any dimension. Of course, the free streaming models of Sec.~\ref{sec:bound} also obey all these properties, but have $v_E\leq v_E^\text{free}$. 

For the construction let us introduce the set of all $(d-1)$-dimensional surfaces $\Phi_\alpha$ on $R^d$, $x^0>0$, that are smooth almost everywhere\footnote{That is, they can contain singular codimension-$1$ sets, where the tangent is not defined.} and have a normal vector $n=(n^0,\vec{n})$, which satisfies  in all points where the normal is well defined
\be
{ \sqrt{\vec{n}^2}\ov |n^0|}\leq {1 \ov \alpha} \,, \qquad 0<\alpha<\infty\,.
\ee
 That is, the slope of the tangent vectors of the surface is bounded above by $\alpha$; $\alpha=1$ means that the maximal slope is given by the speed of light. 

For any $\Sigma\in \Phi_\alpha$  let us define $V(\Sigma)$ as the volume of $\Sigma$ using the metric $ds^2=d\vec{x}^2$, which is degenerate. That is, $V(\Sigma)$ is the volume of the spatial projection of $\Sigma$. Then, let us consider the following entropy function for an arbitrary spatial region
\be
S_\alpha(A)=s \min_{\substack{\Sigma\in \Phi_\alpha, \\ \partial \Sigma=\partial A}} V(\Sigma)\,.
\ee
For this formula we can consider surfaces $\Sigma$ that hit the $t=0$ boundary of the spacetime and end there. Equivalently, we can think that once they hit $t=0$ on the boundary, they run on the $t=0$ plane, but this part of the surface does not have any volume. Because for $ds^2$ only the spatial volume counts, a large number of surfaces give the same entropy. 
The minimal surfaces $\Sigma$ can be thought of as the analog of a minimal cut in the tensor network; as explained in the caption of Fig.~\ref{fig:mincut}, the minimal cut is not unique either.

It is immediate that items (a), (b), (c), (d), (e), and (f) above are satisfied for any $\alpha< \infty$. Regarding the tsunami velocity for small times, we see that the least volume will be given by a surface that runs to the past of $A$ as fast as it can subject to the constraints on the slope to end on the $t=0$ boundary. This gives 
\be
v_E=\alpha\,.
\ee
From the arguments of Sec.~\ref{sec:vproof} it is clear that $v_E>1$ violates SSA, which is condition (h) above. This then restricts the range of $\al$ to $0<\al\leq 1$, and we will show below that for any $\al$ in this range (h) is satisfied. In particular, for $\al=1$ we have a model that gives $v_E=1$ and satisfies all the conditions we listed above.

\begin{figure}[!h]
\begin{center}
\includegraphics[scale=0.68]{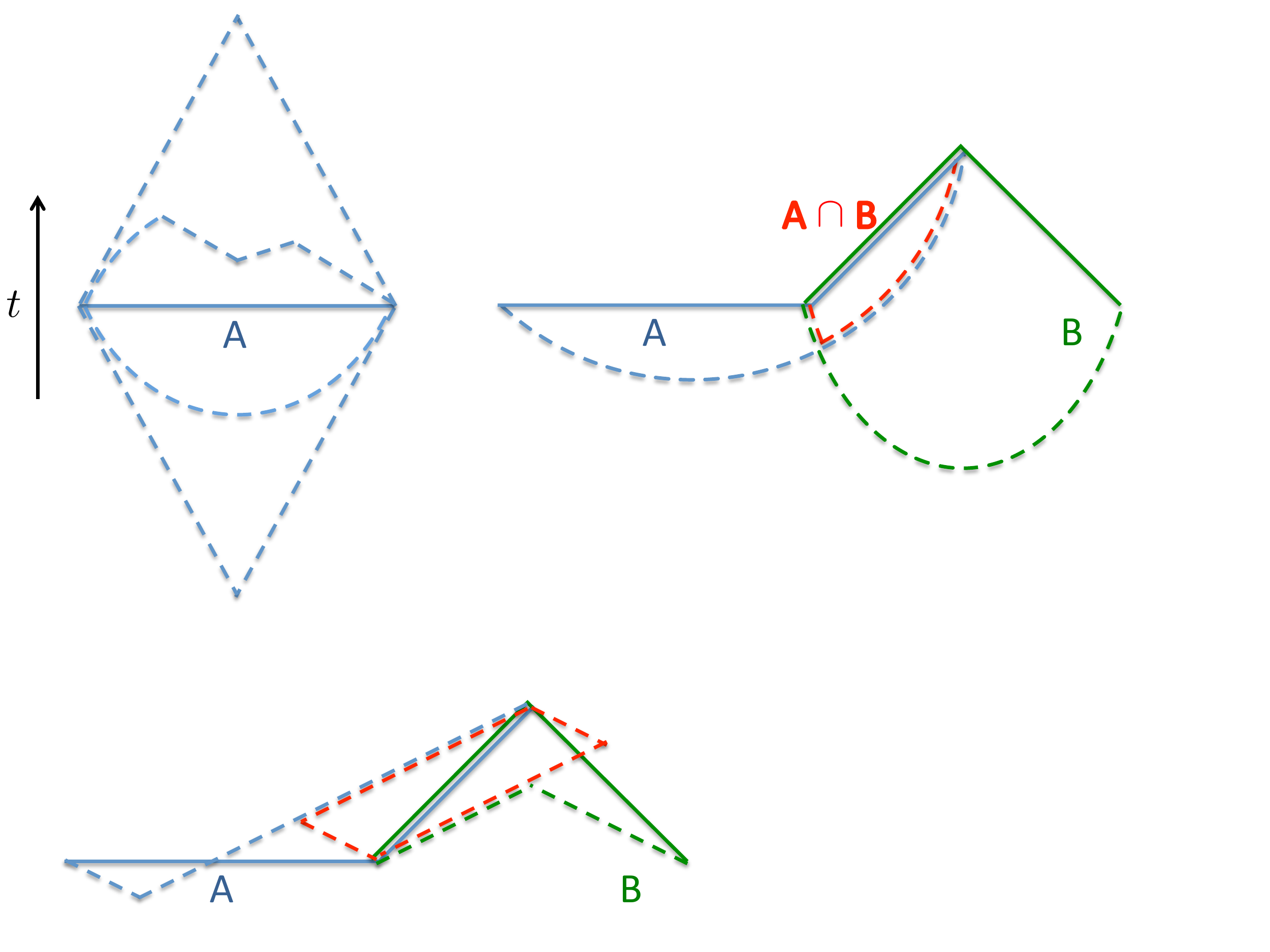}
\end{center}
\caption{{\bf Left:} Minimal surfaces drawn by dashed lines corresponding to a strip $A$. Because $ds^2$ is a degenerate metric only the spatial projection of the surfaces contribute to their volume $V(\Sig)$. Here we chose $\al<1$, so the surfaces from $\Phi_\alpha$ are allowed to be steeper than the light cone. {\bf Right:} Proof of SSA for the case $0<\al\leq 1$. We choose $A$ to consist of a constant time and a lightlike strip (drawn by blue), while $B$ is a union of two lightlike strips (drawn by green). (The proof works for arbitrary $A$ and $B$ on the same Cauchy slice.) Example minimal surfaces corresponding to $A$ and $B$ are drawn by dashed lines and colored in the same colors as $A$ and $B$ respectively. We can reinterpret the intersecting minimal surfaces as surfaces ending on $A\cap B$ (drawn by red) and $A\cup B$ (not drawn). Note that in the present case the red surface is minimal, while the one corresponding to $A\cup B$ may or may not be minimal depending on the position of the $t=0$ boundary.
}
 \label{fig:ABsetup}
\end{figure}

We now prove that SSA holds for $0<\al\leq 1$, but is violated in some circumstances for $\al>1$. Let us take a Cauchy slice containing two overlapping regions $A$ and $B$, and for $0<\al\leq 1$ use the degeneracy of the minimal surfaces to push $\Sigma_A$ and $\Sigma_B$ to the past so that they necessarily intersect, see Fig.~\ref{fig:ABsetup}. We can reinterpret the two surfaces as corresponding to $A\cup B$ and $A\cap B$. They may not be minimal, but minimal surfaces corresponding to $A\cup B$ and $A\cap B$ will just decrease the entropy, so SSA follows. These steps are identical to the proof of SSA in holography for static situations~\cite{Headrick:2007km}. 

\begin{figure}[!h]
\begin{center}
\includegraphics[scale=0.68]{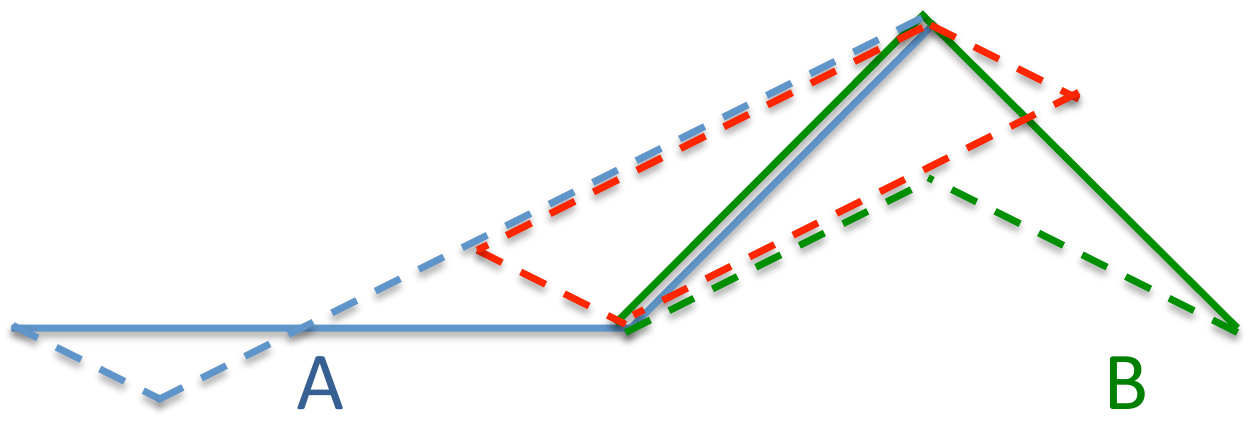}
\end{center}
\caption{Failure of SSA for $\al>1$. We use the same regions $A$ and $B$ as on Fig.~\ref{fig:ABsetup}. However, the minimal surface corresponding to $A$ now cannot be pushed further to the past, while the one corresponding to $B$ cannot be pushed further to the future. Thus, minimal surfaces corresponding to $A$ and $B$ cannot intersect, and the geometric proof of SSA breaks down. SSA is violated because the minimal surfaces corresponding to $A\cap B$ have larger volume than the spatial projection of $A\cap B$.
}
 \label{fig:ABsetup2}
\end{figure}

The above proof fails for $\al>1$, because in some cases the minimal surfaces corresponding to $A$ and $B$ cannot be made to intersect. SSA is violated in the setup of Fig.~\ref{fig:ABsetup2}, as the entropies for $A,\, B$, and $A\cup B$ are all proportional to their spatial projection, while $S(A\cap B)$ is larger than the spatial projection of $A\cap B$ due to the turning point. Note that for a lightlike (or sufficiently boosted) strip the minimal surface is necessarily cuspy in this case, any smoothing would bring the surface out of the set $\Phi_\alpha$; we allowed for codimension-1 singularities on $\Sigma$ in our assumptions.

This model for $v_E=1$ ($\al=1$) in $(1+1)$ dimensions gives exactly the prescription of the tensor network discussed in the last subsection, and it coincides with the holographic 
result. For higher dimensions it again leads to $v_E =1$. Note that in this model in higher dimensions the minimal surface that gives $v_E =1$ corresponds to the situation where entanglement essentially only propagates along one direction (i.e. the direction perpendicular to the boundary), which is clearly a bit peculiar. This aspect is similar to the behavior of the maximal RPS model discussed in the last section. It would be nice to understand whether we are missing some more subtle  isotropy constraint. Without using the isotropy of the system, one cannot hope to prove a more restrictive upper bound than $v_E \leq 1$, as many decoupled $(1+1)$-dimensional CFTs placed next to each other satisfy all our assumptions, and trivially yield $v_E=1$.
Another question we leave for future research is how the present model compares with the holographic results once we set $\alpha=v_E^\text{holo}$.

Finally, we note that the family of models we constructed in this section gives for a region lying on a constant time slice $S(A)\leq s \, {\rm vol}(A)$ for all times, and can give rise to nonzero mutual information at intermediate times. 
    
\pagebreak 
    
\subsection{Holographic theories maximize entanglement spread in $(1+1)$ dimensions} \label{sec:holomax}

The tensor network model seems to include by construction the heuristic idea of a ``maximal spread of entanglement''. If we think that strong interactions are effectively taken into account by random unitaries in the vertices of the tensor network, the model would try to produce locally a random pure state as fast as possible given the causality constraints. 

In fact, we can prove that the holographic entropy formula $S_H(A)$~\eqref{HoloResult} gives the absolute maximum for the entropy $S(A)$, for any number of intervals, and any global quench in relativistic theories in $(1+1)$ dimensions. 
 
In order to show this, let us take regions at a fixed time $t$ and start with a single interval of size $r$. From SSA we have $S^{\prime\prime}(r)\leq 0$ (for any translationally invariant state). From the physical assumptions about the global quench we also have that $S(r)\sim s r$ for small $r$, as small intervals saturate at the volume law, and that the entropy saturates to an $r$ independent constant $S(r)\sim 2 s v_E t$ for large $r$, as large intervals are in the linear regime. It is immediate that the holographic $S_H(r)=s\min(r,2 v_E t)$ is the maximum concave function with these two asymptotic behaviors. Here $v_E=1$, but we write it explicitly for convenience. 

For $n$ intervals the proof is by induction. We start by assuming that $S_H(B)$ is the maximum possible entropy for any set $B$ with less than $n$ intervals. 

For $n$ intervals let us write $A=\le\{(a_1,b_1),(a_2,b_2),\dots,(a_n,b_n)\ri\}$.  
Recall that the geodesics describing the bulk extremal surface for this region in the holographic setup all lie in some spacelike surface, and cannot cross each other~\cite{AW} (otherwise another surface exists with smaller area).\footnote{That $S_H(A)$ gives the absolute maximum for the entropy can be proven directly from the formula~(\ref{HoloResult}) without referring to geodesics, but perhaps it is easier to visualize the proof presented here.} It is not difficult to realize\footnote{This can also be proved by induction: a given geodesic splits the set of geodesics in two, to the ones outside and inside of it, and consequently the problem is mapped to the same problem with less intervals.} that there must be a geodesic in the extremal surface for $A$ that either joins the two end-points of an interval from $A$, say $I_j=(a_j,b_j)$, or there is at least one geodesic that joins the two consecutive points $b_k,a_{k+1}$ for some $k$. This last possibility is equivalent to joining the two end points in an interval from $\bar{A}$, which itself is a union of intervals and two half lines. As these two cases are completely analogous, without loss of generality let us restrict our attention to the first case, i.e., that $I_j=(a_j,b_j)$ determines a geodesic in the extremal surface. Then the rest of the geodesics give the extremal surface for $B=A-I_j=A\cap \bar{I_j}$. We have 
\be
S_H(A)=S_H(I_j)+S_H(B)\ge S(I_j)+S(B)\,,       
\ee
where in this last inequality we used the induction hypothesis. Now, by subadditivity 
\be
S(I_j)+S(B)\ge S(A)\,,
\ee
and this gives the proposed inequality
\be
S_H(A)\ge S(A)\,.
\ee

We note that the assumptions about the function $S(A)$ are minimal: we only used SSA for regions on constant time slices, the volume law for small regions and saturation of entropy for large regions. In this sense the proof can be carried over to nonrelativistic situations, or to systems with $v_E<1$. The proof can easily be generalized to regions not lying on a constant time slice in the relativistic case. 

From another perspective, this result also sheds different light on holographic entanglement entropy itself. For example, essentially the same proof shows that the holographic entanglement entropy given by minimal geodesics in pure AdS$_3$ space is an absolute maximum for the vacuum entanglement entropy of any region and any $(1+1)$-dimensional CFT with the same central charge.\footnote{We need the condition of equal central charge to have the same entropy for a single interval.}   
 
\section{Tsunami velocity and mutual information}  \label{sec:tus}

In this section we further examine the relation~\eqref{ve}, which we copy here for convenience (see also Fig.~\ref{uno}) 
\be
v_E=1-\frac{I(W(t),X)}{s_{\textrm{eq}} A_\Sig \de t} , \label{ve1} 
\ee
in the context of free streaming and interaction RPS models discussed in earlier sections, which sheds new insight onto the values of $v_E$. 

\begin{figure}[!h]
\begin{center}
\includegraphics[scale=0.68]{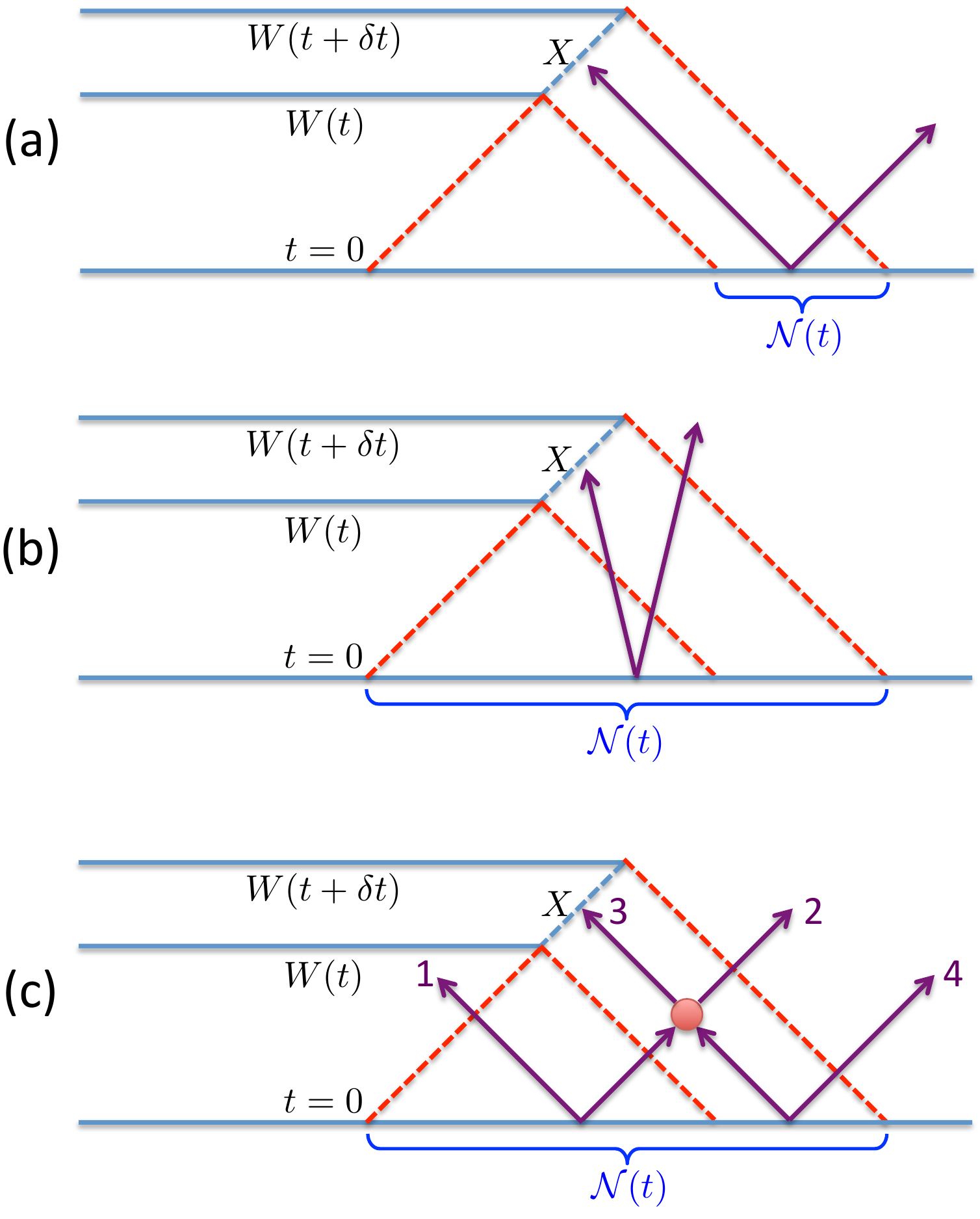}
\end{center}
\caption{Entanglement of $W (t)$ and $X$ in $(1+1)$ dimensions. The definition of $W(t)$ and $X$ are the same as those in Fig.~\ref{uno}. 
Plot (a): In a free propagation model with particles traveling at speed of light, only particles from region labelled 
by $\sN (t)$ can contribute to $S(X)$. Clearly there is no entanglement 
between $W(t)$ and $X$, as no particles from the same point can both reach $X$ and $W(t)$. 
Plot (b):  In a free propagation model which contains also particles traveling smaller than speed of light (i.e.~inside the light cone), there is entanglement between $W(t)$ and $X$. The region $\sN(t)$ which can contribute to $S(X)$ is also much larger. 
Plot (c): Even with particles only traveling at the speed of light, scatterings can also generate entanglement 
between $W(t)$ and $X$. Suppose initially 1 and 2 are entangled. Scattering between 2 and 3 will generate entanglement between 1 and 3, thus leading to entanglement between $W(t)$ and $X$. 
 The region $\sN(t)$ which can contribute to $S(X)$ is also much larger than that of (a) and is the same as that of (b). 
}
 \label{fig:new2d}
\end{figure}

Let us first consider $(1+1)$ dimensions. In a free propagation system with all particles traveling at the speed of light, 
from plot (a) of Fig.~\ref{fig:new2d}, it is clear that there is no entanglement between $W(t)$ and $X$ (see caption). This is consistent with 
$v_E =1$ in such a model. There is, however, entanglement between $W(t)$ and $X$ even in a free propagation system if there are also particles which can travel smaller than the speed of light. See plot (b) of Fig.~\ref{fig:new2d}. 
This means that whenever there is propagation of entanglement inside the light cone we expect $v_E < 1$. 

In a system with only particles traveling at the speed of light, interactions will generically generate entanglement 
between $W(t)$ and $X$, as indicated in plot (c) of Fig.~\ref{fig:new2d}. Thus in $(1+1)$ dimensions, interactions {\it slow} down $v_E$ compared with the free propagation at the speed of light, and for a general interacting theory we should have
$v_E < 1$. This conclusion is consistent with $v_E =1$ from CFT~\cite{Calabrese:2005in}, as well as holographic calculations~\cite{Balasubramanian:2011ur}, where the results apply in the ``infinite scattering limit'' $R, t \gg {1 \ov T}$. In other words, in these theories, we expect 
\be \label{subn}
\fR_\Sig (t) = 1 - {1 \ov t T}  f (t/R) + \cdots
\ee
with $f$ a positive function. For holographic theories, the function $f$ can be read from (5.38) of the second reference of~\cite{Liu:2013iza} and is indeed positive. \cite{Cardy:2015xaa} also found that interactions that break the scaling symmetry slow down the tsunami. It can also be readily checked that in the maximal RPS model $I (W(t), X)=0$. 

The story in higher dimensions is rather different. Even in a free propagation system with all particles traveling at the speed of light, there is entanglement between $W(t)$ and $X$, which can be immediately seen as follows.  Particles from a point $\vec x$ with nonzero velocity in directions perpendicular to $x_1$ will have a velocity smaller than the speed of light in the $x_1$ direction. In other words, when projected to the $x_1$ direction, these particles propagate ``inside'' the light cone. Thus from plot (b) of Fig.~\ref{fig:new2d} they will generate nonzero $I (W(t), X)$. 
This is perfectly consistent with our earlier discussion that $v_E < 1$ for higher dimensional free propagation models. In fact by computing directly $I (W(t), X)$ and then using~\eqref{ve} gives an independent derivation of $v_E$, which by consistency should agree with our earlier expression~\eqref{env}. Below we will check this is indeed the case. 

With already nontrivial entanglement between $W(t)$ and $X$ in free propagations, interactions can 
increase $v_E$ if they can reduce entanglement between $W(t)$ and $X$. It appears hard to imagine 
that the entanglement can be reduced completely to zero. But at the moment we do not have any definite argument to exclude the possibility that in an isotropic system there exists some limit, in which $v_E = 1$ can be approached like in~\eqref{subn}.

\begin{figure}[!h]
\begin{center}
\includegraphics[scale=0.6]{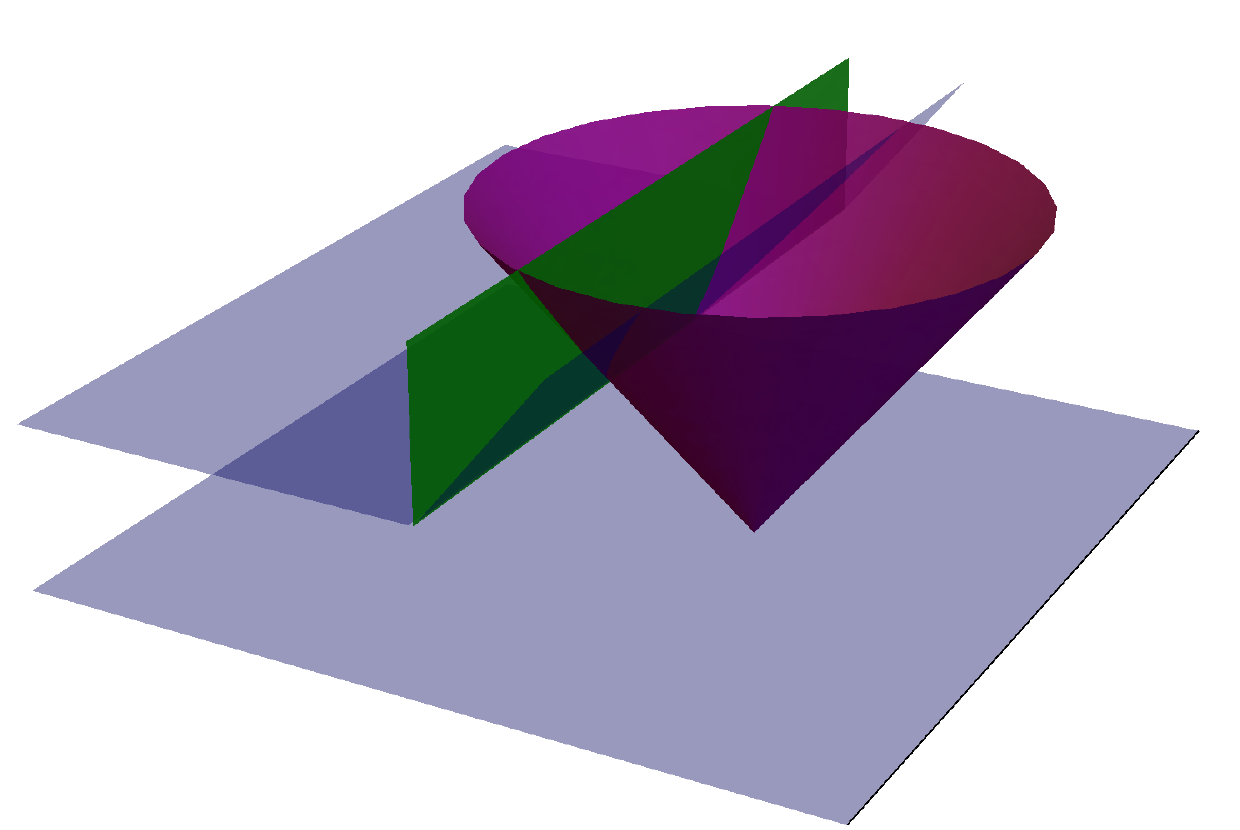}
\includegraphics[scale=0.6]{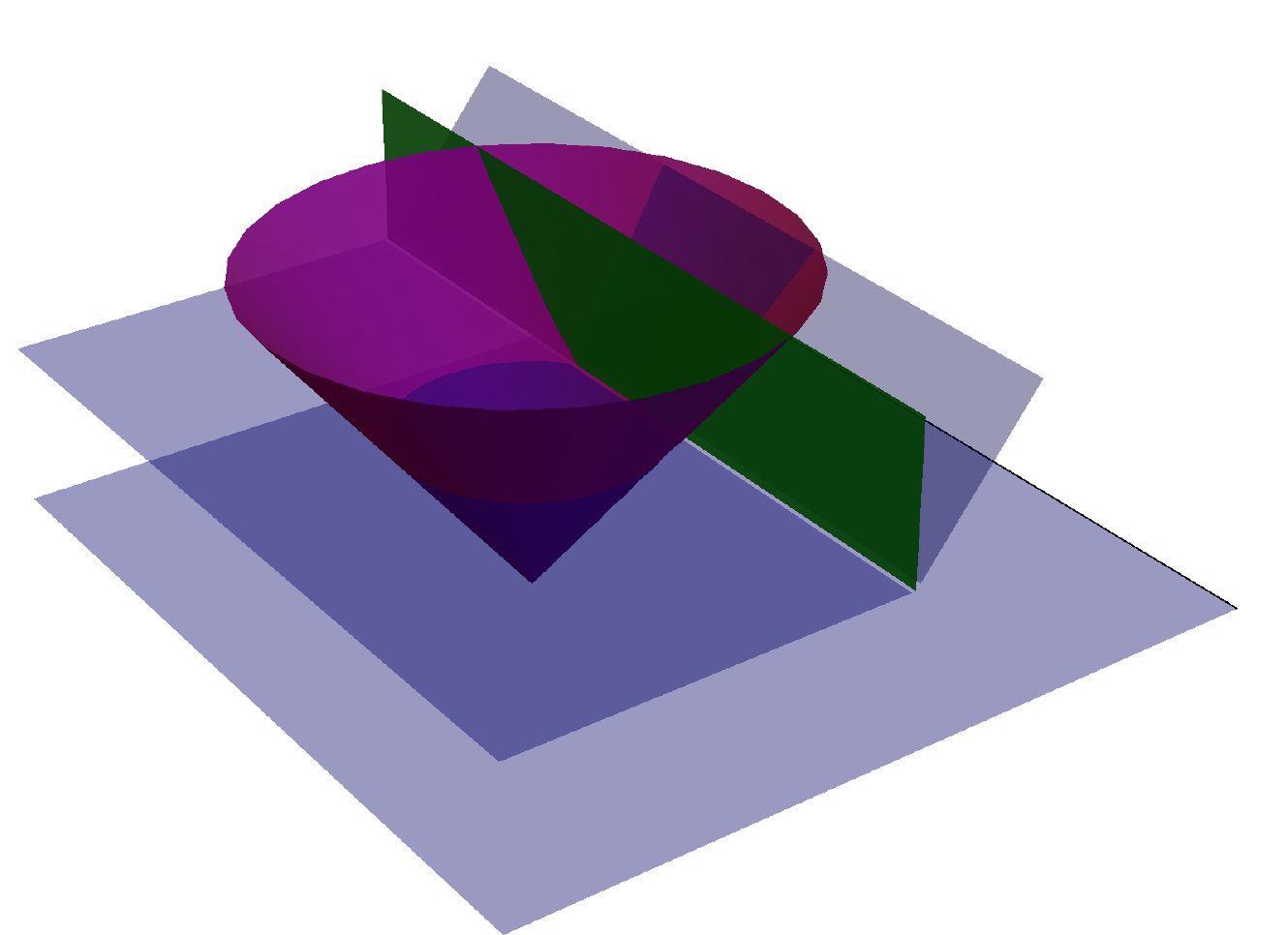}
\end{center}
\caption{$W(t)\cup X$ is drawn by blue, the light cone by purple, and the entangling surface by green. The left figure is for $y>0$, while the right figure depicts $y<0$.   \label{fig:lightsheet}}
\end{figure}

Finally, as promised earlier we show that $v_E$ derived from~\eqref{ve} agrees with~\eqref{env} for a
general free propagating model.  Consider a light cone from a point of distance $y$ from the entangling surface (see e.g. Fig.~\ref{fig:lightsheet}), as in Fig.~\ref{fig:linear}. Positive $y$ is outside $\Sig$, negative is inside $\Sig$.  A light cone centered at $y$ intersects the regions $W(t)$ and $W(t)\cup X$ in spherical caps with opening angles
 \bea
 \theta &=& \arccos\left( {y\ov t} \right)\,,\\
 \theta+\de \theta &=& \arccos\left( {y- \de t\ov t+ \de t}\right)\,,
 \eea   
respectively. The corresponding spherical caps have normalized area $\xi$ and $\xi+\de\xi$ respectively. 

In complete analogy with the entropy, the mutual information can be calculated for every light cone independently, as in~\eqref{masf}. It is given by a formula analogous to~\eqref{eprv}
\be
I(W(t),X)= A_\Sig\int_{-t}^t dy \ \big(\mu_{\rm cap} \le[\xi(y/t)\ri]+s_{\rm eq}\, \de\xi(y/t) - \mu_{\rm cap} \le[\xi(y/t)+\de \xi(y/t)\ri]\big)\,, \label{mutulight}
\ee
where we used that the small annulus region that we get from the difference of the spherical caps has to obey~\eqref{apro}. On Fig.~\ref{fig:lightsheet} this region is the two small arcs between the blue and green planes.
We series expand~\eqref{mutulight} to get
\be
I(W(t),X)= A_\Sig\, t\int_{-1}^1 dx \ \le(s_{\rm eq} - {d\mu_{\rm cap} \le(\xi\ri)\ov d\xi(x)}\ri) \de\xi(x)\,, \label{mutulight2}
\ee
where we introduced $x=y/t$. We calculate the first term using the expressions:
\bea
\de\xi(x)&=& \frac{\omega_{d-3}}{\omega_{d-2}}\,\sin^{d-3}\theta(x)\, \de\theta(x)\,,\\
\de\theta(x)&=&{\de t\ov t}\, {1-\cos\theta(x)\ov \sin\theta(x)}\,.
\eea
Using these expressions we obtain after the change of variables $dx=d\cos\theta$:
\bea
s_{\rm eq}\, A_\Sig\, t\int_{-1}^1 dx \  \de\xi(x)=s_{\rm eq} A_\Sig\,\de t\,.
\eea
For the second term in~\eqref{mutulight2} we use the chain rule
\be
\de\xi(x)={d\xi(x)\ov d\theta(x)}\, \de\theta(x)={d\xi(x)\ov d\theta(x)}\, { d\theta(x)\ov dx} \, {\de t\ov t} \, (1-x)
\ee
to write:
\bea
A_\Sig\, t\int_{-1}^1 dx \  {d\mu_{\rm cap} \le(\xi\ri)\ov d\xi(x)} \de\xi(x)&=& A_\Sig\,\de t \int_{-1}^1 dx \  {d\mu_{\rm cap} \le(\xi(x)\ri)\ov dx}\, (1-x)\\
&=&A_\Sig\,\de t \int_{-1}^1 dx \  d\mu_{\rm cap}\le(\xi(x)\ri)\\
&=& s_{\rm eq}\,A_\Sig\,\de t \, v_E\,,
\eea
where in the second line we did a partial integration and used $\mu_{\rm cap}(0)=\mu_{\rm cap}(1)=0$, while in the third line we used~\eqref{env}. Combining the two terms then gives
\be
I(W(t),X)=s_{\rm eq} A_\Sig\,\de t\le(1-v_E\ri)\,,
\ee
which reproduces the identity~\eqref{ve}.

\vspace{0.2in}   \centerline{\bf{Acknowledgements}} \vspace{0.2in}  We thank V.~Hubeny and J.~S.~Suh for early collaboration on this project and N.~Bao, A.~Bernamonti, J.~Cardy, A.~Chandran, F.~Galli, T.~Grover, T.~Hartman, D.~Huse, A.~Lewkowycz, J.~Maldacena, R.~Myers, M.~Roberts, V.~Rosenhaus, T.~Ugajin, and G.~Vidal for useful discussions. HC is supported by CONICET, CNEA and Univ. Nac. Cuyo, Argentina. The work of HL is supported in part by funds provided by the U.S. Department of Energy (D.O.E.) under cooperative research agreement DE-FG0205ER41360. MM is supported by the Princeton Center for Theoretical Science. We thank the hospitality of the KITP during the program ``Entanglement in Strongly-Correlated Quantum Matter", where we were supported in part by the grant NSF PHY11-25915. 

\appendix

\section{Concavity of the spherical cap entropy function} \label{appendix1}

\begin{center}
\begin{figure}[!h]
\includegraphics[scale=0.6]{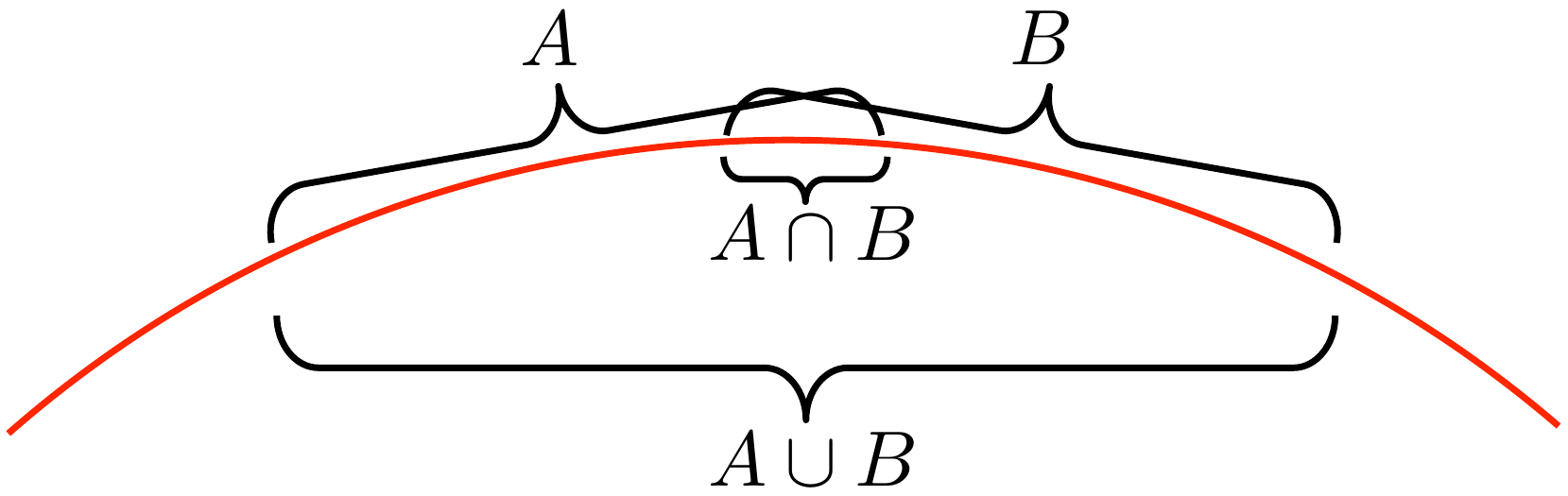}
\caption{ Regions $A, B, A \cap B, A \cap B$ on the light cone drawn by red. \label{fig:SSA2}
}
\end{figure}
\end{center}

First, we consider a situation in $d=3$, where we apply~\eqref{strm} to two regions $A, B$ shown in Fig.~\ref{fig:SSA2}, with all regions $A, B, A \cap B, A \cup B$ singly connected and included in half of the light cone. We then find 
\be 
\mu (\xi_A) + \mu (\xi_B) \geq \mu (\xi_{A\cap B}) + \mu (\xi_{A\cup B})  
\ee
which immediately leads to 
\be 
\mu'' (\xi) \leq 0, 
\ee
if we take $\xi_{A} = \xi, \xi_B = \xi, \xi_{A\cap B} = \xi -  \ep, \xi_{A\cup B} = \xi +  \ep$ with $\ep \to 0$. 
That is,  $\mu (\xi)$ is a concave function. 

Second, we consider spherical caps in $d>3$, where the geometry is more complicated. The simple proof of the concavity of $\mu_{\rm cap}$ in $d=3$ is similar to the entropic proof~\cite{Casini:2004bw} of the C-theorem in $d=2$, whereas the proof in $d>3$ below uses the more complex method of~\cite{hm} that was originally developed to prove the F-theorem in $d=3$.

\begin{figure}[t]
\begin{center}
\includegraphics[scale=0.7]{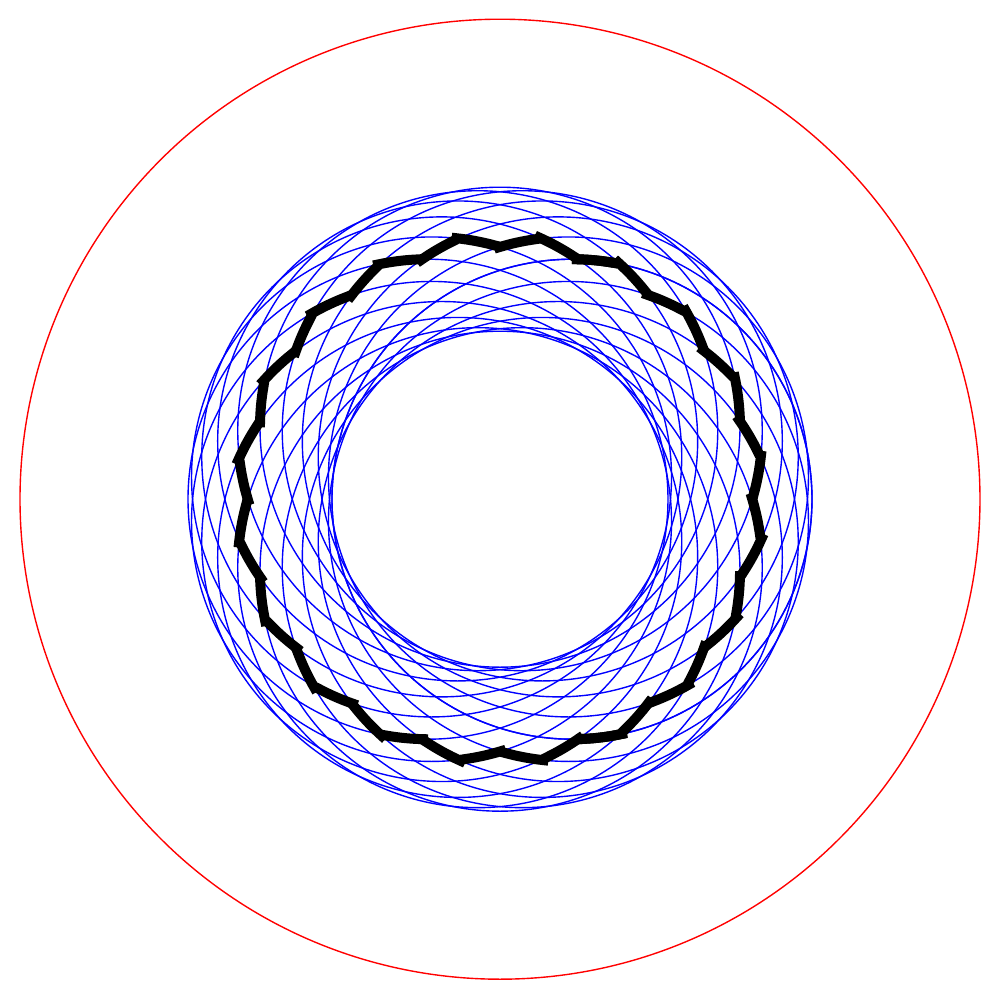}
\caption{$N$ spherical caps which are copies of a single spherical cap rotated around a point different from its center (shown in blue). The unit sphere is shown in red and the region with black contour is one of the regions appearing in the right hand side of (\ref{ecu}). 
  \label{rotated}}
\end{center}
\end{figure}

We are going to use strong subadditivity to prove the concavity of the function $\mu_{\rm cap}(\xi)$ as a function of the normalized area $\xi$. Strong subadditivity (\ref{strm}) for two spherical caps $A$ and $B$ involves the intersection and union of these regions, which are not spherical caps anymore. In order to have an inequality containing only spherical caps we can follow the procedure used in \cite{hm}. We use a large number $N$ of spherical caps $X_i$, $i=1,...,N$, which are copies of one spherical cap rotated around a point in the sphere, see figure \ref{rotated}.  Strong subadditivity leads to   
\begin{equation}
\sum_{i=1}^N S(X_{i}) \geq S(\cup _{i}X_{i})+S(\cup _{\{ij\}}(X_{i}\cap
X_{j}))+S(\cup _{\{ijk\}}(X_{i}\cap X_{j}\cap X_{k}))+...   +S(\cap _{i}X_{i}) \,. \label{ecu}
\end{equation}
The regions on the right hand side of this inequality are not spherical caps, but as shown in figure \ref{rotated} they approach to spherical caps in the limit of large $N$\footnote{We know the wiggly caps of figure \ref{rotated} have an entropy which converges to the one of the spherical cap because, as we have shown in Sec.~\ref{sec:bound1}, variations of entropy are bounded by $s_{\rm eq}$ times variations of area.}. In this limit (\ref{ecu}) will be converted into an inequality involving an integral of spherical caps with sizes varying between a maximum $\xi_{\rm max}$ and a minimum $\xi_{\min}$ (see figure \ref{rotated}). Following \cite{hm} we can then take $\xi_{\rm max}-\xi_{\rm min}=\epsilon$ and expand the inequality for small $\epsilon$ to get an inequality for $\mu_{\rm cap}(\xi)$ and its derivatives for a single $\xi$.

Though we can follow these same steps here, we can obtain directly the final result without doing the explicit calculations by arguing as follows. As a result of this procedure we should get a differential inequality for $\mu_{\rm cap}(\xi)$. Because strong subadditivity is linear in the entropy and involves four regions, it will always lead to linear differential inequalities containing at most second derivatives of the entropies. Hence we should get
\be
f_1(\xi) \mu_{\rm cap}^{\prime\prime}(\xi)+f_2(\xi) \mu_{\rm cap}^\prime(\xi)+ f_3(\xi) \mu_{\rm cap}(\xi)\leq 0\,.\label{ine1}
\ee   
Now, it is evident that the constant function $S(A)={\rm const.}$ is a solution of the strong subadditive equation $S(A)+S(B)=S(A\cap B)+S(A\cup B)$ rather than inequality. Hence, a constant $\mu_{\rm cap}(\xi)={\rm const.}$ must be a solution of  (\ref{ine1}) with equality rather than inequality. This is only possible if $f_3(\xi)\equiv 0$. In the same way, the area function $S(A)= \xi_A$ is a solution of strong subadditivity with equality for any regions. Hence $\mu_{\rm cap}(\xi)=\xi$ should be a solution of (\ref{ine1}) with equality to zero. This implies $f_2(\xi)\equiv 0$. Therefore we get that $\mu_{\rm cap}^{\prime\prime}(\xi)$ is always either positive or negative for $\xi$ according to the sign of $f_1(\xi)$. But we know the examples discussed in this paper all have  $\mu_{\rm cap}^{\prime\prime}(\xi)\leq 0$. This then implies $f_1(\xi)$ is positive, and in turn the general inequality
\be
\mu_{\rm cap}^{\prime\prime}(\xi)\leq 0\,
\ee  
for any entropy function on the sphere.\footnote{It is important though that we are assuming a finite and smooth entropy function~\eqref{apro}, in contrast to the vacuum contribution to the entropy, which would give a divergent area law piece.}

\section{Selected consequences of ballistic propagation of entanglement }  \label{sec:details}

In this appendix we study in more detail various aspects of entanglement propagation using the three examples 
discussed in Sec.~\ref{sec:exk}. We will mostly use spherical regions for illustration, although in some cases we make general remarks valid for all measures and general shapes. More specifically, we will investigate the following issues: 

\begin{itemize}

\item full time evolution and the entanglement rate~\eqref{groR} for simple shapes such as the sphere,

\item for the EPR example, one can further define an entanglement density which can be used to better visualize the spread of 
entanglement,

\item saturation time for generic shapes,

\item finite volume effects, 

\item mutual information.

\end{itemize}
 
Along the way we also compare the results obtained here for ballistic propagation with those of holographic systems.

\subsection{Time evolution for simple shapes}

\subsubsection{The $d=2$ interval}

\begin{figure}[!h]
\begin{center}
\includegraphics[scale=0.6]{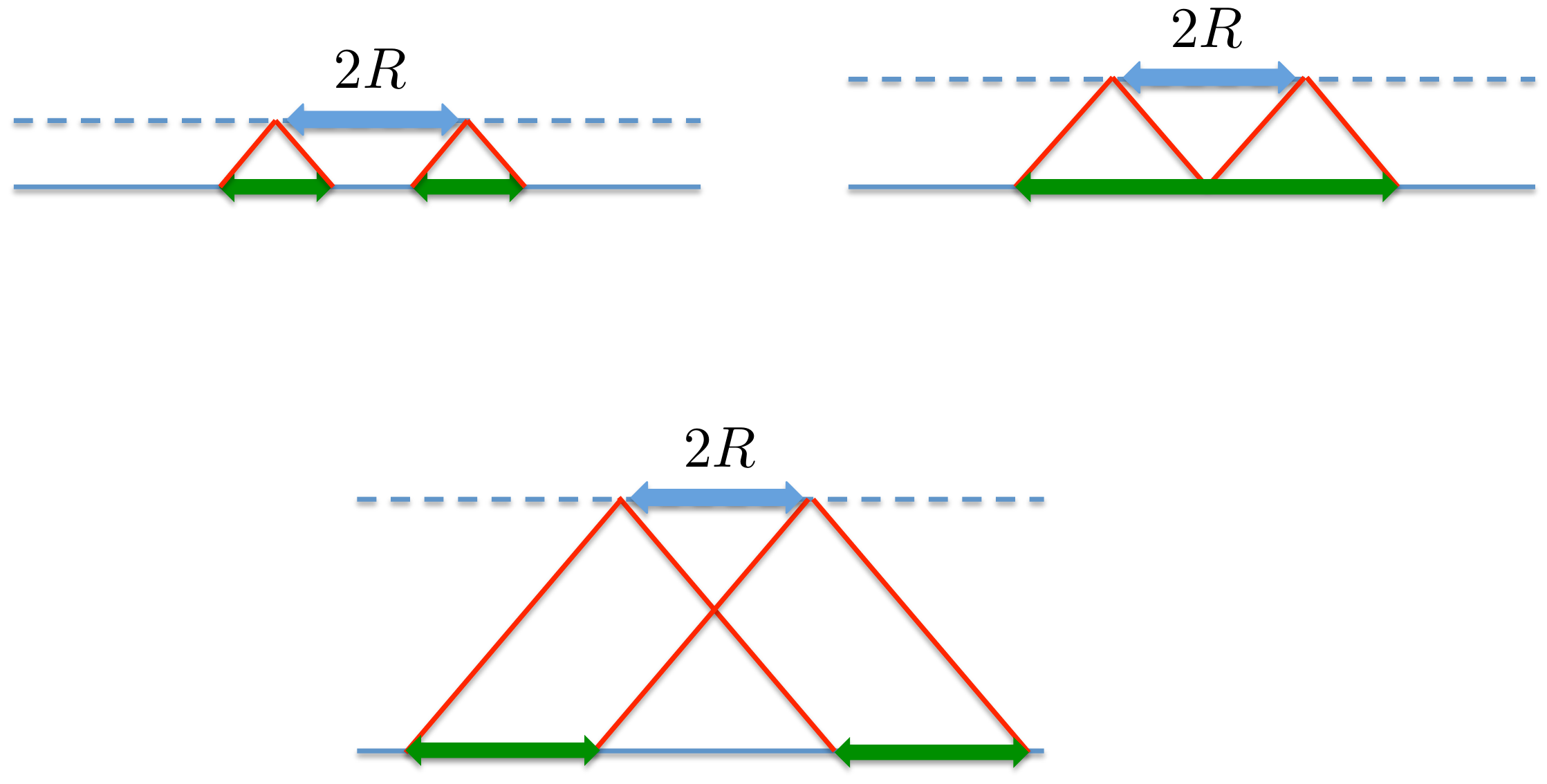}
\caption{$d=2$ spacetime diagram. The first figure is for $t <R$, the second for the saturation time $t= t_s =R$, and the third after $t >R$. The length of the green intervals determines the entanglement entropy. \label{1d}}
\end{center}
\end{figure}

Let us first briefly review the results in $d=2$~\cite{Calabrese:2005in}. 
The entangling region reduces to an interval and the geometry of the problem is presented in fig.~\ref{1d}. On the figure the green region is $\sN(t)$, defined as the region of the space where lie the centers of the light cones that have non-empty intersection with $\Sigma$ at time $t$. In one spatial dimension, because the intersection of a light cone with the region $\Sig$ is a point,  all entanglement measures are equivalent. From the perspective of quasiparticle propagation, since there are only two directions and 
all particles in a direction from a single point propagate side by side, no matter how entanglement is distributed, the entanglement spread  reduces effectively to that of the EPR example. 

We take the width of the interval to be $2R$. From Fig.~\ref{1d} we conclude that the entanglement entropy has the time dependence: 
\be \label{2dresult}
S={s\ov 2}\begin{cases}
4t \qquad (t <R)\\
4R   \qquad (t>R)\ ,
\end{cases}
\ee
where $s$ is given by~\eqref{endi}. We divided $s$ by $2$ as $n/2$ is the density of quasi-particle pairs. Both the slope of the linear growth and
the saturation time agrees with those found from direct field theory calculation~\cite{Calabrese:2005in} and those obtained from holography~\cite{Balasubramanian:2011ur}.

\subsubsection{Full time evolution for a sphere} \label{sec:Fulltime}

Let us now take spherical entangling surfaces of radius $R$, for which the high symmetry enables an analytic treatment of the full time evolution. For a spherical region, the intersection $A$ of any light cone with the region is always a spherical cap, and the EPR, and RPS measures all contribute to $\mu[A]=\mu(\min(\xi_A,\xi_{\bar A}))$. We can conveniently treat all measures at once by working with:
\be \label{ghzm2}
 \mu [A; m] = {s \ov 2m} \le[1- \abs{1-  2 \xi_A}^m \ri] \,,
\ee
where $m=1$ gives the result for the EPR and RPS measures, while $m>1$ corresponds to GHZ blocks. The absolute value comes from combining the relation $\xi_{\bar A}=1 - \xi_A$ with~\eqref{ghzm} valid when $\xi_A<1/2$.

\begin{figure}[!h]
\begin{center}
\includegraphics[scale=0.6]{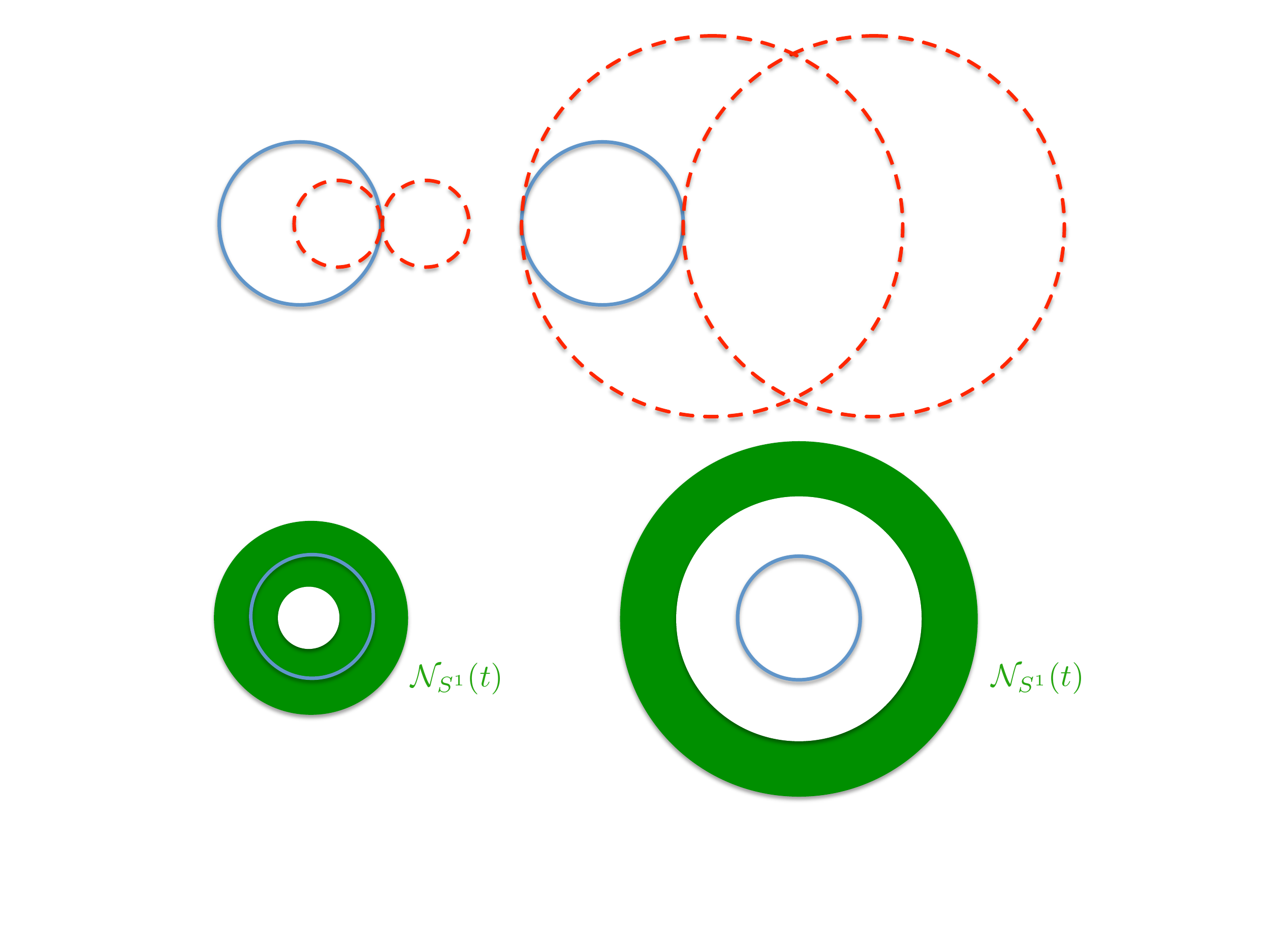}
\caption{The region $\sN_{S^{d-2}}(t)$ for $d=3$. 
The left two figures are for $t <R$ and the right ones are for $t >R$.
The top figure in each column gives two ``critical'' light cones (dashed circles) which just touch the entangling surface (solid circle). They can be used to determine the boundary of  $\sN_{S^{d-2}}(t)$, which is the shaded green region in the bottom row.
 For $t=R$ we would get a filled green disk of radius $2R$. Note that if we restrict to the $x$-axis we get back the $d=2$ picture of fig.~\ref{1d}.  \label{disc}}
\end{center}
\end{figure}

To apply~\eqref{masf} we need to work out first the region $\sN_{S^{d-2}}(t)$, which is explained in Fig.~\ref{disc}.
We then find that 
\be
S(t)= \int_{\sN_{S^{d-2}}(t)}d^{d-1}r\, \mu_{\rm cap}(\xi(r)) 
= \om_{d-2}\int_{\abs{t-R}}^{t+R} dr\, r^{d-2} \mu_{\rm cap}(\xi(r))\,, \label{zdef}
\ee
with
\be
\min(\xi(r),1-\xi(r))=\ha\,  I_z\le({d-2\ov 2},\ha\ri)\,, \qquad z\equiv -{(R+r+t)(R+r-t)(R-r+t)(R-r-t)\ov 4r^2 t^2}\,, \label{xiDef}
\ee
where we used the formula for the area of a spherical cap, and $I_z(a,b)$ is the regularized incomplete beta function. We did not find a way to perform this integral for general $d,\, m$, so we present some examples.

\begin{figure}[!h]
\begin{center}
\includegraphics[scale=0.6]{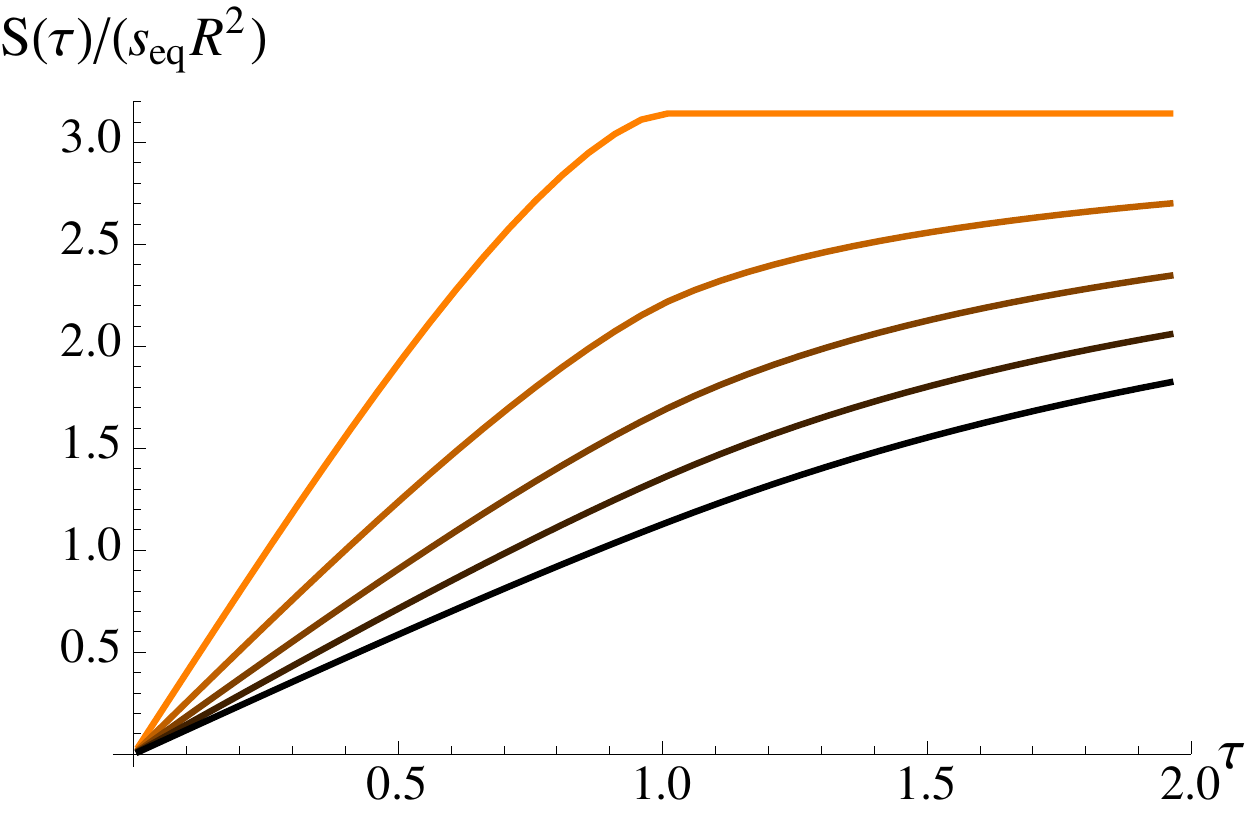}\hspace{1cm}
\includegraphics[scale=0.6]{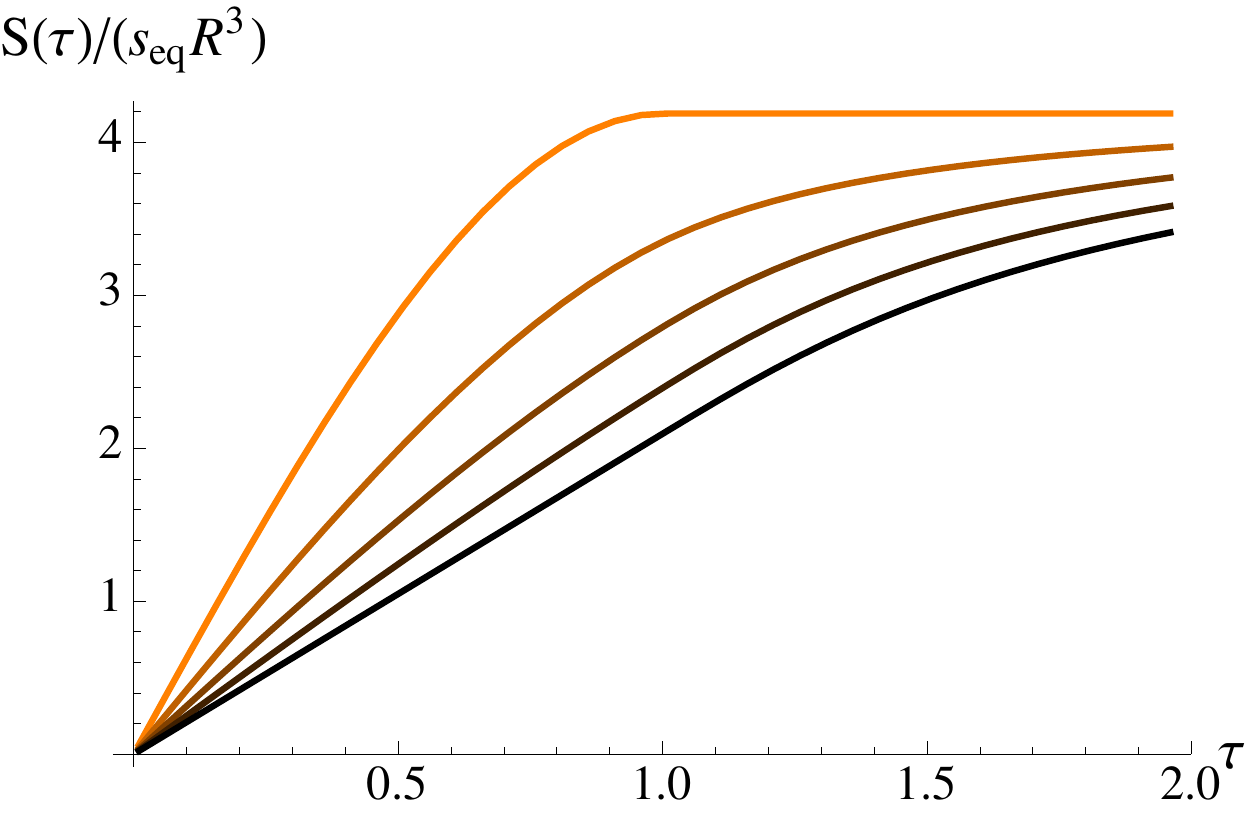}
\includegraphics[scale=0.6]{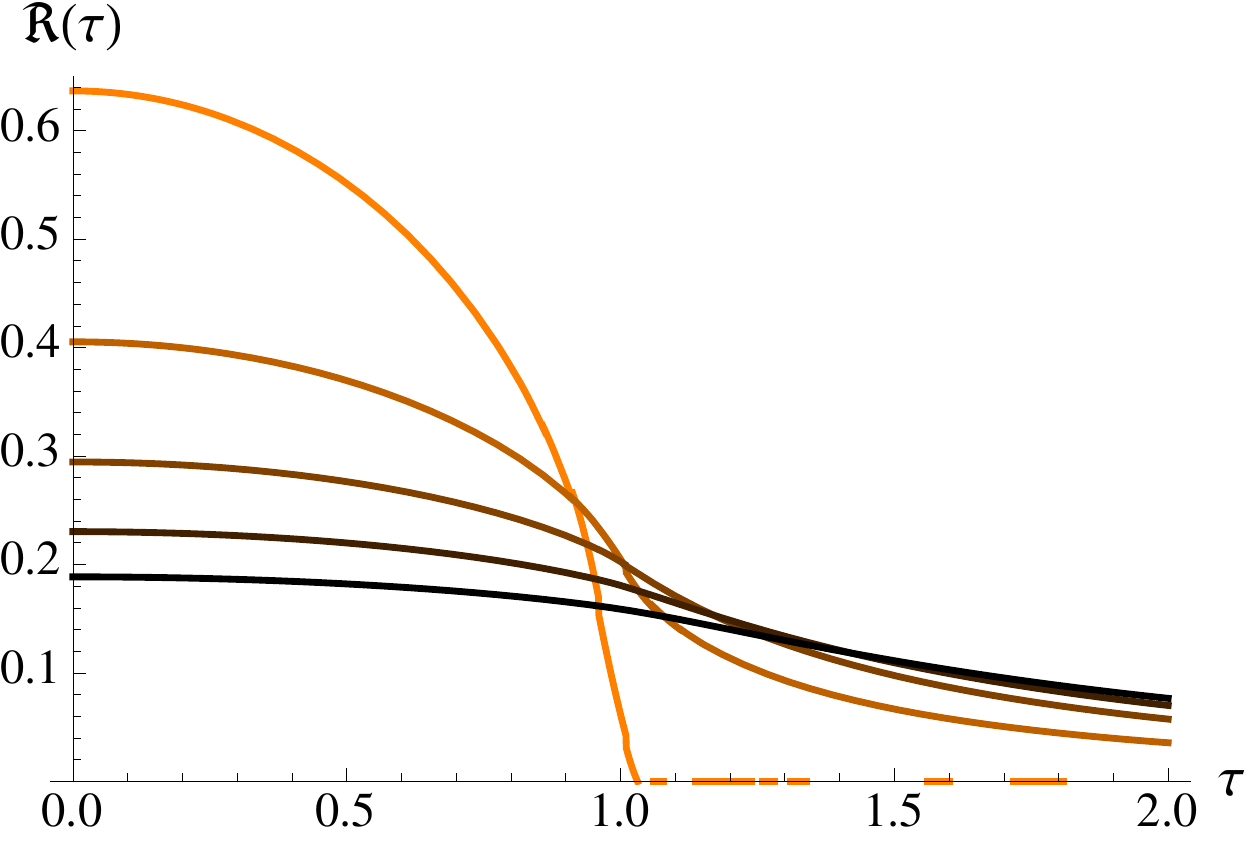}\hspace{1cm}
\includegraphics[scale=0.6]{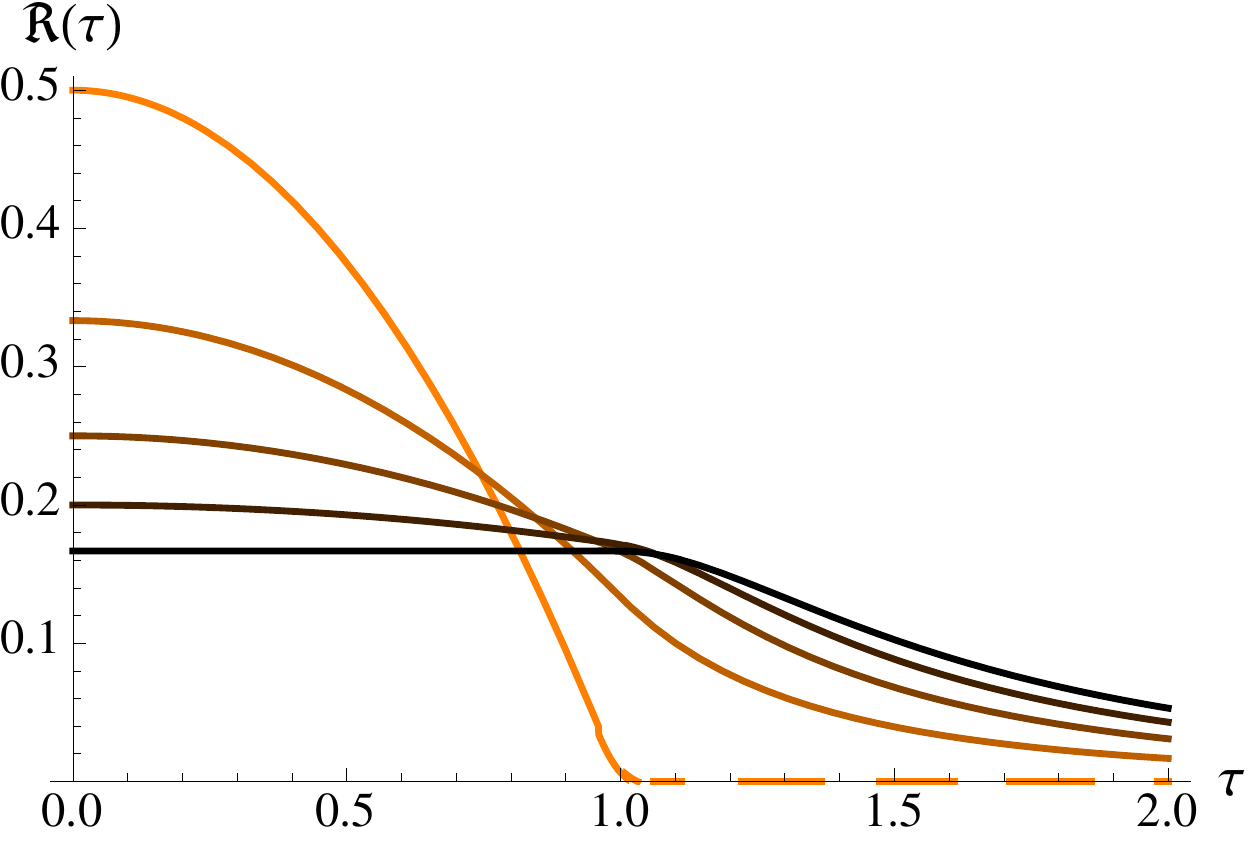}
\caption{ {\bf Top row:} Full time evolution for a sphere in $d=3,4$ for $m=1,\dots, 5$, with larger $m$ corresponding to darker color. 
{\bf Bottom row:} Rate of entanglement growth $\fR_{S^{d-2}}$~\eqref{groR} obtained by taking the time derivative of the functions in the top row. Note that $\fR_{S^{d-2}}$ is monotonically decreasing for all $m$, and hence is bounded by the tsunami velocity $v_E^\text{GHZ}$.
 \label{fulltime}}
\end{center}
\end{figure}

For the EPR case $m=1$, we find (with $\tau \equiv t/R$)
\bea
S_{d=3}(\tau)/R^2&=&\begin{cases}
 2 s\, \le[\tau\sqrt{1-\tau^2}+\arcsin\tau\ri] & (\tau<1)\\
 \pi s & (\tau>1)\,,
\end{cases}
\label{3dTime}\\
S_{d=4}(\tau)/R^3&=&\begin{cases}
2\pi s\, \le( \tau- {\tau^3\ov 3}\ri) & (\tau<1)\\
 {4\pi s\ov 3} &(\tau>1)\,,
\end{cases}\\
S_{d=5}(\tau)/R^4&=&\begin{cases}
\pi s\,\le( {10 \tau - 14 \tau^3 + 4 \tau^5\ov 6 \sqrt{1 - \tau^2}} +\arcsin \tau\ri) & (\tau<1)\\
 {\pi^2 s\ov 2} &(\tau>1)\ .
\end{cases}
\eea
Numerical plots for various $m$ in $d=3,4$ for the GHZ measure are given in Fig.~\ref{fulltime}. Note that while 
for EPR and random state measure, the saturation time is always $t_s = R$, for GHZ the saturation time is infinite. 
It can also be readily checked that for early times
\be
S(t)=s\om_{d-2} \, v_E^\text{GHZ}\, t+\dots
\ee
and at late time the entropy $S_{S^{d-2}} (t \to \infty) = s\, R^{d-1} V_{B^{d-1}}$,
consistent with our general discussion in Sec.~\ref{sec:eqv}.

\subsubsection{Strip in $d=3$}

\begin{figure}[!h]
\begin{center}
\includegraphics[scale=0.8]{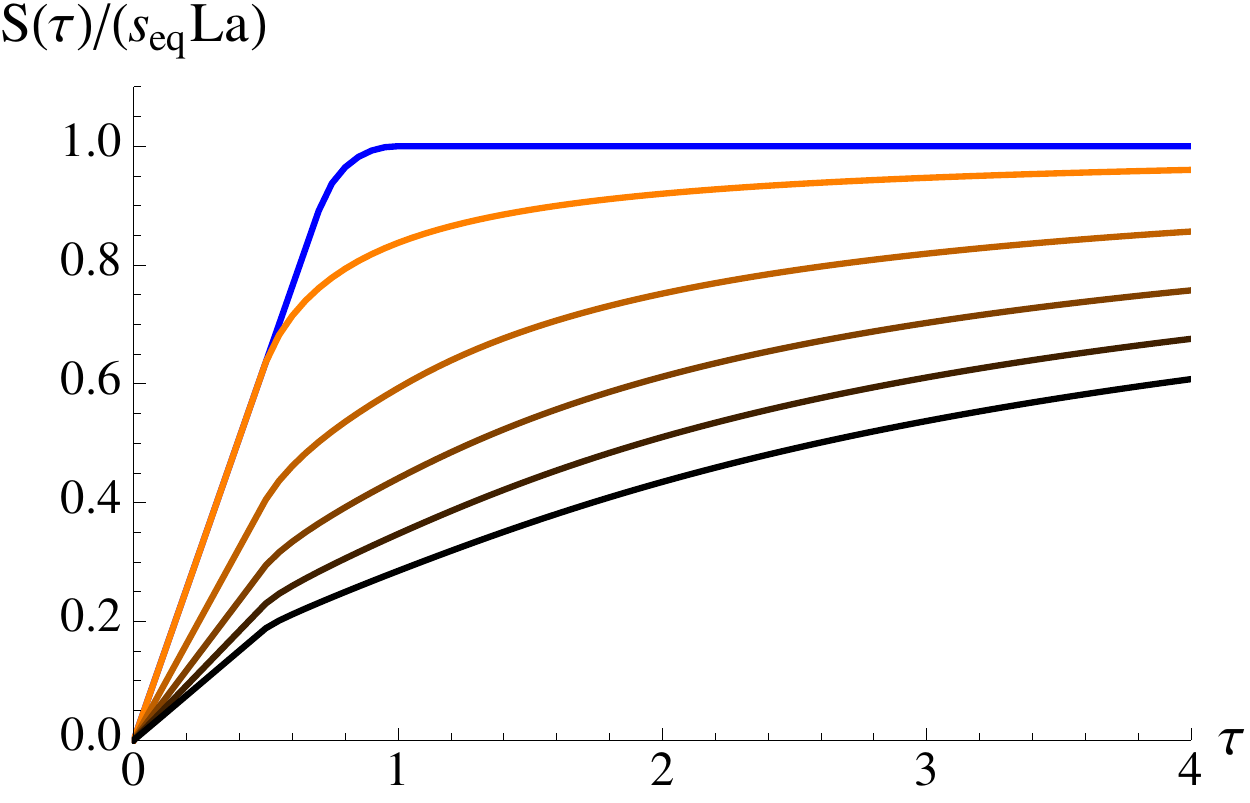}
\caption{ Full time evolution for a strip in $d=3$ for the GHZ block model with $m=1,\dots, 5$, with larger $m$ corresponding to darker shade of orange. The RPS model gives a different time evolution than the EPR model and is plotted with blue.
 \label{fig:strip3d}}
\end{center}
\end{figure}

We consider a strip of width $a$ and length $L$ in $d=3$ as another simple example. Our discussion will be less detailed, than for the sphere case. For early times, $t<a/2$ the time evolution is exactly linear, while for $t>a/2$ the width of the strip starts to play a role. The time dependence becomes more complicated as for some of the light cones $L_\text{strip} (\vx; t)$ becomes a disconnected region, and we have to use~\eqref{ghzmGeneral}. The result for early time valid for arbitrary measure is 
\be
S_\text{strip}(\tau)/(aL)= 2 s\, v_E\,\tau \qquad (\tau<1/2)\,,
\ee
with $\tau=t/a$.
 For the EPR model, we get for later times
\be
S_\text{strip}(\tau)/(aL)={2 s\ov \pi}\,\le[2\tau-\sqrt{4\tau^2-1}+\arccos{1\ov2\tau}\ri] \qquad (\tau>1/2)\,.
\ee
For the case of the strip the RPS model {\it is not equivalent} to the EPR model. The results for the RPS, the EPR, and GHZ block models are plotted in Fig.~\ref{fig:strip3d}. It is remarkable that the measures originating from a quasiparticle picture give an infinite saturation time, while the RPS model saturates in $\tau=1$.

\subsubsection{Two disks and strips in $d=3$} \label{sec:negativeR}

\begin{figure}[!h]
\begin{center}
\includegraphics[scale=0.5]{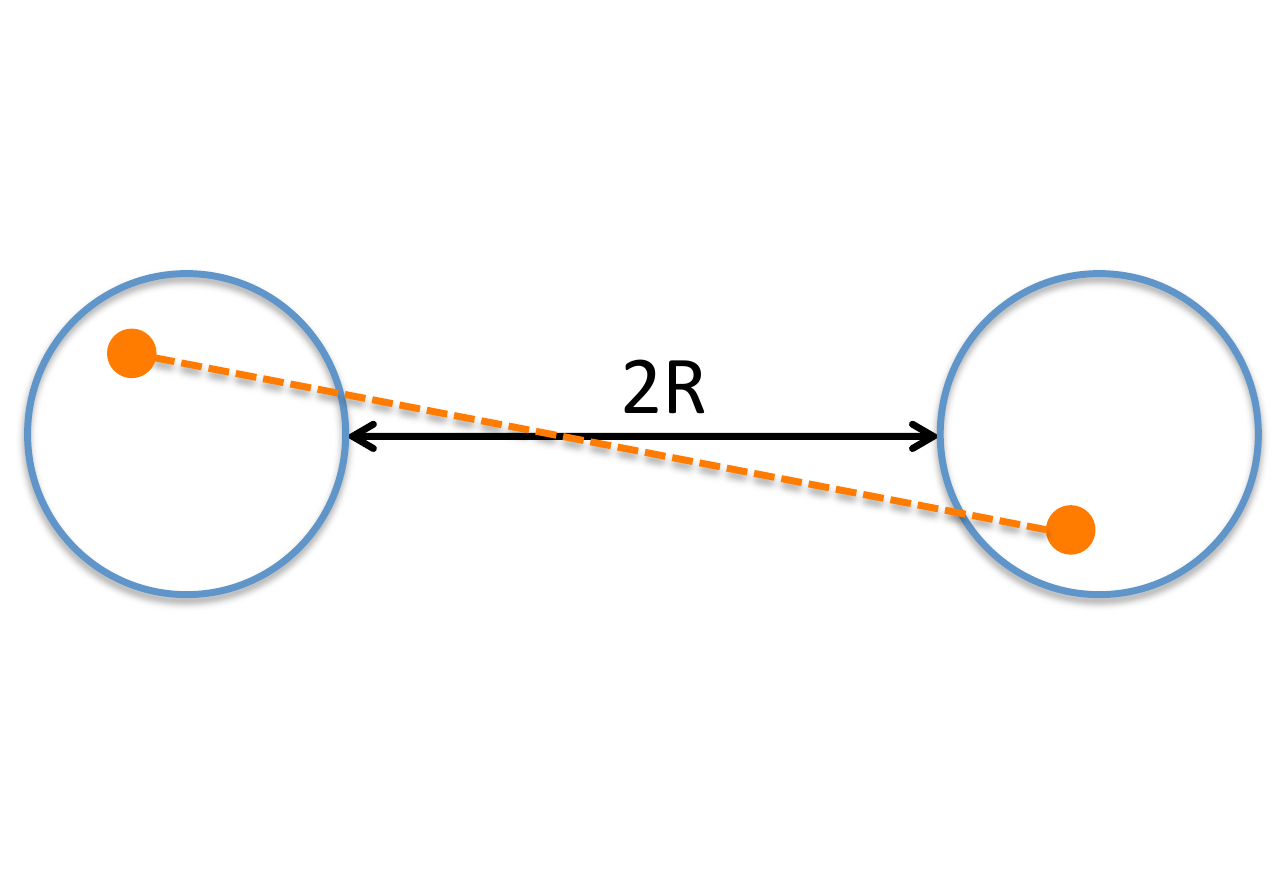}\hspace{1cm}
\includegraphics[scale=0.6]{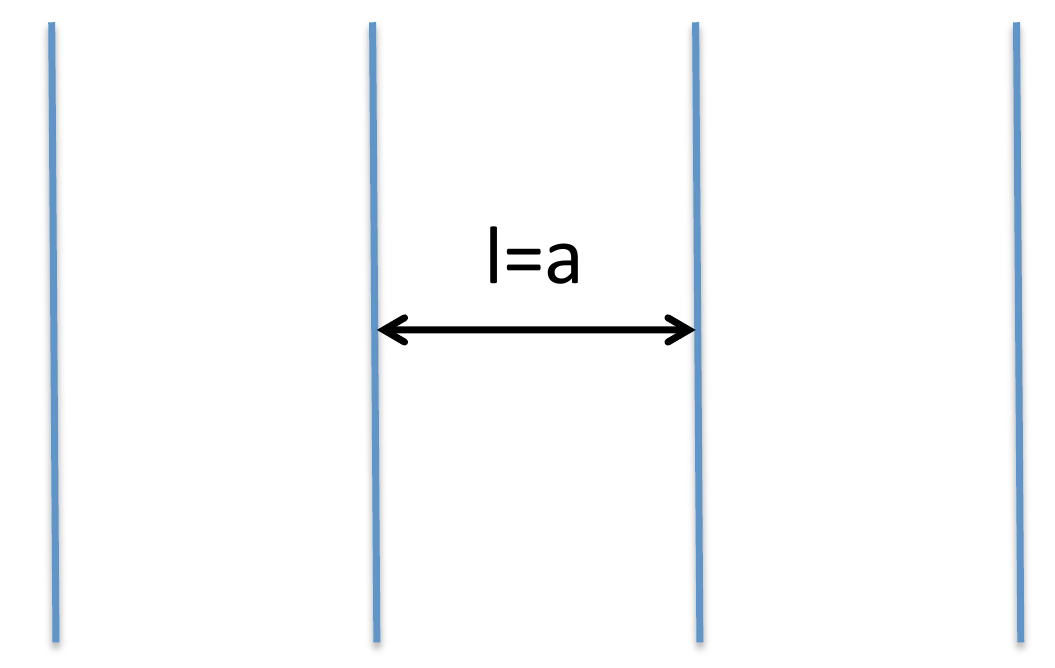}
\includegraphics[scale=0.6]{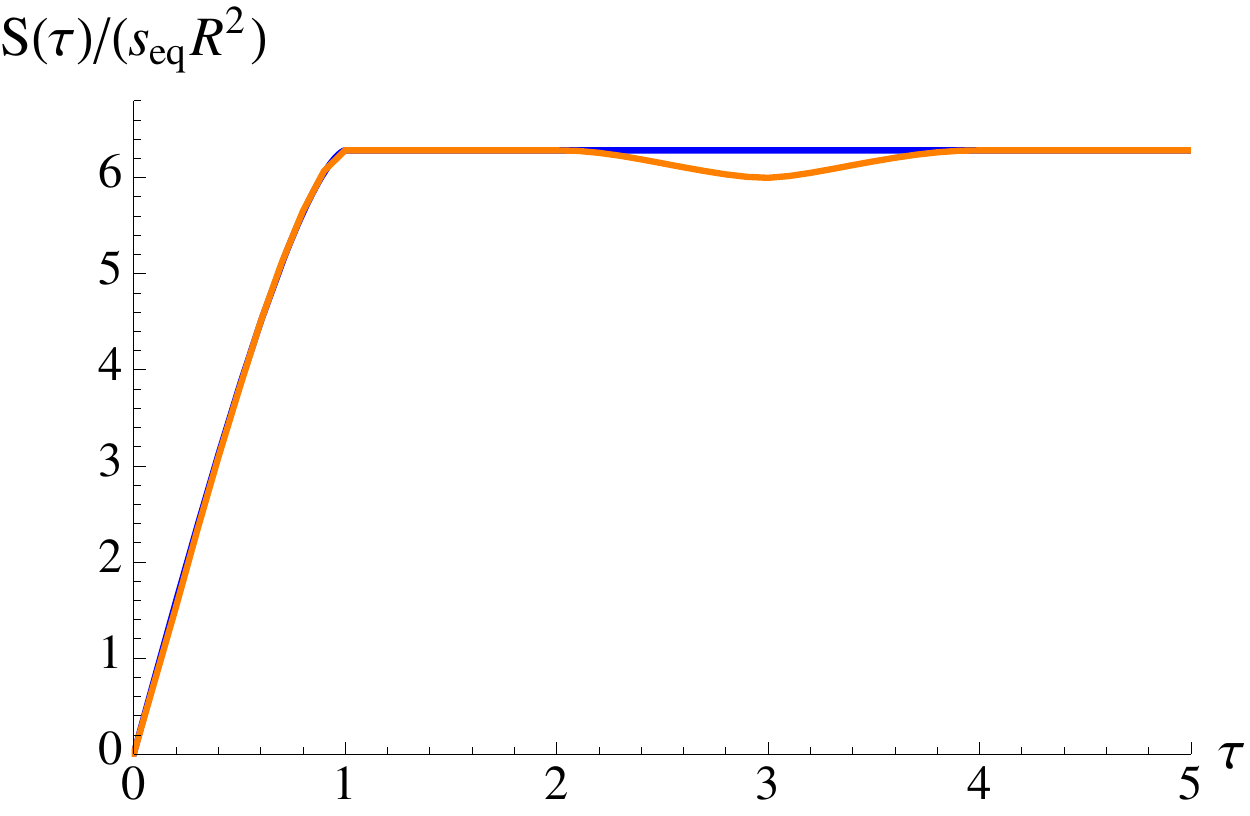}\hspace{1cm}
\includegraphics[scale=0.6]{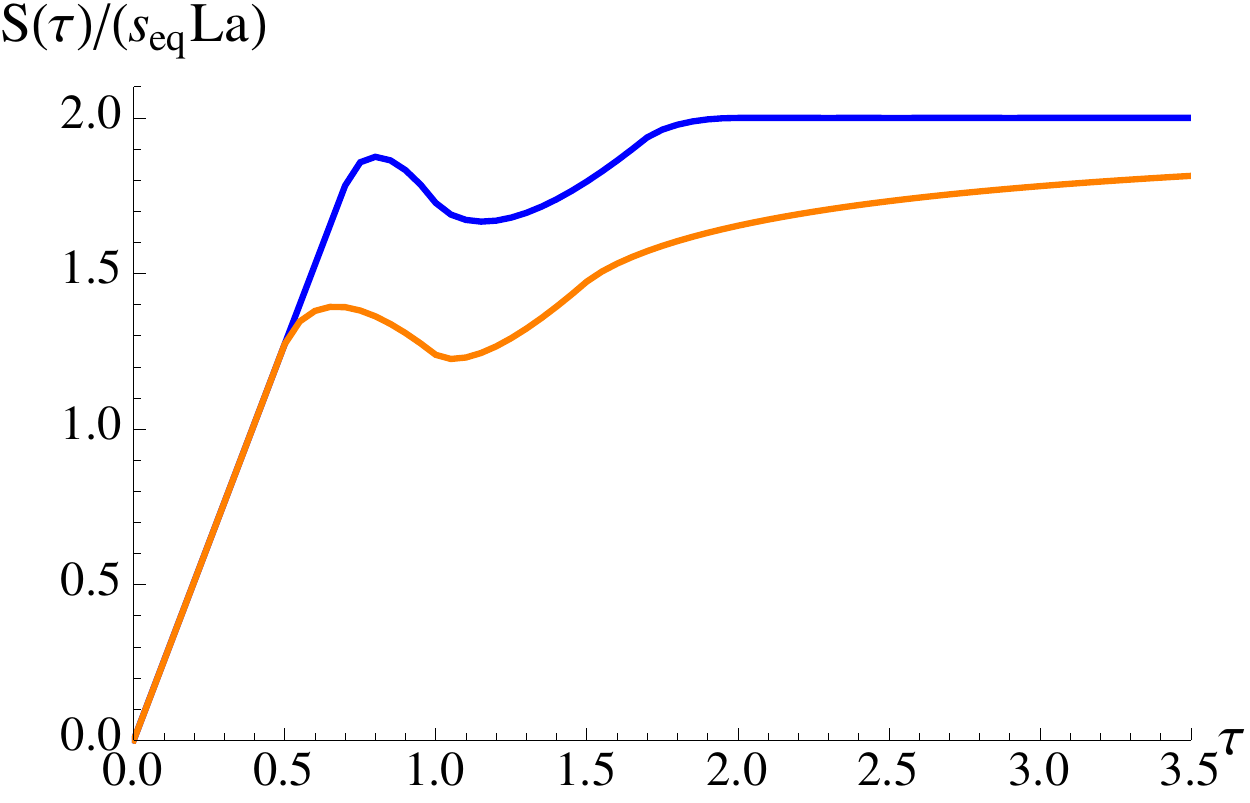}
\caption{ Full time evolution for two disks and strips in $d=3$. The configurations are shown in the top row. The radius of the disks is $R$, and they are separated, by a distance $2R$. The width of the strips is $a$, and they are separated by a distance $l=a$. $\tau$ is the dimensionless time equal to $t/R$ and $t/a$ respectively.
 For the case of two disks we have drawn an EPR pair that contributes to the resonant effect shown in the bottom row. In the bottom row the time evolution is plotted for the RPS (blue) and the EPR (orange) models.  \label{fig:two}}
\end{center}
\end{figure}

We now briefly examine the entanglement entropy of two separated disks and strips. We use the RPS and EPR measures. These examples are complicated enough geometrically to give interesting features in the time evolution: we show on Fig.~\ref{fig:two} that the entanglement growth rate~\eqref{groR} turns negative for them at some intermediate time for the EPR measure.

This phenomenon is the easiest to understand in the case of the two disks. We chose their separation so that at some intermediate time the entropy saturates. However, at some later time some of the quasiparticle pairs that originated from between the two disks do not give entanglement with the outside, as one can ends up in one disk, the other in the other disk. Because the system has already reached the saturation value for the entropy at intermediate times, the entropy must go down. This resonant effect only last for a finite time, and finally we get complete saturation. We discuss aspects of saturation in Appendix~\ref{sec:Saturation}. Similar discussions can be found in~\cite{Calabrese:2005in,Asplund:2013zba,Leichenauer:2015xra,Asplund:2015eha}.

We included the example of two strips, as it also displays negative entanglement growth rate, and this geometry will also play a role in our discussion of mutual information in Appendix~\ref{sec:quD}. In Fig.~\ref{fig:two} we show a setup, where both the EPR and RPS measures exhibit a dip in the entanglement entropy.

\subsection{Entanglement density }

For the case of EPR pairs we can introduce a local, and thus more refined, measure of entanglement: entanglement density. 
The entanglement density $\rho_\Sig (\vec x, t)$ at a given point $\vx \in \sA$ inside $\Sig$ is defined as $\nu$ times 
the density of quasiparticles whose entangled partners lie outside $\Sig$. Recall that $\nu$ was introduced below~\eqref{endi}. It then immediately follows that 
\be 
S_\Sig (t) = \int_{\sA} d^{d-1} x \, \rho_{\Sig} (\vec x, t) \ .
\ee

$\rho_{\Sig} (\vx, t)$ can be readily worked out as follows. In the EPR example, two point $\vx, \vec y$ are  are only entangled for one moment, 
when their distance is exactly $2t$.  One can then introduce an entanglement ``correlation function'' 
\be
s(\vx,\vec y)={s\ov (2t)^{d-2}\,\om_{d-2}}\, \delta\le(\abs{\vec{x}-\vec{y}}-2t\ri)\,.
\ee
The normalization of the above function is determined by requiring 
\be 
\int d^{d-1} y \, s(\vx, \vec y) = s = n \nu \,,
\ee
which is the total amount of entanglement $\vx$ has with the full space. 
The entanglement density $\rho_\Sig (\vx, t)$ can then be obtained by integrating the above expression over all $\vec y$ that lies 
in $\bar \sA$, i.e.~the region outside $\Sig$,  
\be
\rho_{\Sig} (\vec x, t) = \int_{\bar \sA}  d^{d-1} y \, s( \vx, \vec y) =
{s\ov (2t)^{d-2}\,\om_{d-2}}\,\int_{\bar \sA} dy \ \delta\le(\abs{\vec{x}-\vec{y}}-2t\ri)\,. \label{EEdensity}
\ee
The above expression can be easily described in words: draw a sphere of radius $2t$ around $\vx$, and calculate the portion of the surface that falls outside $\Sig$. The simple intuitive picture this is that we are counting the quasiparticles at $\vx$ that have their partners outside, which all lie on the sphere of radius $2t$ drawn around the point.  

\begin{center}
\begin{figure}[!h]
\includegraphics[scale=0.8]{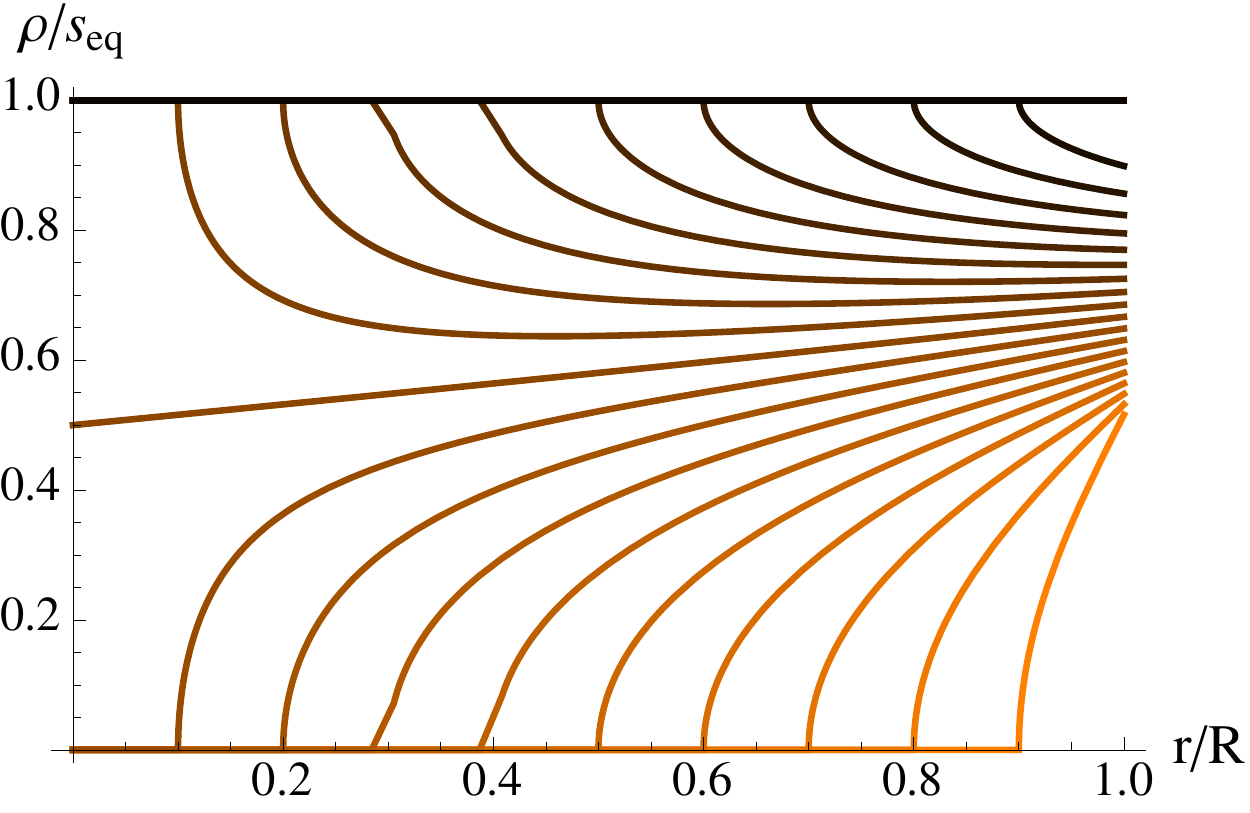}
\caption{Entanglement density as a function of radial distance. Different lines show to the density profiles at different times, and later times correspond to darker colors. The straight  line in the middle is for $t = R/2$. The curves below $t =R/2$ are for $t < R/2$ with time increasing from right to left. The  curves above $t =R/2$ are for $t > R/2$ with time increasing from left to right.
\label{fig:qpdensity}}
\end{figure}
\end{center}

For $\Sig$ a sphere, the integral in~\eqref{EEdensity} can be readily performed. Actually, we have already performed this calculation in Appendix~\ref{sec:Fulltime}, so all we have to do is to replace $t\to 2t$ in~\eqref{xiDef}. For $d=3$ we get 
\be
\rho(r)=s\, \begin{cases}
{\arccos\le(R^2 - r^2 - 4 t^2\ov 4 r t\ri)\ov \pi} & (\abs{R-2t}<r<R)\\
0 & (0<r<R-2t)\\
1&  (0<r<2t-R)\,, 
 \end{cases}
 \label{DiskDensity}
\ee
where the last two cases can only happen before and after $t=R/2$.
The entanglement entropy saturates when $\rho (r)$ reaches $s$ everywhere inside the sphere. 
It is curious that at the center of the sphere $r=0$, the density jumps from $\rho (r=0) =0$ to saturation value $s$ 
at $t = R/2$. We plot~\eqref{DiskDensity} for various times in Fig.~\ref{fig:qpdensity}.

The entanglement density can be used to give a precise definition of the entanglement tsunami introduced in~\cite{Liu:2013iza}: 
we define the tsunami wave front as the boundary between regions of $\rho=0$ and $\rho \neq 0$. 
From~\eqref{EEdensity}, one can immediately conclude that under such a definition, locally the wave front should progress at 
the speed of $2$, i.e.~twice the speed of light. For $\Sig$ being a sphere, the wave front can be visualized from the lower half of the plots 
in Fig.~\ref{fig:qpdensity}, with the wave front reaching $r=0$ at $t = R/2$. Note that when the wave comes in, the region covered by
the wave is only partially entangled with outside, i.e.~with a value smaller than the equilibrium value $s$. The curves in the upper half  
in Fig.~\ref{fig:qpdensity} suggest there is a ``reflected wave'', whose wave front can be defined as the boundary between the region which has reached the equilibrium value $s$ and the region which has not. This reflected wave starts at $ t=R/2$ from the center and moves at speed $2$ outwards, reaching $\Sig$ at $t=R$.  In this example, the linear growth~\eqref{linel} and the associated $v_E$ can be considered
as an average effect. The picture here is very different from that proposed in~\cite{Liu:2013iza} for strongly coupled systems, where
the region covered by the tsunami wave will already have reached their equilibrium value. The difference may be due to that in free theory as we are considering here, there is actually no dynamical process of equilibration. We should also keep in mind, as we will also elaborate below, that generically there is no unambiguous definition for an entanglement density.

\begin{figure}[!h]
\begin{center}
\includegraphics[scale=0.5]{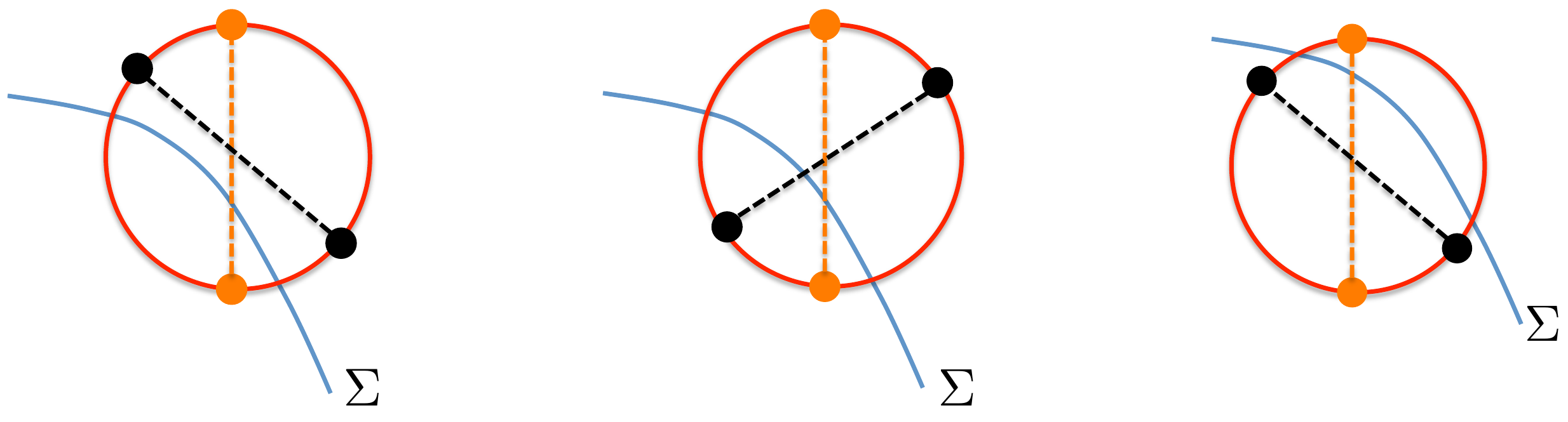}
\caption{Explanation of why the entanglement density cannot be defined for GHZ blocks. In the figure we take $m=2$.
For the left plot we could localize the entanglement to the orange point inside $\Sig$, but in situations depicted in the middle and right 
plots, the entanglement among the block can no longer be localized to the orange point. 
  \label{fig:unloc}}
\end{center}
\end{figure}

The entanglement density discussed above is specific to the EPR example. When there is multipartite entanglement, such a definition 
does not appear to exist even assuming a quasiparticle picture. The basic reason is that with multipartite entanglement, one could no longer 
localize entanglement to a point. This can be readily seen from the GHZ example illustrated in Fig.~\ref{fig:unloc}. We discuss how interactions change the perspective on the tsunami wave front in Sec.~\ref{sec:OnedimInteracting}.

\subsection{Saturation time}\label{sec:Saturation}

In Sec.~\ref{sec:Evolution} we showed that the entanglement entropy has an equilibrium value 
$S_{\rm eq} = s V_\Sig$. In Appendix~\ref{sec:Fulltime} we saw that for a sphere of radius $R$, the EPR model has a finite saturation time 
given by $t_s = R$, while for GHZ block example the saturation time is infinite. In this subsection we make some general remarks 
on the saturation time for general shapes.

\begin{figure}[!h]
\begin{center}
\includegraphics[scale=0.5]{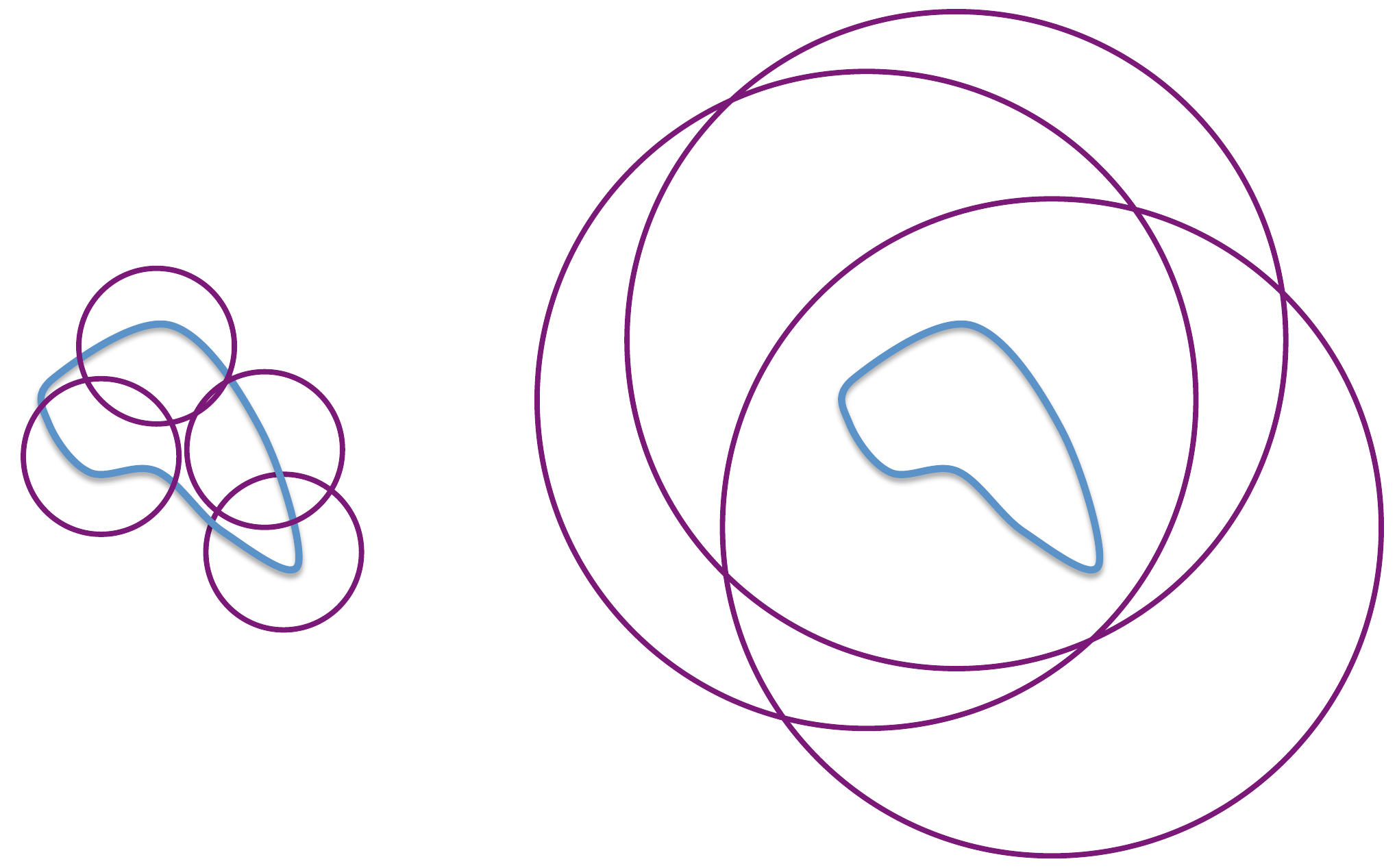}
\caption{The figure on the left is for $t<t_s$, the right one for $t>t_s$.  The purple circles are circles of radius $2t$ centered at some point 
in $\Sig$. At time $t$, all quasiparticles which entangle with those at a point $\vx$ lie on such a circle centered at $\vx$.  
For $t > t_s$ all such circles completely lie outside of $\Sig$. 
\label{shape}}
\end{center}
\end{figure}

Let us first consider the EPR example. From~\eqref{EEdensity} we can immediately conclude that for any $\Sig$ the saturation time $t_s$ equals half of the largest distance between two points on $\Sig$. If $t>t_s$ the entanglement density at any point inside $\Sig$ is $\rho(x)=s$, because the Dirac delta in~\eqref{EEdensity} has support on a circle of radius $2t$ around $\vx$, which now completely lies outside $\Sig$; 
see Fig.~\ref{shape}. Phrased slightly differently, at time $t$, all quasiparticles which entangle with those at $\vx$ lie on the circle of 
radius $2t$ centered at $\vx$, and the total number of them is $n$, the particle density at $\vx$ (which is in fact the same everywhere
due to homogeneity). When this circle lies completely outside $\Sig$, all of them contribute to $S_\Sig$ and the entropy does not change with time. The saturation time is thus $t_s=R$ for spherical $\Sig$ and  $t_s=\infty$ for any non-compact region. In particular, $t_s$ is infinite for a strip.

The RPS measure gives a rather different saturation behavior from the EPR model as evidenced by the finite saturation time even for a non-compact shape; for the example of the strip see Fig.~\ref{fig:strip3d}. The criterion for saturation in the RPS model is that after the saturation time $t_s$ there should not exist any light cone, whose intersection with region $\sA$ has a normalized area grater than $1/2$. 

The results of the EPR and RPS measures should be contrasted with holographic systems, for which $t_s$ is finite for a strip and given by 
\be 
t_s = {R \ov v_E^{\rm holo}}\,,
\ee
where $2R$ is the width of the strip and $v_E$ is the ``tsunami velocity'' which appears in~\eqref{emrp}.
For holographic systems, the saturation time for a 
sphere is given by 
\be 
t_s = {R \ov c_E} > R, \qquad c_E = \sqrt{d \ov 2 (d-1)} < 1 \quad {\rm for} \quad d>2\,,
\ee 
where we have quoted the value of $c_E$ for a neutral system. We thus see that for a strip, holographic systems saturate faster than the 
EPR model ($t_s=\infty$)  and the  RPS model ($t_s=2R$, as can be calculated from the above criterion or read from Fig.~\ref{fig:strip3d}), while for a sphere holographic systems saturate slower than the free streaming models that have $t_s=R$. Note that the velocity $c_E$ has also appeared in~\cite{Roberts:2014isa} as the ``expansion'' velocity 
of the the time evolution of a local operator in a thermal state.

For a general measure from our discussion in Sec.~\ref{sec:eqv} we expect that in the $t \to \infty$ limit, there are generically 
$1/t$ corrections to the leading behavior~\eqref{keke}. This implies that the saturation time is generically infinite. As an example, let us consider 
$\Sig$ a sphere, for which case the shell in Fig.~\ref{fig:LateTime} is also spherical. One can then compute 
the subleading corrections using~\eqref{zdef}  by expanding $\mu_{\rm cap}(\xi)$ to higher orders in $\xi$,
\be \label{eeoo}
\mu_{\rm cap}(\xi)=s_\text{eq}\,\xi+a_2\, \xi^2+a_3 \, \xi^3+\dots\,, 
\ee
where we used~\eqref{hjek}. We will not go into details here, except to mention that whenever the nonlinear term in~\eqref{eeoo}
are non-vanishing one gets subleading corrections in $1/t$ of the form,
\be
S(t)=s_\text{eq} V_{\Sig}+\sum_n {\# \, a_n \ov t^{(n-1)(d-2)}}\,.
\ee
Thus all measures for which there are nonlinear terms in~\eqref{eeoo} have infinite saturation time.

\subsection{Finite volume effects}

So far our discussion assumed the system has an infinite volume. If the system has a finite volume,  the large time 
behavior of the entanglement entropy will be modified when carriers of entanglement can explore the whole volume.

\subsubsection{EPR pairs and GHZ blocks}

Let us first consider the EPR and GHZ examples which can be treated in a unified manner. 
For a system with a finite volume, after a long time there will be no correlation between the positions of the particles that originated from the same point. (This assumptions should hold true except for resonant situations in special geometries.) Then we have a constant density $n$ of quasiparticles, and a total of $n V_\text{system}$ of them, randomly distributed. 

We get entanglement except in cases, when quasiparticles originating from one point are all inside or outside $\Sig$, hence
\be
\label{FiniteVolume}
S_\Sig (t \to \infty) ={s\ov 2m} V_\text{system}\le[1- \le(V_\Sig\ov V_\text{system}\ri)^{2m}-\le(1-{V_\Sig\ov V_\text{system}}\ri)^{2m}\ri]\,.
\ee
This expression is manifestly symmetric under $V_\Sig\to V_\text{system}- V_\Sig$, hence the requirement that the entropy of a region and its complement is equal in a pure state is satisfied. Another consistency check is that in the infinite $V_\text{system}$ limit we get $S_\Sig=s V_\Sig$.
The maximum of~\eqref{FiniteVolume} is achieved for $V_\Sig/ V_\text{system}=1/2$,
\be
S_\Sig \big\vert_\text{max}={s\ov 2m} V_\text{system}\le[1- {1\ov 2^{2m-1}}\ri]\,.
\ee

In the EPR ($m=1$) case the resulting expression is
\be
\label{FiniteVolumeEPR}
S_\Sig (t \to \infty) ={s \ov V_\text{system}} \, V_\Sig \le(V_\text{system}- V_\Sig\ri)\,.
\ee
The maximum entanglement is $S_\Sig\big\vert_\text{max}={s\ov 4} V_\text{system}$.

\subsubsection{Random pure state measure}

According to the discussion in Sec.~\ref{sec:rps} random pure states are expected to give:
\be
\label{FiniteVolumeRandom}
S_\Sig (t \to \infty)=s  \, \min\le(V_\Sig,V_\text{system}- V_\Sig\ri)\,.
\ee
This result is consistent with~\eqref{ran0}; if we followed the time evolution of light cones in a compact geometry, a light cone would become a curve densely filling the whole volume  $V_\text{system}$. The maximum entropy is reached again at $V_\Sig/ V_\text{system}=1/2$, and its value is $S_\Sig\big\vert_\text{max}={s\ov 2} V_\text{system}$, twice the value of the EPR pair model.  Equation~\eqref{FiniteVolumeRandom} is of the form expected from a holographic system.

\subsection{Mutual Information} \label{sec:quD}

We now consider the qualitative behavior of the time dependence of mutual information $I (A, B)$ between two regions $A$ and $B$. 

\subsubsection{EPR pairs}

In the EPR model a pair contributes to the mutual information $I(A,B) = S(A) + S(B) - S(AB) $ for regions $A$ and $B$, if one member of the pair is inside $A$ and the other is in $B$. Let the smallest distance between the two regions be $L_\text{min}$, and the largest $L_\text{max}$. (If one of the regions is non-compact $L_\text{max}$ is infinite.) Because of back to back propagation, the mutual information is nonzero only for times
\be
{L_\text{min}\ov 2}< t <{L_\text{max}\ov 2}\,.
\ee

If the whole system is compact, the mutual information will tend asymptotically to
\be
I (A, B) = {2s\ov V_\text{system}} \, V_A\, V_B\,,
\ee
where we have used~\eqref{FiniteVolumeEPR}.

An interesting feature of the EPR example is that the corresponding mutual information 
is extensive, i.e.~
\be \label{ecb}
I_3^\text{EPR}(A,B,C)=0 , \quad {\rm or} \quad I_{\rm EPR} (A, B) + I_{\rm EPR} (A, C) = I_{\rm EPR} (A, BC) \,,
\ee
where the tripartite information is defined by
\bea
I_3(A,B,C)&\equiv &I(A,B)+I(A,C)-I(A,BC) \nn
&=&S(A)+S(B)+S(C)-S(AB)-S(BC)-S(AC)+S(ABC)\ . \label{I3sym}
\eea
To see this let us consider the contribution of an EPR pair to $I_3 (A, B,C)$ for some regions $A,B,C$. Without loss of generality let us
assume that neither of the quasiparticles is in $C$. This implies
\bea
S(C)=0, 
\qquad S(AC)= S(A) \qquad S(BC)=S(B) \qquad S(ABC)=S(AB) \,,
\eea
from which~\eqref{ecb} immediately follows.

\subsubsection{GHZ blocks}

For GHZ,  $I(A,B)$ becomes non-zero at the time ${L_\text{min}\ov 2}$, as in the EPR case. Here, however, for any time we can always find a light cone that intersects both $A$ and $B$, and we can put some quasiparticles on these intersections without violating momentum conservation.
This leads to a non-vanishing $I(A,B)$ for all times. As $t \to \infty$, as the normalized volume of intersection of a light cone with $A$ and $B$ 
will necessarily go to zero, therefore $I(A,B)$ will asymptote to zero. This phenomenon is reminiscent of the differences in saturation between the GHZ and EPR cases for the entanglement of a single region.

Let us now examine the property of $I_3$ for GHZ. Motivated by the above discussion of the EPR case, we note that in order to get a possibly nonzero $I_3$  in the GHZ case, we need some number of particles in all three regions. From the property that tracing out any number of particles leads to the same amount of entanglement, we conclude that
\be
S(A)=S(B)=S(C)=S(AB)=S(BC)=S(AC)\,. \label{everything}
\ee
From~\eqref{I3sym} this implies
\be
I_3(A,B,C)=S(ABC)\geq0\,.
\ee
$S(ABC)=0$, if there are no particles outside $A,B,C$, while $S(ABC)\neq0$ (and equal to all the rest in~\eqref{everything}), if there are particles in $A,B,C$ and outside of these three regions.

Note that GHZ states are very special, choosing a different multipartite entanglement pattern for the particles would generically lead to a non-definite sign of $I_3$.

\subsubsection{Random pure state measure}

Consider a light cone which intersects with $A$ and $B$, and denote the area of its intersections as $\om_{A,B}$ respectively.
From~\eqref{ran0}, assuming $\om_A>\om_B$, the contribution from this light cone to $I(A,B)$ is:
\bea
I(A,B) & \supset & 2 s \le(\theta\le(\om_A-\ha \om_{d-1}\ri)\, \om_B \ri. \cr
 && \le. + \theta \le(\om_A+\om_B -\ha \om_{d-1} \ri)
\theta\le(\frac{1}{2}\om_{d-1}-\om_A \ri)
\le[\om_A+\om_B-\ha \om_{d-1} \ri] \ri)\,.
\label{RandomMutInf}
\eea
From this expression one can check explicitly that the mutual information is non-extensive and monogamous, i.e.~
\be
I_3 (A, B, C) \leq 0, \quad {\rm or} \quad I (A,B)+I (A,C) \leq I (A,BC)\, . \label{moni}
\ee

In contrast to previous two examples, $I (A,B)$ is nonzero, if there exists a light cone, which intersects both $A$ and $B$, and more than half of its area is inside $AB$. This implies that the mutual information stays zero for all times, if the separation  of the regions is too large compared to their sizes.  Also for large times we always get zero mutual information for compact regions. These results  are reminiscent of holographic examples: holographic mutual information is monogamous~\cite{Hayden:2011ag}, and the time dependence of mutual information shows similar qualitative behavior in holography, see e.g.~\cite{Balasubramanian:2011at}.

In finite volume we  can determine the saturation value of $I(A,B)$ from~\eqref{FiniteVolumeRandom}. For $V_A>V_B$ we get
\bea
I(A,B) &=& 2 s \le(\theta\le(\om_A-\ha V_\text{system}\ri)\, \om_B \ri. \cr
&& \le.  + \theta \le(V_A+V_B -\ha V_\text{system} \ri)
\theta\le(\frac{1}{2}V_\text{system}-V_A \ri)
\le[V_A+V_B-\ha V_\text{system}\ri] \ri)
\label{RandomMutInf1}
\eea
which has the same form as~\eqref{RandomMutInf}. 

\subsubsection{Example of two strips in $d=3$}

\begin{figure}[!h]
\begin{center}
\includegraphics[scale=0.62]{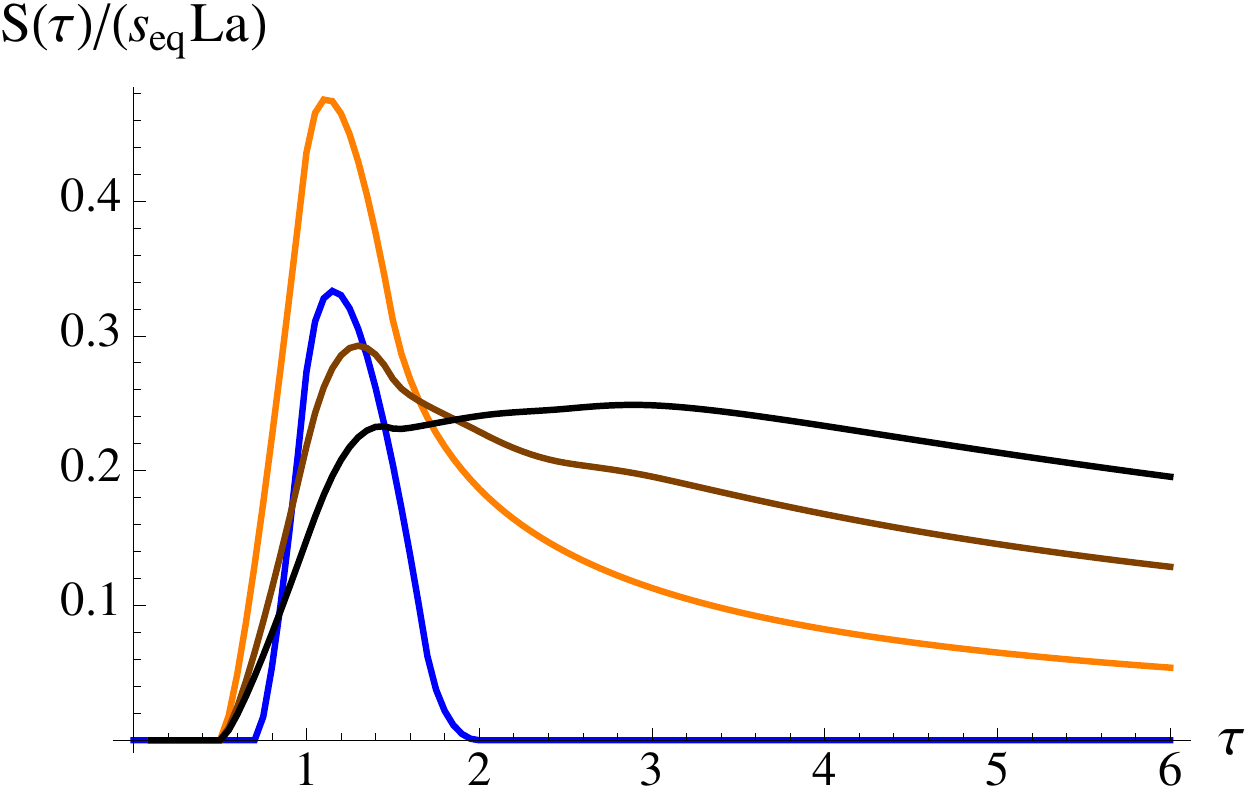}\hspace{0.5cm}
\includegraphics[scale=0.62]{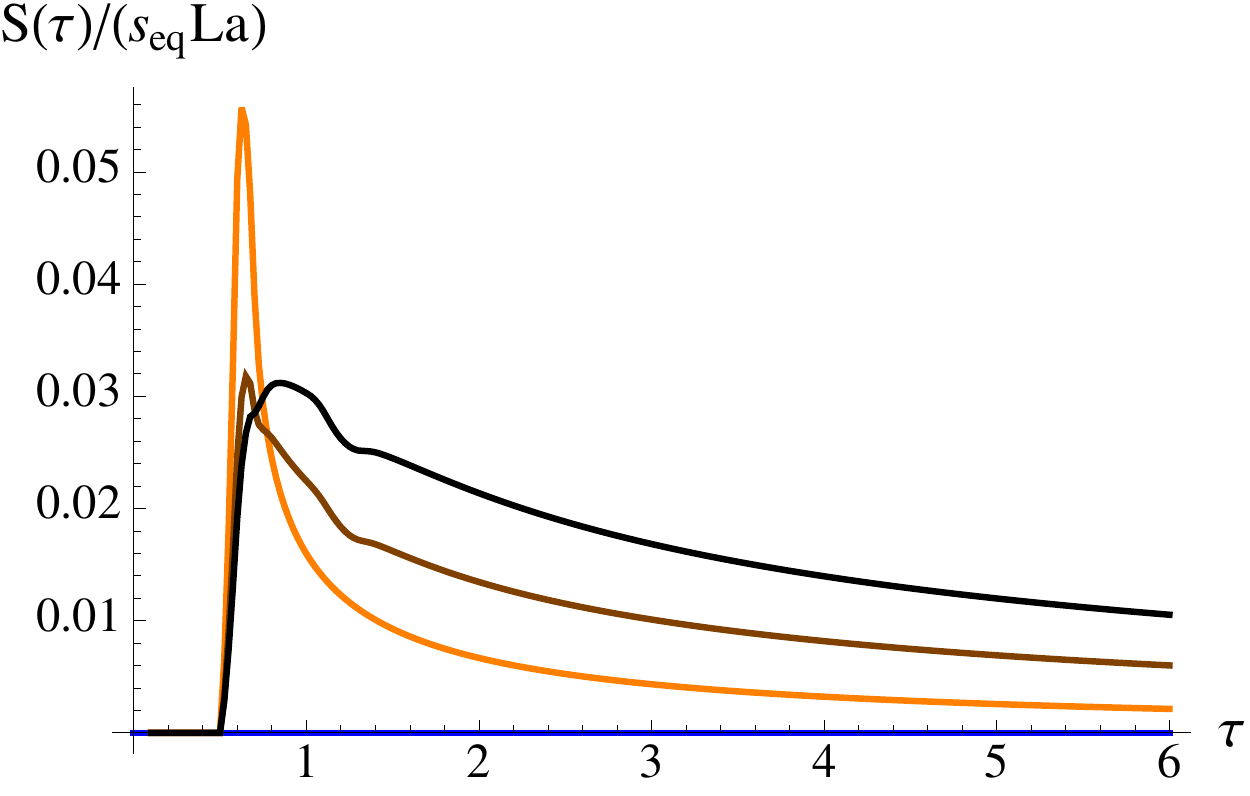}
\caption{Time evolution of mutual information divided $s_{\rm eq}$ and per unit length for two strips of width $a$ separated by a distance $l$ in $d=3$. The left panel is for $a/l=1$, while the right panel is for $a/l=0.2$. The blue curve is the RPS model, and the GHZ curves for $m=1,2,3$ are plotted in increasingly darker shades of orange. Note that for $a/l=0.2$ (right panel) no mutual information is generated for the RPS model. 
 \label{mutuals}}
\end{center}
\end{figure}

As an example consider the mutual information for two parallel strips $A$ and $B$ of width $a$ and length $L$, separated by a 
distance $l$ in $d=3$. In this case the RPS model will have zero mutual 
information unless $\frac{a}{l}\leq \frac{\sqrt{2}-1}{2}\simeq 0.207$. If this holds the mutual information will be non zero starting 
at $t=l/\sqrt{2}$, in contrast to the GHZ and EPR models which have non zero mutual information starting at $t=l/2$ for all 
values of $a/l$. In these latter models the mutual information decays asymptotically to zero with time, while in the RPS model 
is already zero for a finite time. This is illustrated in figure \ref{mutuals} where we plot the numerical evaluation of mutual information
$I(A,B)/(s_{\rm eq}La)$ as a function of time, $\tau=t/a$ for these models. Note that the higher $m$ GHZ models produce slower decay of mutual information, just as they give slower saturation for the entanglement entropy for one strip shown on Fig.~\ref{fig:strip3d}.    

\section{Explicit expressions for entanglement entropies after one scattering event} \label{app:a}

Here we give explicit expressions for $S_{13}^{(f)} $ and $S_{3}^{(f)}$ that we used in the discussion of the effect of a scattering event in Sec.~\ref{sec:singlescat}.  

For $S_{13}$ we have 
\be 
S_{13}^{(f)} = - p_1 \log p_1 - (p_\al -p_1) \log (p_\al-p_1) - p_2 \log p_2 - (1-p_\al- p_2) \log (1-p_\al - p_2)\,,
\ee
where 
\be 
p_1 = {x_1 + x_2 |\al|^2  \ov (1 + |\al|^2)^2} , \qquad p_\al = {1 \ov 1 + |\al|^2}\,, \qquad
p_2 = |\al|^2 {x_3 + x_4 |\al|^2  \ov (1 + |\al|^2)^2}\,,
\ee
and 
\be 
x_1 = |U_{11}|^2 + |U_{13}|^2 , \quad x_2 = |U_{21}|^2 + |U_{23}|^2 , \quad x_3 = |U_{31}|^2 + |U_{33}|^2\,, \quad
x_4 = |U_{41}|^2 + |U_{43}|^2  \,.
\ee
Note that 
\be 
0 \leq x_1, x_2, x_3, x_4 \leq 1, \qquad x_1 + x_2 + x_3 + x_4 = 2 \ .
\ee
For $|\al| =1$ we explained in the main text that get a decrease in entropy. Indeed the above formulas simplify to
\be
S_{13}^{(f)} = -2\le[p_1 \log p_1 + \le(\ha-p_1\ri) \log \le(\ha-p_1\ri)\ri]\leq2\log2=S_{13}^{(i)}\,,
\ee
where we used that in this case $p_1\leq1/2$.

For $S_3^{(f)}$ we have 
\be 
S_3^{(f)} = - p_f \log p_f- (1-p_f) \log (1-p_f) 
\ee
where 
\be 
p_f = p_i + {1 - |\al|^2 \ov (1 + |\al|^2)^2} (x_1 - x_4 |\al|^2) , \quad 
p_i = {|\al|^2 \ov (1 + |\al|^2)^2}  
\ee
and 
\be 
0 \leq x_1 = |U_{11}|^2 + |U_{13}|^2 \leq 1, \qquad 0 \leq x_4 = |U_{41}|^2 + |U_{43}|^2 \leq 1 \ .
\ee
Clearly for $|\al| =1$, $S_3^{(f)}$ reduces to $S_3^{(i)}$, while for other values of $\al$, one can always find values of $x_1$ and $x_4$ which either reduce or enhance it.

\end{document}